\documentclass[aps,prx,reprint,amsmath,amssymb,longbibliography,nofootinbib,superscriptaddress]{revtex4-2}
\usepackage[T1]{fontenc}
\usepackage[utf8]{inputenc}
\usepackage{mathtools,amsthm}
\usepackage{xcolor}
\usepackage[hidelinks]{hyperref}
\makeatletter
\let\frontmatter@footnote@produce\frontmatter@footnote@produce@footnote
\makeatother
\DeclareMathOperator{\Tr}{Tr}
\DeclareMathOperator{\Stab}{Stab}
\newcommand{\STAB}{\mathrm{STAB}}

\newcommand{\CSP}{\mathrm{CSP}}
\newcommand{\CPWP}{\mathrm{CPWP}}
\newcommand{\FW}{\mathcal F_W}
\newcommand{\id}{\operatorname{id}}
\newcommand{\Strange}{\mathbb S}
\newcommand{\ket}[1]{\lvert#1\rangle}
\newcommand{\bra}[1]{\langle#1\rvert}
\theoremstyle{definition}
\newtheorem{definition}{Definition}
\theoremstyle{plain}
\newtheorem{theorem}{Theorem}
\newtheorem{lemma}{Lemma}
\newtheorem{proposition}{Proposition}
\newtheorem{corollary}{Corollary}
\begin{document}

\title{{Magic Manipulation of Random States: From Irreversibility to Reversibility}}
\author{Jicun Li}
\email{lijicun@mail.ustc.edu.cn}
\affiliation{School of Computer Science and Technology, University of Science and Technology of China, Hefei 230027, China}
\author{Wei Xie}
\thanks{\href{mailto:xxieww@ustc.edu.cn}{xxieww@ustc.edu.cn}; Corresponding author}
\affiliation{School of Computer Science and Technology, University of Science and Technology of China, Hefei 230027, China}
\affiliation{Hefei National Laboratory, University of Science and Technology of China, Hefei 230088, China}
\author{Jun Wu}
\affiliation{School of Software Engineering, University of Science and Technology of China, Hefei 230051, China}
\affiliation{Suzhou Institute for Advanced Research, University of Science and Technology of China, Suzhou 215123, China}
\author{Honglin Chen}
\affiliation{School of Computer Science and Technology, University of Science and Technology of China, Hefei 230027, China}
\author{Xiang-Yang Li}
\thanks{\href{mailto:xiangyangli@ustc.edu.cn}{xiangyangli@ustc.edu.cn}; Corresponding author}
\affiliation{School of Computer Science and Technology, University of Science and Technology of China, Hefei 230027, China}
\affiliation{Hefei National Laboratory, University of Science and Technology of China, Hefei 230088, China}

\begin{abstract}
Reversibility means that the resources required for state formation can be fully recovered through distillation, providing a foundation for a unified, thermodynamic description of state conversion. Magic-state manipulation can be irreversible, raising a fundamental question: is irreversibility typical or confined to exceptional states, and how does it depend on the allowed operations? Here we quantify the degree of reversibility using the reversibility ratio, which is defined as the distillation yield divided by the formation cost. We investigate this ratio for random mixed states generated from Haar-random pure states and polynomial-size local random circuits, using the Strange state as the reference resource. We identify a regime in which this ratio approaches one under completely positive-Wigner-preserving operations but vanishes under completely stabilizer-preserving operations. In this reversible regime, conversion rates between typical states are asymptotically given by ratios of their Rényi-2 entropy deficits. These results reveal emergent reversibility in typical mixed state magic and highlight its dependence on operational constraints.
\end{abstract}
\maketitle

\section{Introduction}
\label{sec:introduction}

Magic is a key resource for fault-tolerant quantum computing~\cite{Shor1996,CampbellTerhalVuillot2017}, extending efficiently classically simulable stabilizer computation~\cite{AaronsonGottesman2004,Veitch2012} to universality~\cite{GottesmanChuang1999,BravyiKitaev2005}.
The preparation and consumption of magic resources are important to  many fault-tolerant computing schemes, making their efficient use a central practical concern~\cite{BravyiHaah2012,HowardCampbell2017}.
Measures of magic also help quantify the computational cost of classical simulation methods~\cite{Pashayan2015,HowardCampbell2017}.
These connections motivate the resource theory of magic, which provides a framework for quantifying magic and understanding how it can be manipulated under restricted operations~\cite{Veitch2014,HowardCampbell2017}.

Two widely studied approaches to magic resource theory are distinguished by their choice of free states.
The first is based on the stabilizer formalism, which underlies many schemes for fault-tolerant quantum computation~\cite{Gottesman1996,Gottesman1998FT}.
It regards convex mixtures of stabilizer states as free and identifies magic with nonstabilizerness~\cite{Veitch2014}.
The second, applicable to odd-prime-dimensional systems, is motivated by the classical probabilistic interpretation of nonnegative discrete Wigner representations~\cite{Galvao2005,Veitch2012,MariEisert2012}.
It takes states with nonnegative Wigner functions as free and identifies magic with Wigner negativity~\cite{Gross2006,Veitch2012}.
Both approaches admit classically simulable models~\cite{AaronsonGottesman2004,MariEisert2012}.
Although the two definitions agree on which pure states are free, they differ for mixed states: a state with a nonnegative Wigner function need not be a mixture of stabilizer states~\cite{Gross2006,Veitch2012}.

There are two central operational tasks in magic resource theory: formation and distillation.
Magic-state formation prepares a desired state from standard pure magic states~\cite{Veitch2014,Takagi2022}, whereas magic-state distillation extracts standard pure magic states from noisy resources~\cite{BravyiKitaev2005}.
{Magic-state distillation has been extensively studied~\cite{BravyiHaah2012,CampbellAnwarBrowne2012,HastingsHaah2018,Wang2020,Prakash2020,WillsHsiehYamasaki2025}, alongside complementary work on magic-state dilution and resource formation costs~\cite{CampbellOGorman2016,LiuBuTakagi2019,Beverland2020,Takagi2022}.}
{A central question in quantum resource theories is whether
the resources invested in state formation can be fully
recovered through distillation in the asymptotic limit,
under the same free operations~\cite{Bennett1996,VidalCirac2001,Yang2005,BrandaoPlenio2008,BrandaoPlenio2010Reversibility,BrandaoPlenio2010Stein,BrandaoGour2015,Berta2023Gap,Lami2025Stein,HayashiYamasaki2025Stein}.}

A reversible framework for magic-state manipulation
can be obtained by allowing nonphysical maps that need not
be positive~\cite{ChenGourWang2025}.
However, the resource theory of magic is irreversible under physical free operations~\cite{RegulaLami2024}.
This phenomenon motivates two questions:
Is irreversibility a typical feature of magic, or is it confined to exceptional states?
Do the resource theories of nonstabilizerness and Wigner negativity differ in their reversibility properties?

Random quantum states and random circuits provide tractable models for understanding entanglement, thermalization, and information scrambling in quantum many-body systems~\cite{Page1993,Popescu2006,Nahum2018}.
{Previous studies have quantified the mana and other magic measures
of random states~\cite{WilsonWhite2020,LiuWinter2022,LeoneOlivieroHamma2022,TurkeshiDymarskySierant2025,Iannotti2025,Szombathy2025},
investigated the presence and detectability of magic in random and
noisy mixed states~\cite{LiuLiu2026,HaugTarabunga2026},
and characterized its generation, saturation, and spreading in
random circuits~\cite{TurkeshiTirritoSierant2025,HaugAolitaKim2025,ZhangGu2026}.}
{However, the asymptotic formation, distillation, and reversibility
of magic in typical mixed states remain largely unexplored.}
Here, we address this gap in two complementary settings.
The first consists of Haar-induced states, obtained by tracing out an environment from a globally Haar-random pure state, providing a statistical benchmark for magic retained under partial access to a randomized quantum system~\cite{Zyczkowski2001}.
The second consists of reduced states generated by polynomial-size local random circuits acting on an initial product state, connecting the same operational questions to spatially local dynamics and efficient quantum preparation~\cite{Brandao2016}. For sufficiently long circuits, the resulting ensemble closely approximates the Haar-induced ensemble in moments up to polynomially growing order~\cite{Brandao2016}.

We now summarize our main results. Let $C_{\CSP},D_{\CSP}$ and $C_{\CPWP},D_{\CPWP}$ denote the formation costs and distillation yields under completely stabilizer-preserving (CSP) and completely positive-Wigner-preserving (CPWP) operations, measured in units of the qutrit Strange state.
We compare their {reversibility ratios}, defined as the yield-to-cost ratios ~\cite{HorodeckiSen2003,Horodecki2010Symmetry,GuOlivieroLeone2025,LeoneRizzoEisertJerbi2025} $\eta_{\CSP}:=D_{\CSP}/C_{\CSP}$ and $\eta_{\CPWP}:=D_{\CPWP}/C_{\CPWP}$.
For an $N$-qutrit system and an $(N-\ell_N)$-qutrit environment,
both ensembles exhibit
\begin{equation}
	\eta_{\CSP}\xrightarrow{\Pr}0,
	\qquad
	\eta_{\CPWP}\xrightarrow{\Pr}1
	\label{eq:introduction-recycling}
\end{equation}
when $1\ll\ell_N\ll N$.
For $\ell_N=\Theta(N)$ with {$\ell_N\leq N$},
$\eta_{\CPWP}\xrightarrow{\Pr}1$, whereas
$\eta_{\CSP}\leq1-\varepsilon$ with probability tending to one
for some constant $\varepsilon>0$ independent of $N$.
For bounded $\ell_N$, $\eta_{\CSP}\xrightarrow{\Pr}0$,
while $\eta_{\CPWP}$ remains bounded away from zero
with probability tending to one.
Our results identify a regime of typical {asymptotic reversibility} under CPWP operations despite the absence of universal reversibility.
They reveal that the distinction between nonstabilizerness and Wigner negativity has direct operational consequences, with CPWP operations activating magic recycling.
They further establish the qutrit Strange state as a common resource currency for this typical reversible regime.

The paper is organized as follows.
Section~\ref{sec:setting} introduces the operational framework
and presents the main results.
Sections~\ref{sec:csp} and~\ref{sec:cpwp} establish general
CSP converse bounds and CPWP achievability bounds, respectively.
Section~\ref{sec:random-bounds} develops the random-state estimates
and derives the resulting formation and distillation bounds.
\section{Operational setting and main results}
\label{sec:setting}

For a state $\rho$ on a $d$-dimensional system, we write $S(\rho)=-\Tr(\rho\log_2\rho)$ and $S_2(\rho)=-\log_2\Tr\rho^2$ for its von Neumann and R\'enyi-2 entropies, respectively. The corresponding entropy deficits are $\Delta(\rho)=\log_2d-S(\rho)$ and $\Delta_2(\rho)=\log_2d-S_2(\rho)=\log_2(d\Tr\rho^2)$.

{For a protocol, we use $D_{\rm in}$ and $D_{\rm out}$ to denote
the dimensions of its input and output Hilbert spaces, respectively.}

For the conversion rates measured in units of the Strange state,
we use the constant $\alpha:=\ln3/[2\ln(5/3)]$.

We use $O(\cdot)$, $o(\cdot)$, and $\Omega(\cdot)$ for the standard asymptotic orders. For random sequences, $\xrightarrow{\Pr}$ denotes convergence in probability as $N\to\infty$, and $X_N=o_{\Pr}(a_N)$ means $X_N/a_N\xrightarrow{\Pr}0$ for deterministic $a_N>0$. We write $1\ll\ell_N\ll N$ when $\ell_N\to\infty$ and $\ell_N/N\to0$.

{For a positive integer $q$, let $\mathfrak S_q$ denote the
permutation group on $\{1,\ldots,q\}$, and write
$(x)_q=x(x+1)\cdots(x+q-1)$,
with $(x)_0=1$.}

\subsection{Free operations and conversion tasks}

We consider two resource theories of magic on qutrit systems. The first treats nonstabilizerness as the resource. The second treats Wigner negativity as the resource.

For a single qutrit, let ${\omega_3}=e^{2\pi i/3}$ and define the generalized Pauli operators by
\begin{equation}
 X\ket j=\ket{j+1\bmod3},\qquad
 Z\ket j={\omega_3}^j\ket j.
 \label{eq:qutrit-pauli}
\end{equation}
On $L$ qutrits, the generalized Pauli group is
\[
\mathcal P_L
=
\left\{
\omega_3^c\bigotimes_{j=1}^{L}(X^{a_j}Z^{b_j})
:\ c,a_j,b_j\in\{0,1,2\}
\right\}.
\]
Its elements are called Pauli operators.

For the first theory, the resource theory of nonstabilizerness,
the free states are convex mixtures of pure stabilizer states~\cite{Veitch2014}. A pure stabilizer state is the unique simultaneous $+1$ eigenstate of $L$ independent commuting elements of $\mathcal P_L$. Its stabilizer group $G$ is Abelian, has $|G|=3^L$, and contains no nonidentity scalar operator. The corresponding rank-one projector is
\begin{equation}
 \ket s\!\bra s=\frac1{3^L}\sum_{g\in G}g.
 \label{eq:stabilizer-projector}
\end{equation}
Clifford unitaries are the unitary symmetries of the Pauli group, characterized by
\begin{equation}
 U\mathcal P_LU^\dagger=\mathcal P_L.
 \label{eq:clifford-normalizer}
\end{equation}
Equivalently, pure stabilizer states are precisely the states $U\ket0^{\otimes L}$ with $U$ Clifford~\cite{Heimendahl2022}. We write $\STAB$ for their convex hull, the free-state set of resource theory of nonstabilizerness.

We take the free operations in this theory to be completely
stabilizer-preserving (CSP) quantum channels~\cite{SeddonCampbell2019,Heimendahl2022}.

\begin{definition}[CSP operations]
\label{def:csp-operations}
A completely positive and trace-preserving (CPTP) map $\Lambda_{A\to B}$ is \emph{completely stabilizer-preserving} (CSP) if
\begin{equation}
 (\Lambda_{A\to B}\otimes\id_R)(\STAB)
 \subseteq\STAB
 \label{eq:complete-csp}
\end{equation}
for every qutrit reference system $R$.
\end{definition}

For the second theory, the resource theory of Wigner negativity,
the free states are those with nonnegative discrete Wigner functions~\cite{Gross2006,Veitch2012}. We use the tensor-product Gross representation on a system of dimension $d=3^L$, given by $W_X(u)=d^{-1}\Tr(A_uX)$, where $A_u$ are the phase-point operators and $u\in\mathbb F_3^{2L}$~\cite{Gross2006}. The Wigner-positive free-state set is
\begin{equation}
 \FW=\{\sigma\succeq0:\Tr\sigma=1,\ W_\sigma(u)\geq0\ \forall u\}.
 \label{eq:free-wigner}
\end{equation}
This choice is motivated by the classical probabilistic description of circuits with nonnegative Wigner representations~\cite{MariEisert2012,Wang2019}. The corresponding free operations preserve $\FW$ in the presence of an arbitrary reference system.

\begin{definition}[CPWP operations]
\label{def:cpwp-operations}
A CPTP map $\Lambda_{A\to B}$ is \emph{completely positive-Wigner-preserving} (CPWP)~\cite{Wang2019} if
\begin{equation}
 (\Lambda_{A\to B}\otimes\id_R)(\FW)
 \subseteq\FW
 \label{eq:complete-cpwp}
\end{equation}
for every qutrit reference system $R$.
\end{definition}

We refer to the theory with free states $\STAB$ and free operations $\CSP$ as the \emph{resource theory of nonstabilizerness}, and to the theory with free states $\FW$ and free operations $\CPWP$ as the \emph{resource theory of Wigner negativity}.

For each operation class, we consider asymptotic state conversion, formation, and distillation~\cite{Veitch2014,RegulaLami2024}.

\begin{definition}[{Asymptotic conversion and reversibility ratios}]
\label{def:rates}
For a free-operation class $\mathcal O$, the conversion rate $R_{\mathcal O}(\rho\to\sigma)$ is the supremum of $r\geq0$ for which CPTP maps $\Lambda_n\in\mathcal O$ satisfy
\begin{equation}
 \lim_{n\to\infty}\frac12
 \bigl\|\Lambda_n(\rho^{\otimes n})-\sigma^{\otimes\lfloor rn\rfloor}\bigr\|_1=0.
 \label{eq:rate-error}
\end{equation}
{We define the formation cost and distillable yield by}
\begin{equation}
 C_{\mathcal O}^{\phi}(\rho)=\frac1{R_{\mathcal O}(\phi\to\rho)},
 \qquad
 D_{\mathcal O}^{\phi}(\rho)=R_{\mathcal O}(\rho\to\phi),
 \label{eq:cost-yield}
\end{equation}
with $1/0=\infty$. When $C_{\mathcal O}^{\phi}(\rho)>0$, the {reversibility ratio} ~\cite{HorodeckiSen2003,Horodecki2010Symmetry,GuOlivieroLeone2025,LeoneRizzoEisertJerbi2025} is
\begin{equation}
 \eta_{\mathcal O}^{\phi}(\rho)=
 \frac{D_{\mathcal O}^{\phi}(\rho)}{C_{\mathcal O}^{\phi}(\rho)}.
 \label{eq:recovery}
\end{equation}
\end{definition}
The yield--cost relation gives $D_{\mathcal O}^{\phi}\leq C_{\mathcal O}^{\phi}$~\cite{Takagi2022}.  We consider {any pure magic currency} $\phi$. A special case is the qutrit strange state,
\begin{equation}
 \Strange=\ket{\Strange}\!\bra{\Strange},
 \qquad \ket{\Strange}=\frac{\ket1-\ket2}{\sqrt2}.
 \label{eq:strange}
\end{equation}
The Strange state attains the maximal single-qutrit mana, $\log_2(5/3)$~\cite{Veitch2014,Wang2020}.  The Clifford symmetries of $\Strange$ also allow noisy inputs to be twirled into a one-parameter family~\cite{JainPrakash2020,Prakash2020}. {It has been studied as a distillation target}~\cite{DawkinsHoward2015,Prakash2020}.

We recall stabilizer nullity, an important measure of pure-state magic that will be useful for deriving our conversion bounds.

\begin{definition}[Stabilizer nullity]
\label{def:pure-nullity}
For a pure state $\psi=\ket\psi\!\bra\psi$ on $L$ qutrits, let $\Stab(\psi)=\{P\in\mathcal P_L:P\ket\psi=\ket\psi\}$. Its stabilizer nullity is~\cite{Beverland2020}
\begin{equation}
 \nu(\psi)=L-\log_3|\Stab(\psi)|.
 \label{eq:pure-nullity}
\end{equation}
\end{definition}
Stabilizer nullity counts the qutrits remaining after all pure stabilizer factors are separated off by Clifford decoding~\cite{Looi2011} and constrains the resource cost of state conversion~\cite{Beverland2020}.

{For mixed states, we use the convex extension of stabilizer nullity~\cite{Tarabunga2025}.}

\begin{definition}[Mixed-state stabilizer nullity]
\label{def:nullity}
For a state $\rho$ on $L$ qutrits, its mixed-state stabilizer
nullity is
\begin{equation}
 V(\rho)=\inf_{\rho=\sum_jp_j\psi_j}
          \sum_jp_j\nu(\psi_j),
 \label{eq:nullity}
\end{equation}
where the infimum is over all pure-state decompositions of
$\rho$, with $\psi_j$ denoting pure-state.
\end{definition}

This quantity is the smallest average pure-state nullity
among all ensemble realizations of $\rho$.

\subsection{Typical-state irreversibility and CPWP reversibility}
\label{subsec:typical-results}

Let the retained system contain $N$ qutrits and its environment contain $N-\ell_N$ qutrits, with dimensions
\begin{equation}
 d=3^N,\qquad K=3^{N-\ell_N},\qquad c=K/d=3^{-\ell_N}.
 \label{eq:main-random-dimensions}
\end{equation}
We consider two ensembles of random mixed states:
\begin{enumerate}
 \item[(i)] \emph{Haar-induced states}, obtained by tracing out the $K$-dimensional environment of a Haar-random pure state on $\mathbb C^d\otimes\mathbb C^K$.
 \item[(ii)] \emph{Local-random-circuit states}, obtained by applying a circuit from the fixed finite-gate ensemble of Sec.~\ref{subsec:random-models} to $\ket0^{\otimes(2N-\ell_N)}$ and tracing out the environment. We choose a sufficiently large polynomial circuit length, as specified in that section.
\end{enumerate}
In both models the target is sampled once and held fixed throughout the asymptotic conversion task.

\begingroup
\par\endgroup
\begin{theorem}[{Reversibility ratios of typical states}]
\label{thm:typical}
Let $\rho_N$ be drawn from either ensemble above, with { $0\leq\ell_N\leq N$}, {and let $\phi$ be a pure magic currency}.
\begin{enumerate}
 \item[(i)] \emph{CPWP reversibility.} If $\ell_N\to\infty$ and  $\ell_N\leq N$, then
 \begin{equation}
  \eta_{\CPWP}^{\Strange}(\rho_N)\xrightarrow{\Pr}1.
  \label{eq:typical-cpwp-efficiency}
 \end{equation}
 For a general pure currency $\phi$,
 \begin{equation}
  \eta_{\CPWP}^{\phi}(\rho_N)
  \xrightarrow{\Pr}\eta_{\CPWP}^{\phi}(\Strange)>0.
  \label{eq:typical-pure-currency-efficiency}
 \end{equation}
 \item[(ii)] \emph{{Nonvanishing CPWP reversibility ratio.}}
 If $\ell_N=O(1)$, then there exists a constant $\eta_*>0$,
 independent of $N$, such that
 \begin{equation}
 	\Pr\!\left\{
 	\eta_{\CPWP}^{\phi}(\rho_N)\geq\eta_*
 	\right\}\xrightarrow[N\to\infty]{}1.
 \end{equation}
 \item[(iii)] {\emph{CSP irreversibility.}} Throughout $\ell_N=o(N)$,
 \begin{equation}
  \eta_{\CSP}^{\phi}(\rho_N)\xrightarrow{\Pr}0,
  \label{eq:typical-csp-efficiency}
 \end{equation}
 \begingroup
 If $\ell_N=\Theta(N)$ and $\ell_N\leq N$, there exists a
 constant $\varepsilon_\phi>0$, independent of $N$, such that
 \begin{equation}
  \eta_{\CSP}^{\phi}(\rho_N)\leq1-\varepsilon_\phi
  \label{eq:typical-csp-linear-gap}
 \end{equation}
 with probability tending to one.
 \par\endgroup
\end{enumerate}
\end{theorem}

\begin{proof}
	
	For each part, intersect the high-probability events in
	Theorems~\ref{thm:random-three-rates} and
	\ref{thm:random-cpwp-distillation} as needed. Their intersection
	still has probability tending to one.
	
	\medskip
	\noindent\textbf{(i)} Suppose $\ell_N\to\infty$.
	By Corollary~\ref{cor:random-cpwp-leading-rates},
	the CPWP formation cost and distillation yield in Strange
	states both equal $\alpha\ell_N+o_{\Pr}(\ell_N)$.
	 Therefore,
	\begin{equation}
		\eta_{\CPWP}^{\Strange}(\rho_N)
		=
		\frac{D_{\CPWP}^{\Strange}(\rho_N)/(\alpha\ell_N)}
		{C_{\CPWP}^{\Strange}(\rho_N)/(\alpha\ell_N)}
		\xrightarrow{\Pr}1,
	\end{equation}
	which proves Eq.~\eqref{eq:typical-cpwp-efficiency}.
	
	For {any pure magic currency} $\phi$,
	Lemma~\ref{lem:pure-currency-conversions} in
	Appendix~\ref{app:pure-currency} gives
	$0<D_{\CPWP}^{\phi}(\Strange)
	\leq C_{\CPWP}^{\phi}(\Strange)<\infty$. By
	Eq.~\eqref{eq:currency-efficiency-sandwich},
	\begin{equation}
		\eta_{\CPWP}^{\phi}(\rho_N)
		\xrightarrow{\Pr}
		\frac{D_{\CPWP}^{\phi}(\Strange)}{C_{\CPWP}^{\phi}(\Strange)}
		=\eta_{\CPWP}^{\phi}(\Strange)>0.
	\end{equation}
	This proves part~(i).
	
	\medskip
	\noindent\textbf{(ii)}
	If $\ell_N=O(1)$, choose a fixed $L<\infty$ such that
	$0\leq\ell_N\leq L$ for all sufficiently large $N$.
	{Combining the yield lower bound $D_L>0$ in
	Theorem~\ref{thm:random-cpwp-distillation}(iii) with the formation bound
	in Theorem~\ref{thm:random-three-rates}(3) gives}
	\begin{equation}
		\begin{gathered}
			D_L\leq D_{\CPWP}^{\Strange}(\rho_N)
			\leq C_{\CPWP}^{\Strange}(\rho_N)\leq C_L,\\
			C_L=\frac{\log_2(1+3^L)+1}{2\log_2(5/3)}.
		\end{gathered}
		\label{eq:random-bounded-cost-yield}
	\end{equation}
	Thus
	\begin{equation}
		\Pr\!\left\{
		\eta_{\CPWP}^{\Strange}(\rho_N)\geq D_L/C_L
		\right\}\longrightarrow1.
		\label{eq:random-bounded-recovery}
	\end{equation}
	
	The lower bound in
	Eq.~\eqref{eq:currency-efficiency-sandwich} then implies
	\[
	\Pr\!\left\{\eta_{\CPWP}^{\phi}(\rho_N)\geq\eta_*\right\}
	\longrightarrow1,
	\]
	where
	\[
	\eta_*=
	\frac{D_{\CPWP}^{\phi}(\Strange)}{C_{\CPWP}^{\phi}(\Strange)}
	\frac{D_L}{C_L}>0
	\]
	is independent of $N$. This proves part~(ii).
	
	\medskip
	\noindent\textbf{(iii)}
	The first two bounds of Theorem~\ref{thm:random-three-rates} give
	\begin{equation}
		\begin{aligned}
			C_{\CSP}^{\phi}(\rho_N)&\geq\frac{\kappa N}{\nu(\phi)},\\
			D_{\CSP}^{\phi}(\rho_N)&\leq\frac{\ell_N+O(1)}{\nu(\phi)}.
		\end{aligned}
	\end{equation}
	Whenever the formation cost is positive and finite, it follows that
	\begin{equation}
		0\leq\eta_{\CSP}^{\phi}(\rho_N)
		\leq\frac{\ell_N+O(1)}{\kappa N}\longrightarrow0,
	\end{equation}
	when $\ell_N=o(N)$. This proves the sublinear statement.
	
	Now suppose $\ell_N=\Theta(N)$ and $\ell_N\leq N$.
	Using $\CSP\subseteq\CPWP$, the cost lower bound in
	Eq.~\eqref{eq:currency-rate-sandwiches}, and
	Theorem~\ref{thm:random-cpwp-distillation}(ii), we obtain
	\begin{equation}
		\begin{aligned}
			C_{\CSP}^{\phi}(\rho_N)
			&\geq C_{\CPWP}^{\phi}(\rho_N)\\
			&\geq C_{\CPWP}^{\phi}(\Strange)
			D_{\CPWP}^{\Strange}(\rho_N)\\
			&\geq C_{\CPWP}^{\phi}(\Strange)
			[\alpha\ell_N-O(\log N)]
		\end{aligned}
		\label{eq:csp-linear-formation-from-cpwp}
	\end{equation}
	with probability tending to one. Dividing the CSP yield bound
	above by this cost lower bound, and using
	$\log N/\ell_N\to0$, yields
	\begin{equation}
		\eta_{\CSP}^{\phi}(\rho_N)
		\leq\frac{1}{\alpha\nu(\phi)C_{\CPWP}^{\phi}(\Strange)}+o(1)
		\label{eq:csp-linear-efficiency-bound}
	\end{equation}
	Lemma~\ref{lem:pure-currency-strict-gap} makes the constant
	on the right strictly smaller than one. In particular, choose
	\begin{equation}
		\varepsilon_\phi
		:=\frac12\left[
		1-\frac{1}{\alpha\nu(\phi)C_{\CPWP}^{\phi}(\Strange)}
		\right]>0.
		\label{eq:csp-linear-gap-constant}
	\end{equation}
	 This completes part~(iii)
	and the proof.
\end{proof}

The matching CPWP rates have a further consequence beyond
recovery relative to a pure currency: they determine direct
interconversion rates between typical mixed states. In the
divergent-imbalance regime, the ratio of their $\Delta_2$ values
governs this interconversion, as stated in the following corollary.

\begin{corollary}[Interconversion of typical mixed states]
\label{cor:typical-interconversion}
Let $\rho_N$ and $\sigma_N$ be $N$-qutrit states, each sampled from either ensemble above, with respective environment dimensions $3^{N-\ell_N^\rho}$ and $3^{N-\ell_N^\sigma}$. Assume
\begin{equation}
 \ell_N^\rho,\ell_N^\sigma\longrightarrow\infty,
 \qquad \ell_N^\rho,\ell_N^\sigma \leq N.
 \label{eq:typical-conversion-regime}
\end{equation}
The two ensemble choices may differ, and the samples need not be independent. Then
\begin{align}
 R_{\CPWP}(\rho_N\to\sigma_N)
 &=\frac{\Delta_2(\rho_N)}{\Delta_2(\sigma_N)}
   [1+o_{\Pr}(1)]\notag\\
 &=\frac{\ell_N^\rho}{\ell_N^\sigma}
   [1+o_{\Pr}(1)].
 \label{eq:typical-interconversion}
\end{align}
 In particular, if $\ell_N^\rho=\ell_N^\sigma$, both directional rates converge to one.
\end{corollary}
\begin{proof}
{First, composition of conversion protocols gives}
\begin{equation}
 \begin{aligned}
 \frac{D_{\CPWP}^{\Strange}(\rho_N)}
      {C_{\CPWP}^{\Strange}(\sigma_N)}
 &\leq R_{\CPWP}(\rho_N\to\sigma_N)\\
 &\leq\frac{C_{\CPWP}^{\Strange}(\rho_N)}
               {C_{\CPWP}^{\Strange}(\sigma_N)}.
 \end{aligned}
 \label{eq:typical-conversion-sandwich}
\end{equation}
The lower bound first distills $\rho_N$ into $\Strange$ and then forms $\sigma_N$. For the upper bound, forming $\rho_N$ and converting it into $\sigma_N$ cannot improve on the optimal formation cost of $\sigma_N$.
Applying Eq.~\eqref{eq:random-cpwp-leading-rates}
to both states and substituting into
Eq.~\eqref{eq:typical-conversion-sandwich} gives
\[
R_{\CPWP}(\rho_N\to\sigma_N)
=\frac{\ell_N^\rho}{\ell_N^\sigma}[1+o_{\Pr}(1)].
\]
{For $\tau_N\in\{\rho_N,\sigma_N\}$, with corresponding imbalance
$\ell_N^\tau$, Lemma~\ref{lem:random-purity} gives
$\Delta_2(\tau_N)=\ell_N^\tau\log_2 3+o_{\Pr}(1)$ in this regime.
This proves the corollary.}
\end{proof}
Our results establish reversibility as a typical phenomenon in
the resource theory of Wigner negativity. For $1\ll\ell_N\le N$, typical states admit {asymptotically
reversible} conversion under CPWP operations. Moreover, the R\'enyi-2 entropy deficit
{$\Delta_2$ determines the leading-order conversion rates
within these ensembles: the rate between any two such states is
asymptotically given by the ratio of their entropy deficits.}
This gives $\Delta_2$ the operational role
played by entropy in magic resource theories.

The same typical states exhibit sharply different recovery
behavior in the resource theory of nonstabilizerness.
Under CSP operations, irreversibility is typical throughout
$0\le\ell_N\le N$, whenever the formation cost is finite.
For $\ell_N=o(N)$, the {reversibility ratio}
vanishes in probability.

For $\ell_N=N$, the environment is trivial and both ensembles
consist of pure states.
Even for these typical pure states, the resource theory of
nonstabilizerness exhibits asymptotic irreversibility under
CSP operations.
In the resource theory of Wigner negativity, the same typical
states admit asymptotically reversible formation and
distillation under CPWP operations.

Finally, the Strange state acquires an operational role as an
asymptotic exchange for typical mixed states. It calibrates
both their formation cost and distillable yield, while
$\eta_{\CPWP}^{\phi}(\Strange)$ determines their
{reversibility ratio relative to} any pure currency $\phi$.

\subsection{General formation and distillation bounds}
\label{subsec:general-bounds}

The typical-state separation rests on four general bounds: a formation lower bound and a distillation upper bound for CSP, and achievable formation and distillation bounds for CPWP.

\begingroup
For CPWP distillation, we seek an effect $0\preceq E\preceq I$
whose acceptance probability on $\rho$ exceeds all its nonnegative
phase-point responses. Such an effect yields a distillation protocol based on
likelihood-ratio testing, as described in Sec.~\ref{sec:cpwp}.
Write
\begin{equation}
 p=\Tr(\rho E),\qquad 0\leq\Tr(A_vE)\leq z
 \quad\text{for every }v,
 \label{eq:general-test-parameters}
\end{equation}
and define the binary relative entropy
\begin{equation}
 \begin{aligned}
 d_{\mathrm{bin}}(p\|z)
 ={}&p\ln\frac pz+(1-p)\ln\frac{1-p}{1-z},
 \end{aligned}
 \label{eq:general-test-exponent}
\end{equation}
for $0\leq p\leq1$ and $0<z<1$.
\endgroup
\begin{theorem}[General bounds {on} formation and distillation]
\label{thm:general}
For every $d$-dimensional qutrit state $\rho$ and {any pure magic currency} $\phi$,
\begin{align}
 C_{\CSP}^{\phi}(\rho)&\geq\frac{V(\rho)}{\nu(\phi)},\notag\\
 D_{\CSP}^{\phi}(\rho)&\leq
 \frac{\log_2d-S(\rho)}{\nu(\phi)\log_2 3}.
 \label{eq:csp-bounds}
\end{align}
For the Strange currency,
\begin{equation}
 C_{\CPWP}^{\Strange}(\rho)
 \leq\frac{\Delta_2(\rho)}{2\log_2(5/3)}.
 \label{eq:cpwp-cost}
\end{equation}
\begingroup
Moreover, any effect $E$ satisfying
Eq.~\eqref{eq:general-test-parameters} with $0<z<p\leq1$ gives
\begin{equation}
 D_{\CPWP}^{\Strange}(\rho)
 \geq\frac{d_{\mathrm{bin}}(p\|z)}{\ln(5/3)}>0.
 \label{eq:cpwp-yield}
\end{equation}
\endgroup
\end{theorem}
Their proofs are given in Secs.~\ref{sec:csp} and~\ref{sec:cpwp}, respectively. The CPWP bounds extend to any pure magic
currency by applying the composition inequalities for asymptotic
conversion rates to protocols through $\Strange${.}

\section{General limits under CSP operations}
\label{sec:csp}

We prove the CSP bounds in Theorem~\ref{thm:general} for an arbitrary pure magic currency. Formation is limited by the nullity required in every pure-state decomposition of the target, whereas distillation is limited by its distinguishability from the maximally mixed state. Both arguments use the stabilizer structure of CSP channels, but they constrain different properties of a mixed state.

\subsection{Formation cost from stabilizer nullity}
\label{subsec:csp-formation}

\begingroup

We show that the mixed-state stabilizer nullity $V$ provides
a converse bound on magic-state formation under CSP operations.
Our argument relies on three properties commonly used to
derive asymptotic converse bounds~\cite{Horodecki2000,Lami2020}:
monotonicity, strong superadditivity, and lower semicontinuity.
Explicitly, $V$ is monotone under every CSP channel $\Lambda$,
\begin{equation}
	V(\Lambda(\rho))\leq V(\rho),
\end{equation}
and strongly superadditive across any bipartition,
\begin{equation}
	V(\rho_{AB})\geq V(\rho_A)+V(\rho_B).
\end{equation}
Moreover, $V$ is lower semicontinuous on every fixed
finite-dimensional system: whenever
$\|\rho_j-\rho\|_1\to0$,
\begin{equation}
	V(\rho)\leq\liminf_{j\to\infty}V(\rho_j).
\end{equation}
These properties are proved in Appendix~\ref{app:mixed-nullity}.
Together with the pure-state identity
$V(\phi^{\otimes n})=n\nu(\phi)$~\cite{Beverland2020},
they imply the following lower bound on the asymptotic
formation cost.
\par
\endgroup

\begin{theorem}[Formation converse]
\label{thm:csp-formation}
For every finite-dimensional qutrit state $\rho$ and
{every pure magic currency} $\phi$,
\begin{equation}
 C_{\CSP}^{\phi}(\rho)\geq\frac{V(\rho)}{\nu(\phi)}.
 \label{eq:formation-converse}
\end{equation}
\end{theorem}

\begin{proof}
If $V(\rho)=0$, the bound is immediate. If no positive conversion
rate from $\phi$ to $\rho$ is achievable, then
$C_{\CSP}^{\phi}(\rho)=\infty$, and the bound also holds.
We may therefore assume $V(\rho)>0$ and fix an achievable rate
$r>0$. Set $m_n=\lfloor rn\rfloor$. By
Definition~\ref{def:rates}, there are CSP channels $\Lambda_n$
such that
\begin{equation}
 \begin{aligned}
  \sigma_n&=\Lambda_n(\phi^{\otimes n}),\\
  \epsilon_n&=\frac12\|\sigma_n-\rho^{\otimes m_n}\|_1
  \longrightarrow0.
 \end{aligned}
\end{equation}
For sufficiently large $n$, we have $m_n\geq1$. Let
$\sigma_{n,i}$ denote the marginal of $\sigma_n$ on its $i$th
target system. Contractivity of trace distance under partial
trace gives
\[
 \frac12\|\sigma_{n,i}-\rho\|_1\leq\epsilon_n
 \qquad (1\leq i\leq m_n).
\]
On the fixed single-copy system, define
\[
 v_\epsilon(\rho)
 =\inf_{\frac12\|\omega-\rho\|_1\leq\epsilon}V(\omega).
\]
Equation~\eqref{eq:roof-lsc} implies
$v_{\epsilon_n}(\rho)\to V(\rho)$.

The convex roof agrees with $\nu$ on pure states, and pure-state
nullity is additive under tensor products~\cite{Beverland2020}.
Using CSP monotonicity, Eq.~\eqref{eq:roof-monotonicity}, and
iterating Lemma~\ref{lem:strong-superadditivity}, we obtain
\begin{equation}
 \begin{aligned}
  n\nu(\phi)
  &=V(\phi^{\otimes n})\\
  &\geq V(\sigma_n)\\
  &\geq\sum_{i=1}^{m_n}V(\sigma_{n,i})\\
  &\geq m_n\,v_{\epsilon_n}(\rho).
 \end{aligned}
 \label{eq:formation-nullity-budget}
\end{equation}
Dividing by $n$ and taking $n\to\infty$ gives
$rV(\rho)\leq\nu(\phi)$. Since this holds for every achievable
$r>0$,
\[
 R_{\CSP}(\phi\to\rho)\leq\frac{\nu(\phi)}{V(\rho)}.
\]
The definition $C_{\CSP}^{\phi}(\rho)=1/R_{\CSP}(\phi\to\rho)$
then yields Eq.~\eqref{eq:formation-converse}.
\end{proof}

\subsection{Entropy bound on pure-state distillation}
\label{subsec:csp-distillation}

We next bound the CSP distillation yield in terms of the input
entropy deficit. The proof associates each accurate distillation
protocol with a hypothesis test distinguishing the input from
the maximally mixed state.

To construct this test, we combine the Kraus representation of
CSP channels with an overlap bound for the target currency.

Let $\phi$ be {a pure magic currency} on $L$ qutrits,
and write $k=\nu(\phi)>0$. By magic compression, there exists
a Clifford unitary $U$ such that
\begin{equation}
 U\phi U^\dagger
 =
 (|0\rangle\langle0|)^{\otimes(L-k)}
 \otimes\widetilde\phi,
 \label{eq:currency-decoding}
\end{equation}
where $\widetilde\phi$ is a pure state on $k$ qutrits
with no nonidentity Pauli stabilizer.

The following lemma shows that tensor powers of the compressed
currency have overlap bounded away from one with every proper
stabilizer code, uniformly in the number of copies.

\begin{lemma}[Stabilizer-code overlap on pure magic]
	\label{lem:currency-gap}
	For the compressed state $\widetilde\phi$ in
	Eq.~\eqref{eq:currency-decoding}, there exists a constant
	$\gamma_\phi<1$ such that, for every $m\geq1$ and every
	proper stabilizer-code projector $\Pi\neq I$ on $mk$ qutrits,
	\begin{equation}
		\operatorname{Tr}
		\bigl(\widetilde\phi^{\otimes m}\Pi\bigr)
		\leq\gamma_\phi.
		\label{eq:uniform-code-gap}
	\end{equation}
\end{lemma}

\begin{proof}
	We divide the proof into three steps.
	
	Define
	\begin{equation}
		\lambda_\phi
		=
		\max_{\substack{P\in\mathcal P_k\\P\not\propto I}}
		\left|
		\operatorname{Tr}(\widetilde\phi P)
		\right|.
		\label{eq:currency-pauli-response}
	\end{equation}
	For every nonscalar Pauli operator $P$, we have
	\[
	\left|\operatorname{Tr}(\widetilde\phi P)\right|<1.
	\]
	Indeed, since $P$ is unitary and $\widetilde\phi$ is pure,
	equality in the Cauchy--Schwarz inequality would imply that
	$P|\widetilde\phi\rangle$ is proportional to
	$|\widetilde\phi\rangle$, contradicting the assumption that
	$\widetilde\phi$ has no nontrivial Pauli stabilizer.
	Since $\mathcal P_k$ is finite, the maximum in
	Eq.~\eqref{eq:currency-pauli-response} is attained and
	\begin{equation}
		0\leq\lambda_\phi<1.
		\label{eq:lambda-strict-gap}
	\end{equation}
	
	Let $P$ be any nonscalar Pauli operator on the $mk$-qutrit
	system. It factorizes across the $m$ copies as
	\[
	P=P_1\otimes\cdots\otimes P_m,
	\]
	where each $P_j$ is a Pauli operator on $k$ qutrits and
	at least one $P_j$ is nonscalar. Since the input is a
	product state,
	\begin{align}
		\left|
		\operatorname{Tr}(\widetilde\phi^{\otimes m}P)
		\right|
		&=
		\prod_{j=1}^{m}
		\left|\operatorname{Tr}(\widetilde\phi P_j)\right|
		\nonumber\\
		&\leq \lambda_\phi.
		\label{eq:multicopy-pauli-gap}
	\end{align}
	Here we used Eq.~\eqref{eq:currency-pauli-response} for
	the nonscalar factors and the bound
	$\left|\operatorname{Tr}(\widetilde\phi P_j)\right|\leq1$
	for all remaining factors.

	Let $\Pi$ be a proper stabilizer-code projector. Since its
	code space is a proper subspace, its stabilizer constraints
	include at least one nontrivial Pauli constraint. Thus there
	exist a nonscalar Pauli operator $P$ and an eigenvalue
	${\omega_3}^s$, where ${\omega_3}=e^{2\pi i/3}$ and
	$s\in\{0,1,2\}$, such that
	\[
	\operatorname{ran}\Pi
	\subseteq
	\ker(P-{\omega_3}^s I).
	\]
	The projector onto this eigenspace is
	\begin{equation}
		E_s(P)=\frac13
		\left(I+{\omega_3}^{-s}P+{\omega_3}^{-2s}P^2\right),
		\label{eq:pauli-eigenspace-projector}
	\end{equation}
	and hence
	\[
	0\preceq\Pi\preceq E_s(P).
	\]
	Using the positivity of $\widetilde\phi^{\otimes m}$, we obtain
	\begin{align}
		\operatorname{Tr}(\widetilde\phi^{\otimes m}\Pi)
		&\leq
		\operatorname{Tr}(\widetilde\phi^{\otimes m}E_s(P))
		\nonumber\\
		&=\frac13\left[
		1+2\operatorname{Re}\left(
		{\omega_3}^{-s}
		\operatorname{Tr}(\widetilde\phi^{\otimes m}P)
		\right)\right]
		\nonumber\\
		&\leq\frac{1+2\lambda_\phi}{3},
		\label{eq:currency-code-overlap-bound}
	\end{align}
	where we used
	Eq.~\eqref{eq:multicopy-pauli-gap}.
	
	Setting
	\begin{equation}
		\gamma_\phi:=\frac{1+2\lambda_\phi}{3}<1
		\label{eq:currency-gap-constant}
	\end{equation}
	establishes Eq.~\eqref{eq:uniform-code-gap}, uniformly for
	all $m\geq1$ and all proper stabilizer-code projectors
	$\Pi$.
\end{proof}

Suppose a CSP protocol approximately prepares the desired pure target. Its stabilizer Kraus representation contains full-output-rank branches and branches supported on proper stabilizer codes. Lemma~\ref{lem:currency-gap} bounds the latter branches’ target overlap away from one, so accurate preparation requires their total probability to be small. Thus full-output-rank branches carry almost all the probability.

\begin{lemma}
\label{lem:input-test}
Let $\omega$ have dimension $D_{\rm in}$. Suppose a CSP channel $\Lambda$ outputs $m\geq1$ copies of $\widetilde\phi$ {(so $D_{\rm out}=3^{mk}$)} with
\begin{equation}
 \frac12\bigl\|\Lambda(\omega)-\widetilde\phi^{\otimes m}\bigr\|_1
 \leq\epsilon.
 \label{eq:distillation-error}
\end{equation}
Then there is an effect $0\preceq T\preceq I$ satisfying
\begin{equation}
 \Tr(T\omega)\geq1-\frac{\epsilon}{1-\gamma_\phi},
 \qquad
 \Tr\!\left(T\frac{I}{D_{\rm in}}\right)\leq{\frac1{D_{\rm out}}}.
 \label{eq:test-two-inputs}
\end{equation}
\end{lemma}
\begin{proof}
{Use the Kraus decomposition in Lemma~\ref{lem:csp-kraus}.} Each normalized branch with $\Pi_j\neq I$ is supported in a proper code, hence has overlap at most $\gamma_\phi$ with $\widetilde\phi^{\otimes m}$. Writing $p_{\rm bad}$ for the total weight of these branches, we have
\begin{equation}
 1-\epsilon\leq\Tr[\widetilde\phi^{\otimes m}\Lambda(\omega)]
 \leq1-(1-\gamma_\phi)p_{\rm bad},
 \label{eq:bad-branch-weight}
\end{equation}
so $p_{\rm bad}\leq\epsilon/(1-\gamma_\phi)$. These branches contribute at most $\gamma_\phi p_{\rm bad}$ to the target overlap. The full-output-rank branches therefore contribute at least $1-\epsilon/(1-\gamma_\phi)$.

Set
\begin{equation}
 T=\sum_{j:\,\Pi_j=I}
 K_j^\dagger\widetilde\phi^{\otimes m}K_j,
 \label{eq:full-rank-test}
\end{equation}
where the sum runs over all full-output-rank branches. Since $\widetilde\phi^{\otimes m}\preceq I$ and $\sum_jK_j^\dagger K_j=I$, this is an effect and has the required acceptance on $\omega$. For these branches, $K_jK_j^\dagger=c_j{I_{D_{\rm out}}}$, whereas trace preservation gives
\begin{equation}
 D_{\rm in}=\sum_jc_j\operatorname{rank}\Pi_j
       \geq{D_{\rm out}}\sum_{j:\,\Pi_j=I}c_j.
 \label{eq:rank-trace-budget}
\end{equation}
Therefore
\begin{equation}
 \Tr\!\left(T\frac{I}{D_{\rm in}}\right)
 =\frac1{D_{\rm in}}\sum_{j:\,\Pi_j=I}c_j\leq{\frac1{D_{\rm out}}}.
 \label{eq:test-false-positive}
\end{equation}
\end{proof}

\begin{theorem}[Distillation converse]
\label{thm:csp-distillation}
For every $d$-dimensional qutrit state $\rho$ and {every pure magic currency} $\phi$,
\begin{equation}
 D_{\CSP}^{\phi}(\rho)
 \leq\frac{\log_2d-S(\rho)}{\nu(\phi)\log_2 3}.
 \label{eq:distillation-converse}
\end{equation}
\end{theorem}
\begin{proof}
	Let $r>0$ be an achievable rate for converting $\rho$ to $\phi$
	under CSP operations, and set $m_n=\lfloor rn\rfloor$.
	By definition, there exists a sequence of CSP channels $\Lambda_n$
	such that
	\begin{equation}
		\epsilon_n:=
		\frac12\left\|
		\Lambda_n(\rho^{\otimes n})-\phi^{\otimes m_n}
		\right\|_1
		\longrightarrow0.
	\end{equation}
	Write $k=\nu(\phi)>0$.
	Applying the magic compression in
	Eq.~\eqref{eq:currency-decoding} to each output copy
	and tracing out the stabilizer factors gives a CSP channel
	$\widetilde\Lambda_n$ satisfying
	\begin{equation}
		\frac12\left\|
		\widetilde\Lambda_n(\rho^{\otimes n})
		-\widetilde\phi^{\otimes m_n}
		\right\|_1
		\leq\epsilon_n.
	\end{equation}
	
	For all sufficiently large $n$, we have $m_n\geq1$ and
	$\epsilon_n<1-\gamma_\phi$.
	Applying Lemma~\ref{lem:input-test} to this channel, with input
	$\omega=\rho^{\otimes n}$ and {dimensions $D_{\rm in}=d^n$
	and $D_{\rm out}=3^{m_nk}$},
	yields an effect $0\preceq T_n\preceq I$ such that
	\begin{align}
		p_n:=\Tr(T_n\rho^{\otimes n})
		&\geq1-\frac{\epsilon_n}{1-\gamma_\phi}, \notag\\
		q_n:=\Tr\!\left(T_n\frac{I}{d^n}\right)
		&\leq3^{-m_nk}.
	\end{align}
	Thus the assumed distillation protocol induces a binary test
	that accepts $\rho^{\otimes n}$ with probability approaching one,
	while accepting the maximally mixed state with probability
	at most $3^{-m_nk}$.
	
	Apply the measurement $\{T_n,I-T_n\}$ to
	$\rho^{\otimes n}$ and $I/d^n$.
	Data processing for the quantum relative entropy
	gives~\cite{Lindblad1975}
	\begingroup
\begin{align}
&n[\log_2d-S(\rho)]\notag\\
    &\quad=D\!\left(\rho^{\otimes n}\middle\|I/d^n\right)\notag\\
    &\quad\geq p_n\log_2\frac{p_n}{q_n}
    +(1-p_n)\log_2\frac{1-p_n}{1-q_n}\notag\\
    &\quad=p_n\log_2\frac1{q_n}
    +(1-p_n)\log_2\frac1{1-q_n}-h_2(p_n)\notag\\
    &\quad\geq p_n\log_2\frac1{q_n}-1\notag\\
    &\quad\geq
    \left(1-\frac{\epsilon_n}{1-\gamma_\phi}\right)
    m_nk\log_2 3-1.
    \label{eq:binary-data-processing}
\end{align}
\par\endgroup
	where $h_2$ is the binary entropy.
	The penultimate inequality uses
	$(1-p_n)\log_2[1/(1-q_n)]\geq0$ and $h_2(p_n)\leq1$;
	the last inequality substitutes the bounds on $p_n$ and $q_n$.
	
	Dividing by $n$ and rearranging, we obtain
	\begin{equation}
		\left(1-\frac{\epsilon_n}{1-\gamma_\phi}\right)
		\frac{m_n}{n}\,k\log_2 3
		\leq\log_2d-S(\rho)+\frac1n.
		\label{eq:finite-block-converse}
	\end{equation}
	Since $\epsilon_n\to0$, $m_n/n\to r$, and
	$\gamma_\phi<1$ is independent of $n$, taking $n\to\infty$
	gives
	\begin{equation}
		rk\log_2 3\leq\log_2d-S(\rho).
	\end{equation}
	Taking the supremum over achievable rates $r$ proves
	\begin{equation}
		D_{\CSP}^{\phi}(\rho)
		\leq
		\frac{\log_2d-S(\rho)}{\nu(\phi)\log_2 3}.
	\end{equation}
	If no positive rate is achievable, the bound holds trivially.
\end{proof}

\section{Achievable conversion under CPWP operations}
\label{sec:cpwp}

We now prove the CPWP achievability statements in
Theorem~\ref{thm:general}. For formation, we first construct
protocols from positive Wigner envelopes and then bound their
asymptotic cost using purity. For distillation, we use tests
with high acceptance on the input and uniformly small absolute
phase-point responses. Both constructions use two-outcome
measure-and-prepare channels, whose CPWP property is verified
through their Choi operators. Throughout this section, the
currency is $\Strange$ and $b=5/3$.

We use the phase-point identities~\cite{Gross2006}
\begin{equation}
 A_u^2=I,\qquad \Tr A_u=1,\qquad
 \Tr(A_uA_v)=d\,\delta_{uv},
 \label{eq:cpwp-phase-identities}
\end{equation}
with $X=\sum_uW_X(u)A_u$. For a CPTP map $\Lambda$,
let $J_\Lambda$ denote its normalized Choi operator.
Its bipartite Wigner function is
\begin{equation}
 W_{J_\Lambda}(u,v)
 ={\frac{1}{D_{\rm in}D_{\rm out}}}
   \Tr[(A_u\otimes A_v)J_\Lambda].
 \label{eq:cpwp-choi-wigner}
\end{equation}
The channel $\Lambda$ is CPWP if and only if
$W_{J_\Lambda}(u,v)\geq0$ for every $u,v$~\cite[Theorem 2]{Wang2019}.
We use this characterization to verify the CPWP property
of the channels constructed below.

\subsection{Formation from positive Wigner envelopes}
\label{subsec:cpwp-envelopes}
Inspired by the $\kappa$-entanglement approach to entanglement
cost under completely PPT-preserving operations~\cite{WangWilde2020},
we seek a positive operator whose Wigner function dominates the
absolute Wigner coefficients of the target state.
The trace of this operator will control the number of Strange
states required by our CPWP formation protocol.
\begin{definition}[Positive Wigner envelope]
\label{def:positive-envelope}
A positive Wigner envelope of a state $\rho$ is a Hermitian operator $Q$ such that
\begin{equation}
 Q\succeq0,\qquad W_Q(u)\geq|W_\rho(u)|\quad\text{for every }u.
 \label{eq:positive-envelope}
\end{equation}
Its trace $q=\Tr Q$ is at least one, and $Q/q\in\FW$.
\end{definition}

Choose the negative Wigner coefficient of $\Strange$ at the origin. Then $W_{\Strange}(0)=-1/3$ and $W_{\Strange}(u)=1/6$ elsewhere~\cite{Wang2020}. Introduce
\begin{equation}
 Q_{\Strange}=\frac{2I-\Strange}{3},\qquad
 {\Xi_{\Strange}}=\frac{I+4\Strange}{3}.
 \label{eq:strange-certificates}
\end{equation}
Direct substitution gives
\begin{align}
 Q_{\Strange}&\succeq0,& \Tr Q_{\Strange}&=b,
 &W_{Q_{\Strange}}&=|W_{\Strange}|,\notag\\
 {\Xi_{\Strange}}\Strange&=b\Strange,
 &&\Tr(A_u{\Xi_{\Strange}})\in\{-1,1\}.
 \label{eq:strange-certificate-properties}
\end{align}
The eigenvalues of ${\Xi_{\Strange}}$ are $b,1/3,1/3$.
The operator ${\Xi_{\Strange}}$ allows us to construct an effect
that accepts $\Strange^{\otimes m}$ with certainty and has
controlled phase-point responses. Combining this effect with
the target output $\rho$ and the Wigner-positive state $Q/q$
gives the following formation protocol.

\begin{lemma}
\label{lem:envelope-formation}
If $Q$ is a positive Wigner envelope of $\rho$ with trace $q$, then, for every integer $m\geq1$ satisfying $b^m\geq q+1$, there is a CPTP-CPWP channel $\Lambda_m$ with
\begin{equation}
 \Lambda_m(\Strange^{\otimes m})=\rho.
 \label{eq:envelope-conversion}
\end{equation}
\end{lemma}

\begin{proof}
	Let $b=5/3$ and define
	\begin{equation}
		P_m=
		\frac{{\Xi_{\Strange}}^{\otimes m}-3^{-m}I}
		{b^m-3^{-m}}.
		\label{eq:formation-effect}
	\end{equation}
	Since the eigenvalues of ${\Xi_{\Strange}}$ are $b,1/3,1/3$,
	we have $0\preceq P_m\preceq I$. Moreover,
	${\Xi_{\Strange}}\Strange=b\Strange$, and hence
	\[
	P_m\Strange^{\otimes m}=\Strange^{\otimes m}.
	\]
	Define
	\begin{equation}
		\Lambda_m(X)=
		\operatorname{Tr}(P_mX)\rho+
		\operatorname{Tr}[(I-P_m)X]\frac Qq.
		\label{eq:formation-channel-proof}
	\end{equation}
	Since $P_m$ and $I-P_m$ form a POVM and both output
	operators are density operators, $\Lambda_m$ is CPTP.
	Furthermore,
	\[
	\Lambda_m(\Strange^{\otimes m})=\rho.
	\]
	
	It remains to prove that $\Lambda_m$ is CPWP. Let
	{$D_{\rm in}=3^m$ and $D_{\rm out}=d$} and use the normalized Choi operator
	\begin{equation}
		J_{\Lambda_m}
		={\frac1{D_{\rm in}}}\left[
		P_m^{\mathsf T}\otimes\rho+
		(I-P_m)^{\mathsf T}\otimes\frac Qq
		\right].
		\label{eq:formation-choi}
	\end{equation}
	Its bipartite Wigner function is
	\begin{equation}
		W_{J_{\Lambda_m}}(x,y)
		={\frac1{D_{\rm in}^2}}\left[
		e_xW_\rho(y)+(1-e_x)\frac{W_Q(y)}q
		\right],
		\label{eq:formation-choi-wigner}
	\end{equation}
	where
	\[
	e_x:=\operatorname{Tr}(A_xP_m^{\mathsf T}).
	\]
	Since transposition permutes the phase-point operators,
	the values of $e_x$ are the same as those of
	$\operatorname{Tr}(A_xP_m)$. Using
	$\operatorname{Tr}(A_u{\Xi_{\Strange}})\in\{-1,1\}$, we obtain
	\begin{equation}
		e_x\in
		\left\{
		\frac{1-3^{-m}}{b^m-3^{-m}},
		-\frac{1+3^{-m}}{b^m-3^{-m}}
		\right\}.
		\label{eq:formation-two-responses}
	\end{equation}
	In particular, $1-e_x>0$ for all $x$.
	
	Since $W_Q(y)\geq|W_\rho(y)|$, a sufficient condition for
	the bracket in Eq.~\eqref{eq:formation-choi-wigner} to be
	nonnegative for all $x,y$ is
	\[
	q\leq\frac{1-e_x}{|e_x|}
	\qquad\text{for all }x.
	\]
	Substituting the two possible values in
	Eq.~\eqref{eq:formation-two-responses}, this condition
	reduces to
	\begin{equation}
		q\leq
		\min\left\{
		\frac{b^m-1}{1-3^{-m}},
		\frac{b^m+1}{1+3^{-m}}
		\right\}
		=\frac{b^m-1}{1-3^{-m}},
		\label{eq:formation-envelope-condition}
	\end{equation}
	where the equality follows from $(b/3)^m<1$.
	The hypothesis $b^m\geq q+1$ guarantees
	Eq.~\eqref{eq:formation-envelope-condition}. Consequently,
	$W_{J_{\Lambda_m}}(x,y)\geq0$ for all $x,y$, so
	$\Lambda_m$ is CPWP. This completes the proof.
\end{proof}

Lemma~\ref{lem:envelope-formation} converts a positive Wigner
envelope into a formation protocol whose resource consumption
is controlled by the logarithm of the envelope trace. For
$\rho^{\otimes n}$, an upper bound on the asymptotic formation
cost therefore follows from an upper bound on the exponential
growth rate of this trace. We next construct envelopes for
which this growth is controlled by the purity of $\rho$.

\subsection{Formation cost controlled by purity}
\label{subsec:cpwp-purity}

To bound the asymptotic formation cost, we seek positive Wigner envelopes \(Q_n\) of \(\rho^{\otimes n}\) with controlled exponential trace growth. Each envelope must satisfy both \(Q_n\succeq0\) and \(W_{Q_n}\geq|W_{\rho^{\otimes n}}|\). Indeed,
\(
|\rho|_W:=\sum_u|W_\rho(u)|A_u,
\qquad
W_{|\rho|_W}=|W_\rho|\geq0,
\)
but $|\rho|_W$ need not be positive semidefinite. Conversely,
\(
\bigl||\rho|_W\bigr|\succeq0,
\)
but its Wigner function need not satisfy
\(
W_{\bigl||\rho|_W\bigr|}(u)\geq|W_\rho(u)|
\quad\text{for every }u.
\)

\begin{lemma}[Purity-controlled envelope construction]
	\label{lem:purity-envelope}
	Let $\rho$ be a state on a $d$-dimensional qutrit system.
	Let $\mathcal P$ denote the cone of positive-semidefinite
	operators and $\mathcal W$ the cone of Hermitian operators with
	nonnegative Wigner coefficients. Define
	\[
	|X|_W:=\sum_u|W_X(u)|A_u,
	\qquad |X|:=\sqrt{X^2},
	\]
	and recursively construct
	\begin{equation}
		X_{-1}=\rho,\;
		X_{2j}=|X_{2j-1}|_W,\;
		X_{2j+1}=|X_{2j}|,
		\; j\geq0.
		\label{eq:envelope-alternating-sequence}
	\end{equation}
	For each $J\geq1$, let
	\[
	\lambda_J:=\|(X_{2J})_-\|_\infty,
	\qquad
	Y_J:=X_{2J}+\lambda_J I,
	\]
	where $(X_{2J})_-$ denotes the negative spectral part of
	$X_{2J}$. Then $\lambda_J\to0$, and for every $n\geq1$,
	the operator
	\begin{equation}
		Q_{J,n}:=
		\sum_{j=0}^{J-1}
		\left(X_{2j}^{\otimes n}+X_{2j+1}^{\otimes n}\right)
		+Y_J^{\otimes n}
		\label{eq:purity-envelope-Q}
	\end{equation}
	is a positive Wigner envelope of $\rho^{\otimes n}$. In
	particular,
	\begin{equation}
		Q_{J,n}\succeq0,\qquad
		W_{Q_{J,n}}(u)\geq
		|W_{\rho^{\otimes n}}(u)|
		\quad\text{for all }u,
		\label{eq:purity-envelope-properties}
	\end{equation}
	and
	\begin{equation}
		\operatorname{Tr}Q_{J,n}
		\leq
		(2J+1)\bigl(\sqrt{d\Tr\rho^2}+d\lambda_J\bigr)^n.
		\label{eq:purity-envelope-trace}
	\end{equation}
\end{lemma}

\begin{proof}
	For every Hermitian $X$,
	\[
	\Tr|X|\geq\Tr X,\qquad
	\Tr|X|_W\geq\Tr X.
	\]
	Hence, writing $t_i=\operatorname{Tr}X_i$, we have
	\[
	1=t_{-1}\leq t_0\leq t_1\leq\cdots.
	\]
	{Both operations preserve the Hilbert--Schmidt norm:}
	\[
	\bigl\||X|\bigr\|_2
	=\bigl\||X|_W\bigr\|_2
	=\|X\|_2.
	\]
	Thus $\|X_i\|_2=\|\rho\|_2$, and Cauchy--Schwarz gives
	\[
	t_i\leq\|I\|_2\|X_i\|_2
	=\sqrt d\,\|\rho\|_2=\sqrt{d\Tr\rho^2}.
	\]
	The sequence $(t_i)$ is therefore bounded and nondecreasing,
	so $t_{i+1}-t_i\to0$. Since
	$X_{2J+1}=|X_{2J}|$,
	\[
	\lambda_J
	\leq\operatorname{Tr}(X_{2J})_-
	=\frac{t_{2J+1}-t_{2J}}2
	\longrightarrow0.
	\]
	Moreover,
	\[
	\begin{gathered}
	Y_J\in\mathcal P\cap\mathcal W,\qquad
	Y_J\pm X_{2J-1}\in\mathcal W,\\
	\operatorname{Tr}Y_J\leq\sqrt{d\Tr\rho^2}+d\lambda_J.
	\end{gathered}
	\]
	
	We use the following tensor-order property of either cone
	$\mathcal C\in\{\mathcal P,\mathcal W\}$:
	\begin{equation}
		A\pm B\in\mathcal C
		\quad\Longrightarrow\quad
		A^{\otimes n}\pm B^{\otimes n}\in\mathcal C.
		\label{eq:envelope-tensor-order}
	\end{equation}
	Indeed, this follows by induction from
	\begin{align*}
		A^{\otimes n}\pm B^{\otimes n}
		=\frac12\big[&
		(A+B)\otimes
		(A^{\otimes(n-1)}\pm B^{\otimes(n-1)})\\
		+&(A-B)\otimes
		(A^{\otimes(n-1)}\mp B^{\otimes(n-1)})
		\big],
	\end{align*}
	using closure of both cones under addition and tensor products.
	
	By construction,
	$X_{2j+1}\pm X_{2j}\in\mathcal P$. Equation
	\eqref{eq:envelope-tensor-order} therefore implies
	\[
	X_{2j}^{\otimes n}+X_{2j+1}^{\otimes n}\succeq0.
	\]
	Together with $Y_J\succeq0$, this proves
	$Q_{J,n}\succeq0$.
	
	For Wigner domination, for either sign we regroup
	\begin{align*}
		Q_{J,n}\pm\rho^{\otimes n}
		={}&X_0^{\otimes n}\pm\rho^{\otimes n}
		+\sum_{j=0}^{J-2}
		\left(X_{2j+1}^{\otimes n}
		+X_{2j+2}^{\otimes n}\right)\\
		&+\left(X_{2J-1}^{\otimes n}+Y_J^{\otimes n}\right).
	\end{align*}
	Every term belongs to $\mathcal W$: this follows from
	$X_0\pm\rho\in\mathcal W$,
	$X_{2j+2}\pm X_{2j+1}\in\mathcal W$, and
	$Y_J\pm X_{2J-1}\in\mathcal W$, together with
	Eq.~\eqref{eq:envelope-tensor-order}. Thus
	$Q_{J,n}\pm\rho^{\otimes n}\in\mathcal W$, which proves
	\[
	W_{Q_{J,n}}(u)\geq
	|W_{\rho^{\otimes n}}(u)|
	\quad\text{for all }u.
	\]
	
	Finally, applying these trace bounds gives
	\begin{align*}
		\operatorname{Tr}Q_{J,n}
		&=\sum_{j=0}^{J-1}(t_{2j}^n+t_{2j+1}^n)
		+(\operatorname{Tr}Y_J)^n\\
		&\leq(2J+1)\bigl(\sqrt{d\Tr\rho^2}+d\lambda_J\bigr)^n.
	\end{align*}
	This establishes all the claimed properties.
\end{proof}

Combining the envelope construction with
Lemma~\ref{lem:envelope-formation} gives an upper bound on the
formation cost that depends only on $d\Tr\rho^2$. Expressing
this dependence through the R\'enyi-2 entropy deficit yields
the following result.

\begin{theorem}[Purity-controlled formation]
	\label{thm:cpwp-purity}
	For every fixed $d$-dimensional qutrit state $\rho$,
	\begin{equation}
		C_{\CPWP}^{\Strange}(\rho)
		\leq
		\frac{\Delta_2(\rho)}{2\log_2 b},
		\qquad b=\frac53.
		\label{eq:cpwp-purity-proved}
	\end{equation}
\end{theorem}

\begin{proof}
	By Lemma~\ref{lem:purity-envelope}, for every $J\geq1$ and
	$n\geq1$, there exists a positive Wigner envelope
	$Q_{J,n}$ of $\rho^{\otimes n}$ satisfying
	\[
	\begin{aligned}
	q_{J,n}&:=\operatorname{Tr}Q_{J,n}\leq(2J+1)\bigl(\sqrt{d\Tr\rho^2}+d\lambda_J\bigr)^n,
	\end{aligned}
	\]
	where $\lambda_J\to0$.
	
	Fix $J$ and choose any
	$\kappa>\log_b\bigl(\sqrt{d\Tr\rho^2}+d\lambda_J\bigr)$. For all sufficiently
	large $n$, the preceding bound implies
	\[
	b^{\lceil n\kappa\rceil}\geq q_{J,n}+1.
	\]
	By Lemma~\ref{lem:envelope-formation}, $\rho^{\otimes n}$
	can therefore be prepared exactly from at most
	$\lceil n\kappa\rceil$ copies of $\Strange$. Consequently,
	\[
	C_{\CPWP}^{\Strange}(\rho)
	\leq\kappa.
	\]
	Taking $\kappa\downarrow\log_b\bigl(\sqrt{d\Tr\rho^2}+d\lambda_J\bigr)$ gives
	\[
	C_{\CPWP}^{\Strange}(\rho)
	\leq\log_b\bigl(\sqrt{d\Tr\rho^2}+d\lambda_J\bigr).
	\]
	Finally, sending $J\to\infty$ yields
	\[
	C_{\CPWP}^{\Strange}(\rho)
	\leq\frac12\log_b(d\Tr\rho^2)
	=\frac{\Delta_2(\rho)}{2\log_2 b},
	\]
	as claimed.
\end{proof}

\subsection{Distillation from phase-space tests}
\label{subsec:cpwp-testing}

The formation channel uses the Strange state as an input with a controlled accepting effect. In the distillation, the state $\tau_{\Strange}=Q_{\Strange}/b$ supplies the Wigner-nonnegative output needed to offset the negative coefficients of $\Strange$.

\begin{lemma}
\label{lem:test-distillation}
For an effect $0\preceq T\preceq I$ and an integer $m\geq1$, define
\begin{equation}
 \Lambda(X)=\Tr(TX)\Strange^{\otimes m}
          +\Tr[(I-T)X]\tau_{\Strange}^{\otimes m}.
 \label{eq:distillation-channel}
\end{equation}
This CPTP map is CPWP if and only if
\begin{equation}
 -\frac1{b^m-1}\leq\Tr(A_uT)\leq\frac1{b^m+1}
 \quad\text{for every }u.
 \label{eq:test-phase-interval}
\end{equation}
In particular, $\max_u|\Tr(A_uT)|\leq1/(b^m+1)$ is sufficient. For every input state $\omega$,
\begin{equation}
 \frac12\bigl\|\Lambda(\omega)-\Strange^{\otimes m}\bigr\|_1
 \leq1-\Tr(T\omega).
 \label{eq:distillation-output-error}
\end{equation}
\end{lemma}

\begin{proof}
	We prove the output error and characterize when
	$\Lambda$ is CPWP.
	Since $0\preceq T\preceq I$, the operators $T$ and $I-T$
	form a POVM. The channel $\Lambda$ measures this POVM
	and prepares $\Strange^{\otimes m}$ or
	$\tau_{\Strange}^{\otimes m}$, respectively, and is therefore
	CPTP.
	
	For any input state $\omega$, let
	$p_T=\operatorname{Tr}(T\omega)$. Then
	\[
	\Lambda(\omega)
	=p_T\Strange^{\otimes m}
	+(1-p_T)\tau_{\Strange}^{\otimes m}.
	\]
	Consequently,
	\begin{align}
		\frac12\left\|
		\Lambda(\omega)-\Strange^{\otimes m}
		\right\|_1
		&=\frac{1-p_T}{2}
		\left\|\tau_{\Strange}^{\otimes m}
		-\Strange^{\otimes m}\right\|_1
		\nonumber\leq 1-p_T,
	\end{align}
	which proves Eq.~\eqref{eq:distillation-output-error}.
	
	Let {$D_{\rm in}$} be the input dimension and
	{$D_{\rm out}=3^m$} the output dimension. The normalized Choi
	operator is
	\[
	J_{\Lambda}
	={\frac1{D_{\rm in}}}
	\left[
	T^{\mathsf T}\otimes\Strange^{\otimes m}
	+(I-T)^{\mathsf T}\otimes
	\tau_{\Strange}^{\otimes m}
	\right]
	\]
	Its bipartite Wigner function is
	\[
	W_{J_{\Lambda}}(u,v)
	={\frac1{D_{\rm in}D_{\rm out}}}
	\operatorname{Tr}\!\left[
	(A_u\otimes A_v)J_{\Lambda}
	\right].
	\]
	Define
	\[
	e_u:=\operatorname{Tr}(A_uT^{\mathsf T}).
	\]
	Since $\operatorname{Tr}A_u=1$, we obtain
	\begin{align}
		W_{J_{\Lambda}}(u,v)
		={\frac1{D_{\rm in}^2}}
		\bigg[
		&e_u W_{\Strange^{\otimes m}}(v)
		\nonumber
		+(1-e_u)
		W_{\tau_{\Strange}^{\otimes m}}(v)
		\bigg].
	\end{align}
	Using
	\[
	W_{\tau_{\Strange}^{\otimes m}}(v)
	=\frac{|W_{\Strange^{\otimes m}}(v)|}{b^m},
	\qquad b=\frac53,
	\]
	we arrive at
	\begin{equation}
		W_{J_{\Lambda}}(u,v)
		={\frac1{D_{\rm in}^2}}
		\left[
		e_u W_{\Strange^{\otimes m}}(v)
		+(1-e_u)\frac{
			|W_{\Strange^{\otimes m}}(v)|}{b^m}
		\right].
		\label{eq:distillation-choi-wigner}
	\end{equation}
	Transposition permutes the phase-point operators, so as
	$u$ ranges over phase space, the values of $e_u$ are exactly
	the values of $\operatorname{Tr}(A_uT)$, up to a permutation
	of phase-space labels.
	
$\Lambda$ is CPWP
	if and only if
	$W_{J_{\Lambda}}(u,v)\geq0$ for every $u,v$.
	Since $\Strange^{\otimes m}$ has both positive and negative
	Wigner coefficients, Eq.~\eqref{eq:distillation-choi-wigner}
	implies the two conditions
	\[
	e_u+\frac{1-e_u}{b^m}\geq0,
	\qquad
	-e_u+\frac{1-e_u}{b^m}\geq0.
	\]
	Equivalently,
	\[
	e_u\geq-\frac1{b^m-1},
	\qquad
	e_u\leq\frac1{b^m+1}.
	\]
	Both conditions are necessary and jointly sufficient.
	Consequently, $\Lambda$ is CPWP if and only if
	\[
	-\frac1{b^m-1}
	\leq\operatorname{Tr}(A_uT)
	\leq\frac1{b^m+1}
	\qquad\text{for every }u,
	\]
	which proves Eq.~\eqref{eq:test-phase-interval}.
	
	Finally, since $b^m-1<b^m+1$, the condition
	\[
	\max_u|\operatorname{Tr}(A_uT)|
	\leq\frac1{b^m+1}
	\]
	is sufficient for $\Lambda$ to be CPWP.
\end{proof}

\begingroup
The following proposition constructs the tests required by
Lemma~\ref{lem:test-distillation} from a single-copy effect with
nonnegative phase-point responses. A target acceptance strictly
larger than all phase-point responses gives a positive distillation rate.

\begin{proposition}[\textbf{CPWP distillation bound}]

\label{prop:cpwp-distillation}
Let $\rho$ be a state and let $0\preceq E\preceq I$, with
$p=\Tr(\rho E)$. Suppose that, for some $0<z<p\leq1$,
\[
0\leq\Tr(A_vE)\leq z\qquad\text{for every }v.
\]
Then
\begin{equation}
 D_{\CPWP}^{\Strange}(\rho)
 \geq\frac{d_{\mathrm{bin}}(p\|z)}{\ln(5/3)}>0,
 \label{eq:cpwp-distillation-bound}
\end{equation}
where $d_{\mathrm{bin}}$ is the binary relative entropy defined in
Eq.~\eqref{eq:general-test-exponent}.
\end{proposition}

\begin{proof}

Set $b=5/3$ and fix any $z<\theta<p$. Let $M_1=E$ and
$M_0=I-E$, and define the threshold test
\begin{equation}
 T_{n,\theta}
 =\sum_{|x|\geq\lceil n\theta\rceil}
 M_{x_1}\otimes\cdots\otimes M_{x_n},
 \label{eq:cpwp-threshold-test}
\end{equation}
where $x\in\{0,1\}^n$ and $|x|$ is its Hamming weight.
This is an effect, since it is a sum of outcomes of the product
POVM. On $\rho^{\otimes n}$, the number of outcomes corresponding
to $E$ is binomial with mean $np$. Hoeffding's inequality gives
\begin{equation}
 1-\Tr(T_{n,\theta}\rho^{\otimes n})
 \leq e^{-2n(p-\theta)^2}.
 \label{eq:cpwp-test-acceptance}
\end{equation}

For a product phase point $A_{\boldsymbol v}$, write
$e_i=\Tr(A_{v_i}E)\in[0,z]$. Its response to the same test is
the upper-tail probability of a sum of independent Bernoulli
variables with parameters $e_1,\ldots,e_n$. For $t\geq0$,
\[
\prod_{i=1}^n(1-e_i+e_i e^t)\leq(1-z+ze^t)^n.
\]
Consequently, the Chernoff bound gives
\begin{equation}
 \begin{aligned}
 0&\leq\Tr(A_{\boldsymbol v}T_{n,\theta})\\
  &\leq\inf_{t\geq0}
  \exp\!\left\{n\left[\ln(1-z+ze^t)-t\theta\right]\right\}\\
  &=\exp[-n d_{\mathrm{bin}}(\theta\|z)].
 \end{aligned}
 \label{eq:cpwp-test-chernoff}
\end{equation}
The infimum is attained at
$t=\ln[\theta(1-z)/(z(1-\theta))]>0$.

Write $s_\theta=d_{\mathrm{bin}}(\theta\|z)>0$. Fix any
\[
0<r<\frac{s_\theta}{\ln b},\qquad m_n=\lfloor rn\rfloor.
\]
Since $r\ln b<s_\theta$, for all sufficiently large $n$,
\[
0\leq\Tr(A_{\boldsymbol v}T_{n,\theta})
\leq e^{-ns_\theta}\leq\frac1{b^{m_n}+1}
\]
uniformly in $\boldsymbol v$. Lemma~\ref{lem:test-distillation}
therefore gives a CPWP channel producing $m_n$ Strange states
with output error at most $e^{-2n(p-\theta)^2}$, which tends to
zero. Taking the supremum over such $r$ yields
\[
D_{\CPWP}^{\Strange}(\rho)
\geq\frac{d_{\mathrm{bin}}(\theta\|z)}{\ln b}.
\]
Finally, let $\theta\uparrow p$. Continuity of
$d_{\mathrm{bin}}(\cdot\|z)$, including at $p=1$, proves the claim.
\end{proof}
\endgroup

These results complete the CPWP part of Theorem~\ref{thm:general}. The purity bound controls the attainable formation cost, while the testing bounds require additional phase-space structure to guarantee a yield.

\section{Formation and distillation of random mixed states}
\label{sec:random-bounds}

\begingroup
This section establishes formation and distillation bounds throughout
$0\leq\ell_N\leq N$, including the typical-state conversion results of
Theorem~\ref{thm:typical} for Haar-induced states and states generated
by local random circuits. CSP formation requires at least a linear
resource budget in $N$. When $\ell_N=o(N)$, the CSP distillation
yield is sublinear, leading to {a vanishing reversibility ratio}.
\par\endgroup
Under CPWP operations, recovery remains bounded away
from zero for bounded $\ell_N$, and conversion becomes asymptotically
reversible in Strange units when $\ell_N\to\infty$.
The proofs combine concentration estimates for the random states
with the general conversion bounds and explicit distillation tests.

\subsection{Random-state ensembles and concentration bounds}
\label{subsec:random-models}

Let $A$ contain $N$ qutrits and $E$ contain $N-\ell_N$ qutrits, where
{$(\ell_N)$ is a deterministic integer sequence satisfying
{ $0\leq\ell_N\leq N$}.} Write
\begin{equation}
 d=3^N,\qquad K=3^{N-\ell_N},\qquad c=\frac Kd=3^{-\ell_N}.
 \label{eq:random-dimensions}
\end{equation}

\begin{definition}[Haar-induced random states]
\label{def:haar-induced}
Sample a unit vector $\ket{\Psi_H}$ from the unitarily invariant
probability measure on $\mathbb C^d\otimes\mathbb C^K$, and define
\begin{equation}
 \Psi_H=\ket{\Psi_H}\!\bra{\Psi_H},\qquad
 \rho_H=\Tr_E\Psi_H.
 \label{eq:haar-induced-state}
\end{equation}
The distribution of $\rho_H$ is the \emph{induced measure}
$\mu_{d,K}$~\cite{Zyczkowski2001,Nechita2007}; we call $\rho_H$ a
Haar-induced random state.
\end{definition}

\begin{definition}[Local random circuits]
\label{def:local-random-circuits}
Arrange $n_{\mathrm{tot}}=2N-\ell_N$ qutrits on a one-dimensional chain,
with $A$ comprising the first $N$ sites and $E$ the remaining sites.
Initialize the chain in $\ket0^{\otimes n_{\mathrm{tot}}}$.
Fix a set $\mathcal G\subset\mathrm{SU}(9)$ of gates acting on two
neighboring qutrits, subject to the following conditions:
\begin{enumerate}
 \item $\mathcal G$ is finite, independent of $N$, contains the identity,
 and is closed under taking inverses.
 \item Every matrix entry of every gate in $\mathcal G$ is algebraic.
 \item $\mathcal G$ is universal: the group it generates is dense in
 $\mathrm{SU}(9)$.
\end{enumerate}
At each step $j$, independently choose a bond
$i_j\in\{1,\ldots,n_{\mathrm{tot}}-1\}$ and a gate $g_j\in\mathcal G$,
both uniformly, and apply $g_j$ to sites $i_j,i_j+1$. A circuit of
length $L$ is the ordered product of these $L$ gates. Its output and
reduced state are
\begin{equation}
 \begin{aligned}
 \Psi_C&=U_C(\ket0\!\bra0)^{\otimes n_{\mathrm{tot}}}U_C^\dagger,\\
 \rho_C&=\Tr_E\Psi_C.
 \end{aligned}
 \label{eq:local-circuit-state}
\end{equation}
This is the $\mathcal G$-local random circuit model of
Ref.~\cite[Sec.~II.B]{Brandao2016}; $L$ counts elementary gate
applications.
\end{definition}

For a concrete choice, take $\mathcal G$ to consist of all two-qutrit
Clifford unitaries of determinant one, together with
$\operatorname{diag}(-1,-1,1)\otimes I_3$.
This finite set lies in $\mathrm{SU}(9)$, has algebraic entries,
contains the identity, and is closed under taking inverses.
It is universal by the universality of the metaplectic
Clifford-plus-reflection basis~\cite{Bocharov2018}, since the added
gate differs from the axial reflection on the first qutrit only
by a global phase.

\paragraph{Basic properties.}
The Haar-induced ensemble has the standard normalized Wishart
representation~\cite{Zyczkowski2001,Nechita2007}
\begin{equation}
 \rho_H\overset{\mathrm{law}}{=}
 \frac{GG^\dagger}{\Tr GG^\dagger},
 \label{eq:random-wishart}
\end{equation}
where $G$ is a $d\times K$ complex Ginibre matrix: its entries are
independent complex Gaussian variables with mean zero and
$\mathbb E|G_{ia}|^2=1$. At $K=d$, $\mu_{d,d}$ is the complex
Hilbert--Schmidt measure. Both ensembles satisfy
$\operatorname{rank}\rho\leq K$ because they are reductions of pure
states with a $K$-dimensional environment.

\begingroup
We choose a polynomial circuit length $L_N=\operatorname{poly}(N)$
sufficiently large that both the joint and reduced output moments
approximate their Haar counterparts:
\begin{equation}
    \begin{aligned}
        \left\|\mathbb E_C\Psi_C^{\otimes q}
        -\mathbb E_H\Psi_H^{\otimes q}\right\|_1
        &\leq\delta_N,\\
        \left\|\mathbb E_C\rho_C^{\otimes q}
        -\mathbb E_H\rho_H^{\otimes q}\right\|_1
        &\leq\delta_N,
        \qquad 1\leq q\leq t_N.
    \end{aligned}
    \label{eq:random-moment-condition}
\end{equation}
\par\endgroup

Let $t_N=N^3$ and $\delta_N=e^{-N^5}$.
By the approximate-design theorem for local random circuits~\cite[Corollary~7]{Brandao2016},
under the conditions in Definition~\ref{def:local-random-circuits},
the above moment bounds can be achieved with a circuit length
\[
L_N=\operatorname{poly}\!\left(
n_{\mathrm{tot}},t_N,\ln(1/\delta_N)\right)
=\operatorname{poly}(N).
\]
Indeed, the corresponding bound on the $t_N$th pure-state moment
 implies
the reduced-state moment bounds for every $1\le q\le t_N$,  by tracing out the environment
and $t_N-q$ copies.
Throughout the following estimates, we choose a sufficiently large
polynomial circuit length $L_N=\operatorname{poly}(N)$ so that
Eq.~\eqref{eq:random-moment-condition} holds.

\begingroup
For either choice $\omega=\Psi$ or $\omega=\rho$, let $O_q$ be some
operator on the corresponding $q$-copy space.
Trace-norm duality and Eq.~\eqref{eq:random-moment-condition} give
\begin{equation}
	\left|
	\mathbb E_C\Tr(O_q\omega_C^{\otimes q})
	-\mathbb E_H\Tr(O_q\omega_H^{\otimes q})
	\right|
	\leq\delta_N\|O_q\|_\infty,
	\; 1\leq q\leq t_N.
	\label{eq:random-moment-transfer}
\end{equation}
In particular, let $V_q$ be a cyclic permutation satisfying
$\Tr(V_q\rho^{\otimes q})=\Tr(\rho^q)$.
Since $\|V_q\|_\infty=1$, we obtain
\begin{equation}
    \begin{aligned}
        \left|
        \mathbb E_C\Tr[(K\rho_C)^q]
        -\mathbb E_H\Tr[(K\rho_H)^q]
        \right|
        \leq K^q\delta_N,
        \; 1\leq q\leq t_N.
    \end{aligned}
    \label{eq:random-trace-moment-transfer}
\end{equation}
\par\endgroup

Below, $\rho_N$ denotes either ensemble, and the error terms containing
$\delta_N$ are omitted for the Haar ensemble.

To evaluate the bounds in
Eqs.~\eqref{eq:formation-converse},
\eqref{eq:distillation-converse},
\eqref{eq:cpwp-purity-proved}, and
\eqref{eq:cpwp-distillation-bound}
for random states, we establish the following lemmas.
\begin{lemma}[Spectral control]
\label{lem:random-spectrum}
For either random-state ensemble, with { $0\leq\ell_N\leq N$} and the circuit
moment condition~\eqref{eq:random-moment-condition}, there are absolute
constants $C_1,C_2>0$ such that
\begin{equation}
 \Pr\!\left\{K\|\rho_N\|_\infty
       \leq1+C_1\sqrt c+C_2/N\right\}\longrightarrow1.
 \label{eq:random-spectrum}
\end{equation}
In particular, with $L_{\rm sp}=1+C_1+C_2$, the same event implies
\[
 K\|\rho_N\|_\infty\leq L_{\rm sp}.
\]
\end{lemma}

\begin{proof}
	Let $|\psi\rangle\in\mathbb C^d\otimes\mathbb C^K$ be Haar distributed,
	and let $\rho_H=\Tr_{\mathbb C^K}|\psi\rangle\langle\psi|$.
	Since $K\le d$, Proposition~6.36 of Ref.~\cite{AubrunSzarek2017}
	gives the following bound on its largest eigenvalue:
	\begingroup
\begin{equation}
\begin{aligned}
        &\Pr_H\!\left\{
        \lambda_{\max}(\rho_H)
        \geq\left(\frac1{\sqrt K}
        +\frac{1+\varepsilon}{\sqrt d}\right)^2
        \right\}\\
        &\quad\leq e^{-K\varepsilon^2},
        \qquad\varepsilon>0.
    \end{aligned}
    \label{eq:random-haar-spectral-tail}
\end{equation}
\par\endgroup
	Writing $c=K/d$ and $(x)_+=\max\{x,0\}$, integration of this tail
	bound yields, for every integer $q\ge1$,
	\begin{equation}
		\begin{aligned}
			&\mathbb E_H
			\left[
			\left(\sqrt{K\|\rho_H\|_\infty}-1-\sqrt c\right)_+^{2q}
			\right]\\
			&\qquad\le
			2q\,c^q\int_0^\infty
			\varepsilon^{2q-1}e^{-K\varepsilon^2}\,d\varepsilon
			=\frac{q!}{d^q}.
		\end{aligned}
	\end{equation}
	Consequently, Minkowski's inequality and $q!\le q^q$ imply
	\begin{equation}
		\left(
		\mathbb E_H[(K\|\rho_H\|_\infty)^q]
		\right)^{1/(2q)}
		\le 1+\sqrt c+\sqrt{\frac qd}.
		\label{eq:random-haar-operator-moment}
	\end{equation}
	It follows that
	\begin{equation}
		\begin{aligned}
			\mathbb E_H\Tr[(K\rho_H)^q]
			&\le d\left(1+\sqrt c+\sqrt{\frac qd}\right)^{2q}\\
			&\le d(1+\sqrt c)^{2q}
			\exp\!\left(\frac{2q^{3/2}}{\sqrt d}\right)\\
			&\le 2d(1+\sqrt c)^{2q},
			\qquad 1\le q\le N^3,
		\end{aligned}
		\label{eq:random-refined-haar-moment}
	\end{equation}
	Since $\rho_H$ is positive semidefinite,
	\[
	(K\|\rho_H\|_\infty)^q
	\le \Tr[(K\rho_H)^q].
	\]
	Thus, taking $q=N^3$, Markov's inequality and
	Eq.~\eqref{eq:random-refined-haar-moment} give
	\begin{equation}
		\begin{aligned}
			&\Pr_H\!\left\{
			K\|\rho_H\|_\infty
			>(1+N^{-1})(1+\sqrt c)^2
			\right\}\\
			&\quad\le
			\frac{\mathbb E_H\Tr[(K\rho_H)^q]}
			{(1+N^{-1})^q(1+\sqrt c)^{2q}}\\
			&\quad\le 2d(1+N^{-1})^{-N^3}\\
			&\quad=\exp[-N^2+O(N)]=o(1).
		\end{aligned}
	\end{equation}
	For circuit states, Eq.~\eqref{eq:random-trace-moment-transfer}
	adds at most $\delta_NK^q$ to the Haar moment bound.
	Thus, taking $q=N^3$, Markov's inequality gives
	\begin{equation}
		\begin{aligned}
			&\Pr\!\left\{
			K\|\rho_N\|_\infty
			>(1+N^{-1})(1+\sqrt c)^2
			\right\}\\
			&\qquad\le
			2d(1+N^{-1})^{-N^3}
			+\delta_NK^{N^3}
			=o(1).
		\end{aligned}
		\label{eq:random-refined-spectral-tail}
	\end{equation}
	Under
	Eq.~\eqref{eq:random-moment-condition}, the two terms are bounded by
	$\exp[-N^2+O(N)]$ and $\exp[-N^5+O(N^4)]$, respectively.
	Finally, $0\le\sqrt c\le1$ implies
	\[
	(1+N^{-1})(1+\sqrt c)^2
	\le 1+3\sqrt c+\frac4N.
	\]
	This proves the claim with $C_1=3$ and $C_2=4$.
\end{proof}

\begin{lemma}[Purity concentration]
\label{lem:random-purity}
\begingroup
For either ensemble with $0\leq\ell_N\leq N$ and the circuit
moment condition~\eqref{eq:random-moment-condition},
\par\endgroup
\begin{equation}
 \Pr\!\left\{
 \left|K\Tr\rho_N^2-(1+c)\right|\leq N^{-1}
 \right\}\longrightarrow1.
 \label{eq:random-purity}
\end{equation}
On this event, the entropy deficit satisfies
\begin{equation}
 \Delta_2(\rho_N)
 =\ell_N\log_2 3+\log_2(1+c)+O(N^{-1}).
 \label{eq:random-renyi-deficit}
\end{equation}
\end{lemma}

\begin{proof}
The exact Haar purity moments are~\cite[Eq.~(14)]{Giraud2007}
\begin{align}
 \mathbb E_H\Tr\rho_H^2
 &=\frac{d+K}{dK+1},\notag\\
 \operatorname{Var}_H(\Tr\rho_H^2)
 &=\frac{2(d^2-1)(K^2-1)}
 {(dK+1)^2(dK+2)(dK+3)}.
 \label{eq:random-purity-moments}
\end{align}
For $Y=K\Tr\rho_N^2$ and $\mu_H=\mathbb E_HY$, these imply
\begin{equation}
 \mu_H=1+c-\frac{1+c}{dK+1},\qquad
 \operatorname{Var}_H(Y)\leq\frac2{d^2}.
 \label{eq:random-scaled-purity}
\end{equation}
By Eq.~\eqref{eq:random-moment-transfer},
\begin{equation}
	\begin{aligned}
		\left|\mathbb E_CY-\mathbb E_HY\right|
		&\le \delta_NK,\\
		\left|\mathbb E_CY^2-\mathbb E_HY^2\right|
		&\le \delta_NK^2.
	\end{aligned}
\end{equation} 
Consequently, Eq.~\eqref{eq:random-moment-transfer} gives
\begin{equation}
 \mathbb E_C(Y-\mu_H)^2
 \leq\frac2{d^2}+\delta_N(K^2+4K),
 \label{eq:random-purity-transfer}
\end{equation}
where we used $\mu_H\leq2$.
\begingroup
Since $1\leq K\leq d$,
$|\mu_H-(1+c)|\leq2/(d+1)\leq(2N)^{-1}$ for large $N$.
Chebyshev's inequality
bounds the failure probability in Eq.~\eqref{eq:random-purity} by
\begin{equation}
 \begin{aligned}
 &4N^2\left[\frac2{d^2}+\delta_N(K^2+4K)\right]\\
 &\quad\leq4N^2\left[\frac2{d^2}+\delta_N(d^2+4d)\right]
 =o(1).
 \end{aligned}
 \label{eq:random-purity-tail}
\end{equation}
This estimate is uniform throughout $1\leq K\leq d$.
\par\endgroup
Finally, $\Delta_2(\rho_N)=\ell_N\log_2 3+\log_2(K\Tr\rho_N^2)$; since $1+c\geq1$, the logarithm changes by $O(N^{-1})$ on the event in Eq.~\eqref{eq:random-purity}, proving Eq.~\eqref{eq:random-renyi-deficit}.
\end{proof}

\begin{lemma}[Uniform separation from low-nullity pure states]
\label{lem:random-low-nullity}
\begingroup
For either ensemble with $0\leq\ell_N\leq N$ and the circuit
moment condition~\eqref{eq:random-moment-condition}, with probability
tending to one,
\begin{equation}
 \sup_{\substack{\psi\ \mathrm{pure}\\
                  \nu(\psi)\leq\lfloor N/2\rfloor}}
 \Tr(K\rho_N\psi)\leq c+\frac14.
 \label{eq:random-low-nullity}
\end{equation}
\par\endgroup
\end{lemma}

\begin{proof}
Set $s=\lceil N/2\rceil$ and $R=3^{N-s}=3^{\lfloor N/2\rfloor}$.
Every pure state with $\nu(\psi)\leq N-s$ has at least $s$ independent
Pauli stabilizers. Selecting $s$ of them gives a rank-$R$ stabilizer-code
projector $\Pi$ with $\Pi\psi\Pi=\psi$. Let $\mathcal C_s$ denote
the set of all rank-$3^{N-s}$ stabilizer-code projectors on $N$ qutrits.
The stabilizer-code counting formula
\cite[Corollary~21]{Gross2006} implies
\begin{equation}
|\mathcal C_s|
=3^s\prod_{j=0}^{s-1}
\frac{3^{2(N-j)}-1}{3^{s-j}-1}
\le 3^{(2N+1)s}.
\end{equation}

\begingroup
For $W=GG^\dagger$, where $G\in\mathbb C^{d\times K}$ is
standard complex Gaussian,
\cite[Proposition~3.3]{CollinsNechita2011} gives
\[
\mathbb E\Tr(W^q)
=\sum_{\pi\in\mathfrak S_q}
K^{\#\pi}d^{\#(\gamma_q^{-1}\pi)},
\]
where $\gamma_q=(1\,2\,\cdots\,q)$ and $\#\pi$ counts cycles. Each $\Pi\in\mathcal C_s$ has rank $R=3^{N-s}$ and is unitarily
equivalent to $\operatorname{diag}(I_R,0)$.
Gaussian unitary invariance therefore gives the same formula
for $\mathbb E\Tr[(\Pi W\Pi)^q]$ with $d$ replaced by $R$.
Using
$\#\pi+\#(\gamma_q^{-1}\pi)\leq q+1$ and
$\sum_{\pi\in\mathfrak S_q}x^{\#\pi}=(x)_q$, we obtain
\begin{align}
	\mathbb E_H\Tr[(K\Pi\rho_H\Pi)^q]
	&=\frac{K^q}{(dK)_q}
	\sum_{\pi\in\mathfrak S_q}
	K^{\#\pi}R^{\#(\gamma_q^{-1}\pi)}\notag\\
	&\leq\frac{K^qR^{q+1}(K/R)_q}{(dK)_q}\notag\\
	&=R\prod_{j=0}^{q-1}
	\frac{c+jR/d}{1+j/(dK)}\notag\\
	&\leq R(c+qR/d)^q.
	\label{eq:random-compressed-bound}
\end{align}

To estimate the moments of $\Pi\rho_C\Pi$, let
$B_q:=K^qV_q\Pi^{\otimes q}$, where $V_q$ cyclically permutes
the $q$ copies. Then
\[
\Tr[(K\Pi\rho\Pi)^q]=\Tr(B_q\rho^{\otimes q}),
\qquad \|B_q\|_\infty\leq K^q.
\]
By the moment condition~\eqref{eq:random-moment-condition},
\begin{align}
	&\left|
	\mathbb E_C\Tr[(K\Pi\rho_C\Pi)^q]
	-\mathbb E_H\Tr[(K\Pi\rho_H\Pi)^q]
	\right|\notag\\
	&\quad\leq
	\|B_q\|_\infty
	\left\|\mathbb E_C\rho_C^{\otimes q}
	-\mathbb E_H\rho_H^{\otimes q}\right\|_1
	\leq K^q\delta_N.
\end{align}

Markov's inequality followed by a union bound over
$\mathcal C_s$ yields
\begin{equation}
\begin{aligned}
 &\Pr\!\left\{
 \max_{\Pi\in\mathcal C_s}\|K\Pi\rho_N\Pi\|_\infty>c+\frac14
 \right\}\\
 &\quad\leq3^{(2N+1)s}R
 \left(\frac{c+qR/d}{c+1/4}\right)^q\\
 &\qquad+3^{(2N+1)s}\delta_N
 \left(\frac{K}{c+1/4}\right)^q.
\end{aligned}
 \label{eq:random-code-union-bound}
\end{equation}
Since $q=N^3$ and $R/d=3^{-s}$, we have $q^2R/d=o(1)$.
For all sufficiently large $N$, $qR/d<1/4$, so $0<c\leq1$ implies
\begingroup
\[
    \begin{aligned}
        \left(\frac{c+qR/d}{c+1/4}\right)^q
        &\leq\left(\frac45\right)^q\exp(q^2R/d)\\
        &=\exp[-N^3\ln(5/4)+o(1)].
    \end{aligned}
\]
\par\endgroup
Together with $K/(c+1/4)\leq4d$, this bounds the right-hand side
of Eq.~\eqref{eq:random-code-union-bound} by
\begingroup
\[
    \begin{aligned}
        &\exp[-N^3\ln(5/4)+O(N^2)]\\
        &\quad+\exp[-N^5+O(N^4)]=o(1),
    \end{aligned}
\]
\par\endgroup
uniformly for $1\leq K\leq d$.

On this event, $K\|\Pi\rho_N\Pi\|_\infty\leq c+1/4$ holds
simultaneously for all $\Pi\in\mathcal C_s$, giving
\begin{equation}
 \Tr(K\rho_N\psi)
 \leq K\|\Pi\rho_N\Pi\|_\infty
 \leq c+\frac14.
 \label{eq:random-code-to-pure-state}
\end{equation}
This inequality holds simultaneously for all such pure states.
\par\endgroup
\end{proof}

\subsection{Consequences for formation and distillation}
\label{subsec:random-three-rates}

We apply the preceding random-state estimates to three general resource
bounds, in the order of CSP formation, CSP distillation, and CPWP
formation. {The bounds hold for any pure magic currency $\phi$.}

\begin{theorem}[Typical formation and distillation bounds]
\label{thm:random-three-rates}
Let $\rho_N$ be sampled from either ensemble in
Definitions~\ref{def:haar-induced} and~\ref{def:local-random-circuits},
with { $0\leq\ell_N\leq N$} and the circuit length chosen to satisfy
Eq.~\eqref{eq:random-moment-condition}.
{Let $\phi$ be a pure magic currency.}
There is an absolute constant $\kappa>0$ such that, with probability
tending to one, the following bounds hold simultaneously:
\begin{enumerate}
\item \emph{CSP formation cost:}
\begin{equation}
 C_{\CSP}^{\phi}(\rho_N)
 \geq\frac{\kappa N}{\nu(\phi)}.
 \label{eq:random-csp-cost}
\end{equation}
\item \emph{CSP distillation yield:}
\begin{equation}
 \begin{aligned}
 D_{\CSP}^{\phi}(\rho_N)
 &\leq\frac{\Delta_2(\rho_N)}{\nu(\phi)\log_2 3}\\
 &\leq\frac{\ell_N+\log_3(1+c+N^{-1})}{\nu(\phi)}.
 \end{aligned}
 \label{eq:random-csp-yield}
\end{equation}
\item \emph{CPWP formation cost:}
\begin{equation}
 \begin{aligned}
 C_{\CPWP}^{\phi}(\rho_N)
 &\leq C_{\CPWP}^{\phi}(\Strange)
 \frac{\Delta_2(\rho_N)}{2\log_2(5/3)}\\
 &\leq C_{\CPWP}^{\phi}(\Strange)\left[
 \alpha\ell_N+
 \frac{\log_2(1+c+N^{-1})}{2\log_2(5/3)}\right].
 \end{aligned}
 \label{eq:random-cpwp-cost}
\end{equation}
\begingroup
In particular, if $\ell_N\to\infty$, then
\begin{equation}
 C_{\CPWP}^{\phi}(\rho_N)
 \leq C_{\CPWP}^{\phi}(\Strange)
 [\alpha\ell_N+o(1)].
 \label{eq:random-cpwp-cost-divergent}
\end{equation}
\par\endgroup
\end{enumerate}
{One may take}
$\kappa=1/(4L_{\rm sp})$ with $L_{\rm sp}$ from
Lemma~\ref{lem:random-spectrum}.
\end{theorem}

\begin{proof}
Intersect the events in Lemmas~\ref{lem:random-spectrum}--\ref{lem:random-low-nullity};
their intersection still has probability tending to one. We prove the
three bounds on this common event.

\medskip
\noindent\emph{1. CSP formation cost.}
The general formation bound in Theorem~\ref{thm:general} is
\begin{equation}
 C_{\CSP}^{\phi}(\rho_N)
 \geq\frac{V(\rho_N)}{\nu(\phi)}.
 \label{eq:random-csp-formation-input}
\end{equation}
It therefore suffices to show $V(\rho_N)\geq N/(4L_{\rm sp})$.
Set $H=K\rho_N$. Lemmas~\ref{lem:random-spectrum}
and~\ref{lem:random-purity} give $\|H\|_\infty\leq L_{\rm sp}$ and
$\Tr(H\rho_N)\geq1+c-N^{-1}\geq c+3/4$ for $N\geq4$.

Consider an arbitrary decomposition
$\rho_N=\sum_jp_j\psi_j$, and let $w$ be the total weight of components
with $\nu(\psi_j)>\lfloor N/2\rfloor$. By
Lemma~\ref{lem:random-low-nullity},
\begingroup
\begingroup
\begin{align}
c+\frac34
    &\leq\sum_jp_j\Tr(H\psi_j)\notag\\
    &\leq(1-w)\left(c+\frac14\right)+wL_{\rm sp}\notag\\
    &\leq c+\frac14+wL_{\rm sp}.
    \label{eq:random-high-nullity-weight}
\end{align}
\par\endgroup
This gives $wL_{\rm sp}\geq1/2$.
\par\endgroup
Thus $w\geq1/(2L_{\rm sp})$, and every such decomposition obeys
\begin{equation}
 \sum_jp_j\nu(\psi_j)
 \geq\bigl(\lfloor N/2\rfloor+1\bigr)w
 \geq\frac{N}{4L_{\rm sp}}.
 \label{eq:random-roof-lower-bound}
\end{equation}
Taking the infimum over decompositions gives the same lower bound on
$V(\rho_N)$. Substitution into
Eq.~\eqref{eq:random-csp-formation-input} proves
Eq.~\eqref{eq:random-csp-cost} with $\kappa=1/(4L_{\rm sp})$.

\medskip
\noindent\emph{2. CSP distillation yield.}
The general distillation converse in
Theorem~\ref{thm:csp-distillation} gives
\begin{equation}
 D_{\CSP}^{\phi}(\rho_N)
 \leq\frac{\log_2d-S(\rho_N)}{\nu(\phi)\log_2 3}.
 \label{eq:random-csp-distillation-input}
\end{equation}
Using $S(\rho_N)\geq S_2(\rho_N)$, this becomes
\begin{equation}
 D_{\CSP}^{\phi}(\rho_N)
 \leq\frac{\Delta_2(\rho_N)}{\nu(\phi)\log_2 3}.
 \label{eq:random-csp-distillation-purity}
\end{equation}
Lemma~\ref{lem:random-purity} supplies
$K\Tr\rho_N^2\leq1+c+N^{-1}$. Since $d/K=3^{\ell_N}$,
\begin{equation}
 \begin{aligned}
 \Delta_2(\rho_N)
 &=\ell_N\log_2 3+\log_2(K\Tr\rho_N^2)\\
 &\leq\ell_N\log_2 3+\log_2(1+c+N^{-1}).
 \end{aligned}
 \label{eq:random-purity-resource-bound}
\end{equation}
Substituting this estimate into
Eq.~\eqref{eq:random-csp-distillation-purity} yields
\begin{equation}
 D_{\CSP}^{\phi}(\rho_N)
 \leq\frac{\ell_N+\log_3(1+c+N^{-1})}{\nu(\phi)},
\end{equation}
which is Eq.~\eqref{eq:random-csp-yield}.

\medskip
\noindent\emph{3. CPWP formation cost.}
The purity-controlled formation bound in
Theorem~\ref{thm:cpwp-purity} is stated in Strange units:
\begin{equation}
 C_{\CPWP}^{\Strange}(\rho_N)
 \leq\frac{\Delta_2(\rho_N)}{2\log_2(5/3)}.
 \label{eq:random-cpwp-formation-input}
\end{equation}
First prepare $\Strange$ from $\phi$, and then
prepare $\rho_N$ from $\Strange$. Composition of asymptotic protocols
{gives}
\begin{equation}
 C_{\CPWP}^{\phi}(\rho_N)
 \leq C_{\CPWP}^{\phi}(\Strange)\,
 C_{\CPWP}^{\Strange}(\rho_N).
 \label{eq:random-cpwp-currency-composition}
\end{equation}
Combining
Eqs.~\eqref{eq:random-cpwp-formation-input}
and~\eqref{eq:random-purity-resource-bound} gives
\begin{equation}
 \begin{aligned}
 C_{\CPWP}^{\phi}(\rho_N)
 &\leq C_{\CPWP}^{\phi}(\Strange)
 \frac{\Delta_2(\rho_N)}{2\log_2(5/3)}\\
 &\leq C_{\CPWP}^{\phi}(\Strange)\left[
 \alpha\ell_N+
 \frac{\log_2(1+c+N^{-1})}{2\log_2(5/3)}\right],
 \end{aligned}
\end{equation}
which proves Eq.~\eqref{eq:random-cpwp-cost}.
\begingroup
When $\ell_N\to\infty$, we have $c=3^{-\ell_N}\to0$, so
$\log_2(1+c+N^{-1})=o(1)$, proving
Eq.~\eqref{eq:random-cpwp-cost-divergent} throughout this range.
\par\endgroup
\end{proof}
\begingroup
The main separation between CSP and CPWP revealed by these bounds
lies in formation. Throughout $0\leq\ell_N\leq N$, for any pure currency $\phi$ with
$C_{\CPWP}^{\phi}(\Strange)<\infty$, the CSP formation cost is at
least $\Omega_\phi(N)$, whereas the CPWP formation cost is at most
$O_\phi(\ell_N+1)$. In particular, the latter is $o(N)$ when
$\ell_N=o(N)$. We next establish the CPWP distillation bounds.
\par\endgroup

\subsection{CPWP distillation for random mixed states}
\label{subsec:random-cpwp-distillation}

\begingroup

We now establish achievable CPWP distillation rates for random
mixed states, with $\Strange$ as the currency. The extension to any pure magic  $\phi$ follows
by concatenating with a CPWP conversion from $\Strange$ to $\phi$,
as detailed in Appendix~\ref{app:pure-currency}. When $\ell_N\to\infty$ with $\ell_N=o(N)$,
the distillation rate asymptotically matches the formation cost.
When $\ell_N=\Theta(N)$ and $\ell_N\leq N$, we obtain the same
leading distillation rate with an $O(\log N)$ correction.
When $\ell_N=O(1)$, the distillation rate is bounded below by a
positive constant independent of $N$. These bounds hold
with probability tending to one.

\begin{theorem}[\textbf{CPWP distillation for random mixed states}]

\label{thm:random-cpwp-distillation}
Let $\rho_N$ be drawn from either ensemble in
Sec.~\ref{subsec:random-models}, with $0\leq\ell_N\leq N$ and the circuit
moment condition~\eqref{eq:random-moment-condition}. Then:
\begin{enumerate}
\item[(i)] If $\ell_N\to\infty$ and $\ell_N=o(N)$, there is an absolute constant
$C_0<\infty$ such that, with probability tending to one,
\begin{equation}
 D_{\CPWP}^{\Strange}(\rho_N)
 \geq\alpha\ell_N-C_0-o(1){.}
 \label{eq:random-distillation-rate}
\end{equation}
\item[(ii)] If $\ell_N=\Theta(N)$ and $\ell_N\leq N$, then, with
probability tending to one,
\begin{equation}
 D_{\CPWP}^{\Strange}(\rho_N)
 \geq\alpha\ell_N-O(\log N).
 \label{eq:random-distillation-theta-rate}
\end{equation}
\item[(iii)] For every fixed $L<\infty$, if $0\leq\ell_N\leq L$, then
\begin{equation}
 \Pr\{D_{\CPWP}^{\Strange}(\rho_N)\geq D_L\}
 \longrightarrow1,
 \label{eq:random-bounded-distillation-rate}
\end{equation}
where
\begin{equation}
 \begin{aligned}
 D_L&=\frac{\delta^2}{2M_L^2\ln(5/3)}>0,\\
 {\delta}&{=
 \frac{2e^{-1/2}-1-e^{-2}}8>0,}
 \end{aligned}
 \label{eq:random-bounded-rate-constant}
\end{equation}
and $M_L\geq1$ is the constant in Lemma~\ref{lem:bounded-witness}.
\end{enumerate}
\end{theorem}

\begin{proof}

\emph{(i) $\ell_N\to\infty$ and $\ell_N=o(N)$.}
{Set $\lambda=K\|\rho_N\|_\infty$ and
$a=e^{-1/2}$.} Define $B_{\rho_N}:=({\sqrt{c}}/d)\sum_{v\in{\boldsymbol{\mathbb F_3^{2N}}}}\sin(x_v)A_v$,
where $x_v=\sqrt{dK}\Tr(A_v\rho_N)-{\sqrt{c}}$.
Since $\Tr(A_vA_w)=d\,\delta_{vw}$, we have
$\Tr(A_vB_{\rho_N})={\sqrt{c}}\sin(x_v)$, and hence
\[
\max_v|\Tr(A_vB_{\rho_N})|\leq {\sqrt{c}}.
\]
Let
\begin{equation}
	R_{\rho_N}=B_{\rho_N}-a(K\rho_N-cI).
	\label{eq:random-distillation-residual}
\end{equation}
By Lemma~\ref{lem:random-centered-residual} in
Appendix~\ref{app:centered-residual}, there is an absolute
constant $M\geq1$ such that, with probability tending to one,
\begin{equation}
	\|R_{\rho_N}\|_\infty\leq M{\sqrt{c}}.
	\label{eq:random-distillation-input}
\end{equation} 

On this event, define
\begin{equation}
	\begin{aligned}
		E_{\rho_N}
		&=\frac{B_{\rho_N}+(ac+M{\sqrt{c}})I}{a\lambda+2M{\sqrt{c}}}\\
		&=\frac{aK\rho_N+R_{\rho_N}+M{\sqrt{c}}I}{a\lambda+2M{\sqrt{c}}}.
	\end{aligned}
\end{equation}
Lemma~\ref{lem:random-centered-residual} gives
\begingroup
\[
    \begin{aligned}
        0\preceq aK\rho_N
        &\preceq B_{\rho_N}+(ac+M\sqrt{c})I\\
        &\preceq(a\lambda+2M\sqrt{c})I,
    \end{aligned}
\]
\par\endgroup
so $E_{\rho_N}$ is an effect. Since
$\operatorname{rank}\rho_N\leq K$, we have
$K\Tr\rho_N^2\geq1$ and $\lambda\geq1$. Therefore,
\begin{equation}
	p:=\Tr(\rho_NE_{\rho_N})
	\geq\frac{a}{a\lambda+2M{\sqrt{c}}}.
\end{equation}
Intersecting with the spectral event of
Lemma~\ref{lem:random-spectrum}, on which
$\lambda\leq1+C_1{\sqrt{c}}+C_2/N$, yields
\[
p\geq1-C({\sqrt{c}}+N^{-1}).
\]
Moreover, $M\geq1$, $\Tr A_v=1$, and ${c\leq\sqrt{c}}$ imply
\begin{equation}
 \begin{aligned}
 \Tr(A_vE_{\rho_N})
 &\geq\frac{ac+(M-1){\sqrt{c}}}{a\lambda+2M{\sqrt{c}}}\geq0,\\
 \Tr(A_vE_{\rho_N})
 &\leq\frac{ac+(M+1){\sqrt{c}}}{a\lambda+2M{\sqrt{c}}}\leq Z_*{\sqrt{c}},\\
 Z_*&=\frac{1+a+M}{a}.
 \end{aligned}
\end{equation}
Set $z=Z_*{\sqrt{c}}$. Since ${\sqrt{c}}\to0$,
\[
p-z\geq1-(C+Z_*){\sqrt{c}}-C/N>0
\]
for all sufficiently large $N$. We may therefore apply
Proposition~\ref{prop:cpwp-distillation}. Its exponent satisfies
\begin{align}
 d_{\mathrm{bin}}(p\|z)
 &\geq p\ln\frac1{Z_*{\sqrt{c}}}-h_2(p)\ln2\notag\\
 &\geq\frac{\ell_N\ln3}{2}-\ln Z_*-o(1),
\end{align}
where $h_2$ is the binary entropy in bits. Indeed,
$\ell_N{\sqrt{c}}\to0$ and $\ell_N/N\to0$, so the acceptance error
contributes only $o(1)$ to the exponent. The proposition gives
\[
D_{\CPWP}^{\Strange}(\rho_N)
\geq\alpha\ell_N-\frac{\ln Z_*}{\ln(5/3)}-o(1).
\]
This proves part~(i) with $C_0=\ln Z_*/\ln(5/3)$.

\medskip
\noindent\emph{(ii) $\ell_N=\Theta(N)$ and $\ell_N\leq N$.}
{Set $H=K\rho_N$ and $\lambda=\|H\|_\infty$.} We first establish the properties of $H$. Since $\operatorname{rank}\rho_N\leq K$, we have
$\lambda=K\|\rho_N\|_\infty\geq1$.
By Lemma~\ref{lem:random-spectrum}, with probability tending to one,
\begin{equation}
	0\preceq H\preceq\lambda I,\qquad
	1\leq\lambda\leq1+3\sqrt c+\frac4N.
	\label{eq:theta-H-spectrum}
\end{equation}
Also, $\operatorname{rank}\rho_N\leq K$ gives
\begin{equation}
 \Tr(\rho_NH)=K\Tr\rho_N^2\geq1.
 \label{eq:theta-H-acceptance}
\end{equation}

To estimate $\Tr(A_vH)$, write
$x_v=[\Tr(A_vH)-{c}]/{\sqrt{c}}$.
The Haar moment bound~\eqref{eq:residual-phase-moments}
and the circuit estimate~\eqref{eq:residual-circuit-phase-event}
remain valid throughout $1\leq K\leq d$, under the circuit
moment condition~\eqref{eq:random-moment-condition}.
Consequently, for either ensemble, there is an absolute
constant $B>0$ such that
\[
\Pr\!\left\{\max_v|x_v|\leq B\sqrt N\right\}
\longrightarrow1.
\]
Thus, setting $t={c}+B\sqrt N\,{\sqrt{c}}$, we have
\begin{equation}
	|\Tr(A_vH)|\leq t\qquad\text{for every }v
	\label{eq:theta-H-phase}
\end{equation}
with probability tending to one.

On the intersection of these events, define
\begin{equation}
 E=\frac{H+tI}{\lambda+t}.
 \label{eq:theta-test-effect}
\end{equation}
Equation~\eqref{eq:theta-H-spectrum} implies $0\preceq E\preceq I$.
By Eq.~\eqref{eq:theta-H-acceptance}, its target acceptance satisfies
\begin{equation}
\begin{aligned}
 p:=\Tr(\rho_NE)
 &\geq\frac{1+t}{\lambda+t}
 =1-\frac{\lambda-1}{\lambda+t}\\
 &\geq1-3{\sqrt{c}}-\frac4N.
\end{aligned}
 \label{eq:theta-test-acceptance}
\end{equation}
Using $\Tr A_v=1$ and Eq.~\eqref{eq:theta-H-phase}, we also obtain
\begin{equation}
 0\leq\Tr(A_vE)
 \leq\frac{2t}{\lambda+t}
 \leq2(B+1)\sqrt N\,{\sqrt{c}}=:z.
 \label{eq:theta-test-phase}
\end{equation}
Since $\ell_N=\Theta(N)$, we have $p\to1$ and $z\to0$, so $z<p$
for all sufficiently large $N$. Proposition~\ref{prop:cpwp-distillation}
therefore applies. Its exponent obeys
\begin{align}
 d_{\mathrm{bin}}(p\|z)
 &\geq p\ln\frac1z-\ln2\notag\\
 &\geq\frac{\ell_N\ln3}{2}-\frac{\ln N}{2}-O(1).
 \label{eq:theta-test-exponent}
\end{align}
Here
$\ln(1/z)=\ell_N\ln3/2-\ln N/2-\ln[2(B+1)]$,
and $p\to 1$. It follows that
\[
D_{\CPWP}^{\Strange}(\rho_N)
\geq\alpha\ell_N-\frac{\ln N}{2\ln(5/3)}-O(1),
\]
which proves part~(ii).

\medskip
\noindent\emph{(iii) $\ell_N=O(1)$.}
Write $\rho=\rho_N$. On the event of Lemma~\ref{lem:bounded-witness}, set
\begin{equation}
 E_\rho=\frac{M_LI-F_\rho}{2M_L},\qquad
 \varepsilon=\frac{\delta}{2M_L}.
 \label{eq:bounded-physical-effect}
\end{equation}
The three witness properties give
\begin{equation}
 \begin{gathered}
 0\preceq E_\rho\preceq I,\\
 \Tr(\rho E_\rho)
 =\frac12-\frac{\Tr(\rho F_\rho)}{2M_L}
 \geq\frac12+\varepsilon,\\
 \Tr(A_vE_\rho)
 =\frac12-\frac{\Tr(A_vF_\rho)}{2M_L}
 \in\left[\frac{M_L-1}{2M_L},\frac12\right].
 \end{gathered}
 \label{eq:bounded-physical-test-bounds}
\end{equation}
Here the first line uses the norm bound, and the last uses
$\Tr A_v=1$ and $M_L\geq1$.

Set $p=\Tr(\rho E_\rho)$ and $z=1/2$. Then
$p\geq1/2+\varepsilon>z$, so
Proposition~\ref{prop:cpwp-distillation} gives
\begin{equation}
	\begin{aligned}
		D_{\CPWP}^{\Strange}(\rho)
		&\geq\frac{d_{\mathrm{bin}}(p\|1/2)}{\ln(5/3)}\\
		&\geq\frac{2\varepsilon^2}{\ln(5/3)}
		=D_L>0.
	\end{aligned}
	\label{eq:bounded-relative-entropy-rate}
\end{equation}
The second inequality follows from Pinsker's inequality,
$d_{\mathrm{bin}}(p\|1/2)\geq2(p-1/2)^2$,
and $p\geq1/2+\varepsilon$.
These estimates hold with probability tending to one,
proving part~(iii).
\end{proof}

Combining Theorem~\ref{thm:random-three-rates}(3)
with Theorem~\ref{thm:random-cpwp-distillation}(i)--(ii) and \(D_{\CPWP}\leq C_{\CPWP}\), we have
\begin{corollary}
	\label{cor:random-cpwp-leading-rates}
	Under the assumptions of
	Theorem~\ref{thm:random-cpwp-distillation},
	if $\ell_N\to\infty$, then
	\begin{equation}
		\begin{aligned}
			C_{\CPWP}^{\Strange}(\rho_N)
			&=\alpha\ell_N+o_{\Pr}(\ell_N),\\
			D_{\CPWP}^{\Strange}(\rho_N)
			&=\alpha\ell_N+o_{\Pr}(\ell_N){.}
		\end{aligned}
		\label{eq:random-cpwp-leading-rates}
	\end{equation}
\end{corollary}

\endgroup

\section{Conclusion}
We have established typical {asymptotic reversibility} under
CPWP operations in random mixed-state models, with conversion
rates determined by ratios of R\'enyi-2 entropy deficits
when the system--environment imbalance diverges.
Our results further reveal an operational separation between
the resource theories of Wigner negativity and nonstabilizerness:
under CSP operations, typical states remain irreversible
throughout $0\le\ell_N\le N$ whenever the formation cost is finite.
The Strange state serves as a common currency for these
{asymptotically reversible} conversions.
{For any pure magic currency $\phi$, the limiting CPWP
reversibility ratio equals
$\eta_{\CPWP}^{\phi}(\Strange)$.}

These findings give the entropy deficit an operational meaning
in the conversion of typical quantum resources.
They show that a simple reversible description emerges
under CPWP operations, while the same states retain
irreversible losses under CSP operations.
The recoverability of magic therefore depends on both the
state and the operational constraints.
A natural question is how broadly the entropy-deficit
conversion law extends beyond the random-state ensembles
considered here.
More broadly, our results motivate the search for typical
reversible regimes in other quantum resource theories,
where simple conversion laws may emerge even when
universal reversibility is unavailable.
\label{sec:conclusion}

\begin{acknowledgments}
This work was partially supported by the Quantum Science and Technology–
National Science and Technology Major Project (Grant
No. 2021ZD0302901) and the National Natural Science Foundation of China (Grant No. 62672467 and
No. 62102388).

 GPT-6 Astra assisted with the random-state analysis and the proofs, particularly the proof of Lemma~\ref{lem:random-centered-residual}.
GPT-6 Astra also assisted in drafting and revising the manuscript.
\end{acknowledgments}

\appendix
\begingroup

\section{Properties of mixed-state stabilizer nullity}
\label{app:mixed-nullity}

We establish the properties of the mixed-state stabilizer nullity $V$
used in the CSP formation converse.

\subsection{CSP monotonicity}

Pure-state nullity is known to be nonincreasing under stabilizer
operations~\cite{Beverland2020}. To extend this
monotonicity to mixed states under the full CSP class, we first
derive the following Kraus representation of CSP channels.

\begin{lemma}[Stabilizer Kraus representation]
	\label{lem:csp-kraus}
	Every CSP channel admits a Kraus decomposition
	\begin{equation}
		\Lambda(X)=\sum_j K_j X K_j^\dagger,
		\qquad
		K_j=\sqrt{c_j}\,U_jP_j,
		\label{eq:kraus-polar}
	\end{equation}
	where $c_j>0$, each $P_j$ is a stabilizer-code
	projector, and {each $U_j$ is a Clifford isometry}. Let $\Pi_j$
	denote the projector onto the output support of branch $j$.
	It is a stabilizer-code projector and satisfies
	\[
	K_jK_j^\dagger=c_j\Pi_j.
	\]
	For every pure input $\psi$ and every nonzero branch,
	\begin{equation}
		\nu\!\left(
		\frac{K_j\psi K_j^\dagger}
		{\Tr(K_j\psi K_j^\dagger)}
		\right)
		\leq \nu(\psi).
		\label{eq:branch-nullity}
	\end{equation}
\end{lemma}

\begin{proof}
{The Choi characterization of CSP channels
\cite[Lemma~2]{Heimendahl2022}, together with the bipartite
stabilizer-state normal form \cite[Theorem~6]{Looi2011},
gives the Kraus representation}
\[
\Lambda(X)=\sum_j K_j X K_j^\dagger,
\qquad K_j=\sqrt{c_j}\,U_jP_j,
\]
where {$U_j$ is the clifford isometry specified above}, $P_j$ is a stabilizer-code
projector, and $c_j>0$. Trace preservation implies
$\sum_j c_jP_j=I$.
	
	Now let $\ket{\psi}$ be a pure input for which
	$P_j\ket{\psi}\neq0$. The normalized output of branch $j$ is
	\[
	\frac{K_j\ket{\psi}}{\|K_j\ket{\psi}\|}
	=
	U_j\frac{P_j\ket{\psi}}{\|P_j\ket{\psi}\|}.
	\]
	{The code isometry $U_j$ preserves pure-state stabilizer nullity.}
	Moreover, $P_j$ is a stabilizer-code
	projector. {It cannot increase
	pure-state stabilizer nullity.} Therefore,
	\[
	\nu\!\left(
	\frac{K_j\ket{\psi}}{\|K_j\ket{\psi}\|}
	\right)
	=
	\nu\!\left(
	\frac{P_j\ket{\psi}}{\|P_j\ket{\psi}\|}
	\right)
	\leq\nu(\psi).
	\]
	This is equivalent to Eq.~\eqref{eq:branch-nullity}.
\end{proof}

The branchwise inequality immediately implies CSP monotonicity
of the mixed-state nullity. Indeed, applying
Eq.~\eqref{eq:branch-nullity} to each component of a pure-state
decomposition of $\rho$ produces an ensemble of $\Lambda(\rho)$
with no larger average nullity. Taking the infimum over input
decompositions gives
\begin{equation}
 V(\Lambda(\rho))\leq V(\rho).
 \label{eq:roof-monotonicity}
\end{equation}

\subsection{Lower semicontinuity}

We next establish lower semicontinuity of $V$.
Pure-state nullity can change discontinuously under arbitrarily
small perturbations, so we use the equivalent formulation given
in the following lemma. This formulation replaces the optimization
over pure-state ensembles by one over positive semidefinite operators
$X_P$ supported on stabilizer codes. On a fixed system, the code
projectors form a finite set, while the objective is linear in
the operators $X_P$, allowing a compactness argument to control
the limit.

\begin{lemma}
	\label{lem:roof-code-program}
	For every quantum state $\rho$ on $N$ qutrits,
	\begin{equation}
		\begin{aligned}
			V(\rho)
			&=\min_{\{X_P\}}
			\sum_P\log_3(\operatorname{rank}P)\,\Tr X_P,\\
			&X_P\succeq0,\qquad
			X_P=PX_PP,\qquad
			\sum_PX_P=\rho,
		\end{aligned}
		\label{eq:roof-code-program}
	\end{equation}
	where $P$ ranges over all stabilizer-code projectors on $N$
	qutrits, including the identity.
\end{lemma}

\begin{proof}
	For a pure state $\psi$, let $P_\psi$ be the code projector
	defined by its full stabilizer group:
	\[
	P_\psi:=\frac1{|\Stab(\psi)|}
	\sum_{g\in\Stab(\psi)}g.
	\]
	Since
	\[
	\operatorname{rank}P_\psi
	=\frac{3^N}{|\Stab(\psi)|}
	=3^{\nu(\psi)},
	\]
	we have
	\begin{equation}
		\nu(\psi)=\log_3\operatorname{rank}P_\psi.
		\label{eq:nullity-code-rank}
	\end{equation}
	
	Consider an arbitrary pure-state decomposition
	\[
	\rho=\sum_a p_a\psi_a.
	\]
	Grouping its terms according to their full stabilizer-code
	projectors, define
	\[
	X_P=\sum_{a:P_{\psi_a}=P}p_a\psi_a.
	\]
	Then
	\[
	X_P\succeq0,\qquad X_P=PX_PP,\qquad
	\sum_PX_P=\rho,
	\]
	and
	\[
	\sum_P\log_3(\operatorname{rank}P)\Tr X_P
	=\sum_a p_a\nu(\psi_a).
	\]
	Taking the infimum over pure-state decompositions gives
	\begin{equation}
		\min_{\{X_P\}}
		\sum_P\log_3(\operatorname{rank}P)\Tr X_P
		\leq V(\rho).
		\label{eq:code-program-upper}
	\end{equation}
	
	Conversely, let $\{X_P\}$ be any feasible family and take a
	spectral decomposition over the nonzero spectrum,
	\[
	X_P=\sum_{\alpha:x_{P,\alpha}>0}
	x_{P,\alpha}\varphi_{P,\alpha},
	\]
	where $\varphi_{P,\alpha}$ denotes a normalized pure-state
	projector. Since $X_P=PX_PP$,
	\[
	P\varphi_{P,\alpha}P=\varphi_{P,\alpha}.
	\]
	If $\operatorname{rank}P=3^k$, the stabilizer group defining
	$P$ supplies $N-k$ independent Pauli stabilizers of every
	vector in its range. Therefore
	\[
	\nu(\varphi_{P,\alpha})
	\leq k
	=\log_3\operatorname{rank}P.
	\]
	The spectral decompositions of all $X_P$ together form a
	pure-state decomposition of $\rho$, so
	\[
	\begin{aligned}
		V(\rho)
		&\leq
		\sum_{P,\alpha}
		x_{P,\alpha}\nu(\varphi_{P,\alpha})\\
		&\leq
		\sum_P\log_3(\operatorname{rank}P)\Tr X_P.
	\end{aligned}
	\]
	Taking the minimum over {feasible families} and combining this
	inequality with Eq.~\eqref{eq:code-program-upper} proves
	Eq.~\eqref{eq:roof-code-program}.
 There are finitely many stabilizer-code
projectors {in a fixed dimension}, and the feasible set in
Eq.~\eqref{eq:roof-code-program} is closed and bounded
because $\sum_P\Tr X_P=1$. Hence its minimum is attained.
\end{proof}

\begin{lemma}[Lower semicontinuity]
	\label{lem:roof-properties}
	The convex roof $V$ is lower semicontinuous on every fixed system:
	\begin{equation}
		\lim_{\epsilon\downarrow0}
		\inf_{\frac12\|\sigma-\rho\|_1\leq\epsilon}V(\sigma)
		=V(\rho).
		\label{eq:roof-lsc}
	\end{equation}
\end{lemma}

\begin{proof}
	Let $\rho_n\to\rho$ and pass
	to a subsequence such that
	\[
	V(\rho_n)\longrightarrow\liminf_mV(\rho_m).
	\]
	For each $n$, let $\{X_P^{(n)}\}$ be a minimizing family in
	Lemma~\ref{lem:roof-code-program}. Since there are finitely
	many $P$ and
	\[
	\sum_P\Tr X_P^{(n)}=1,
	\]
	compactness gives a further subsequence for which
	$X_P^{(n)}\to X_P$ for every $P$. The limiting family
	satisfies
	\[
	X_P\succeq0,\qquad X_P=PX_PP,\qquad
	\sum_PX_P=\rho.
	\]
	It is therefore feasible for $\rho$, and
	\[
	\begin{aligned}
		V(\rho)
		&\leq
		\sum_P\log_3(\operatorname{rank}P)\Tr X_P\\
		&=\lim_n V(\rho_n)
		=\liminf_mV(\rho_m).
	\end{aligned}
	\]
	Thus $V$ is lower semicontinuous.
	
	The neighborhood characterization of lower semicontinuity
	\cite[Theorem~1.6]{RockafellarWets1998} gives
	Eq.~\eqref{eq:roof-lsc}.
\end{proof}

\subsection{Strong superadditivity}

To bound the formation cost, we need to estimate the nullity
of many-copy outputs. A standard approach is to use strong
superadditivity.

\begin{lemma}[Strong superadditivity of $V$]
	\label{lem:strong-superadditivity}
	For every bipartite state $\rho_{AB}$,
	\begin{equation}
		V(\rho_{AB})\geq V(\rho_A)+V(\rho_B),
		\label{eq:strong-superadditivity}
	\end{equation}
	where $\rho_A=\operatorname{Tr}_B\rho_{AB}$ and
	$\rho_B=\operatorname{Tr}_A\rho_{AB}$.
\end{lemma}

\begin{proof}
	We first establish the pure-state inequality
	\begin{equation}
		V(\psi_A)+V(\psi_B)\leq \nu(\psi_{AB})
		\label{eq:pure-strong-superadditivity}
	\end{equation}
	for every pure state $|\psi\rangle_{AB}$.
	Let $N=N_A+N_B$ be the total number of qutrits and set
	\[
	k=\nu(\psi_{AB}).
	\]
	Let $P$ be the stabilizer-code projector defined by the full
	Pauli stabilizer group of $|\psi\rangle_{AB}$. Since this group
	has $N-k$ independent generators,
	\[
	\operatorname{rank}P=3^k,
	\qquad
	P|\psi\rangle=|\psi\rangle.
	\]
	Choose a logical computational basis
	$\{|x_L\rangle_{AB}:x\in\mathbb F_3^k\}$ for the code space
	$P$. Introduce
	a reference system $R$ of $k$ qutrits and define
	\begin{equation}
		|\chi\rangle_{ABR}
		=
		\frac{1}{\sqrt{3^k}}
		\sum_{x\in\mathbb F_3^k}
		|x_L\rangle_{AB}|x\rangle_R.
		\label{eq:logical-stabilizer-purification}
	\end{equation}
	This is a {stabilizer purification} of $\frac{P}{3^k}$. Tracing out $R$ gives
	\begin{equation}
		\chi_{AB}
		=
		\operatorname{Tr}_R|\chi\rangle\langle\chi|
		=
		\frac{1}{3^k}
		\sum_x |x_L\rangle\langle x_L|
		=
		\frac{P}{3^k}.
		\label{eq:chi-projector}
	\end{equation}
	By the tripartite stabilizer-state normal form, there exist local
	Clifford unitaries $U_A,U_B,U_R$ such that
	\begin{equation}
		\begin{aligned}
			|\widetilde\chi\rangle
			&=(U_A\otimes U_B\otimes U_R)|\chi\rangle\\
			&=
			|\mathrm{fix}\rangle_{A_sB_s}
			\otimes|\Phi\rangle_{A_0R_A}^{\otimes a}
			\otimes|\Phi\rangle_{B_0R_B}^{\otimes b}\\
			&\qquad\otimes
			|\mathrm{GHZ}\rangle_{A_cB_cR_C}^{\otimes t},
		\end{aligned}
		\label{eq:tripartite-stabilizer-normal-form}
	\end{equation}
	where
	\[
	|\Phi\rangle
	=\frac{1}{\sqrt3}\sum_{j=0}^2|j\rangle|j\rangle,
	\qquad
	|\mathrm{GHZ}\rangle
	=\frac{1}{\sqrt3}\sum_{j=0}^2|j\rangle|j\rangle|j\rangle.
	\]
	Here $|\mathrm{fix}\rangle_{A_sB_s}$ contains the local pure
	stabilizer factors and the Bell pairs between $A$ and $B$.
	The systems $A_0,B_0$ contain $a,b$ qutrits, respectively,
	and $A_c,B_c$ contain $t$ qutrits. The reference system
	decomposes as $R=R_A R_B R_C$. Since $\chi_R=I_R/3^k$ is
	maximally mixed, there are no pure local factors
	on $R$, and hence
	\begin{equation}
		k=a+b+t.
		\label{eq:reference-count}
	\end{equation}
	
	Write
	\[
	\widetilde P=(U_A\otimes U_B)P(U_A\otimes U_B)^\dagger.
	\]
	Equation~\eqref{eq:chi-projector} implies
	$\widetilde\chi_{AB}=\widetilde P/3^k$. Taking the partial trace
	over $R$ in Eq.~\eqref{eq:tripartite-stabilizer-normal-form}
	gives
	\begin{equation}
		\begin{aligned}
			\frac{\widetilde P}{3^k}
			={}&
			|\mathrm{fix}\rangle\langle\mathrm{fix}|_{A_sB_s}
			\otimes\frac{I_{A_0}}{3^a}
			\otimes\frac{I_{B_0}}{3^b}\\
			&\otimes
			\frac{1}{3^t}
			\sum_{z\in\mathbb F_3^t}
			|z\rangle\langle z|_{A_c}
			\otimes |z\rangle\langle z|_{B_c}.
		\end{aligned}
		\label{eq:projector-partial-trace}
	\end{equation}
	Multiplying by $3^k=3^{a+b+t}$, we obtain the structure of the
	entire code projector:
	\begin{equation}
		\begin{aligned}
			\widetilde P
			={}&
			|\mathrm{fix}\rangle\langle\mathrm{fix}|_{A_sB_s}
			\otimes I_{A_0}\otimes I_{B_0}\\
			&\otimes
			\sum_{z\in\mathbb F_3^t}
			|z\rangle\langle z|_{A_c}
			\otimes |z\rangle\langle z|_{B_c}.
		\end{aligned}
		\label{eq:code-projector-normal-form}
	\end{equation}

	Since $|\psi\rangle\in\operatorname{ran}P$, its locally
	Clifford-transformed state
	$|\widetilde\psi\rangle=(U_A\otimes U_B)|\psi\rangle$
	belongs to $\operatorname{ran}\widetilde P$. Consequently,
	there exist unnormal states
	$|\varphi_z\rangle_{A_0B_0}$ such that
	\begin{equation}
		|\widetilde\psi\rangle
		=
		|\mathrm{fix}\rangle_{A_sB_s}
		\otimes
		\sum_{z\in\mathbb F_3^t}
		|z\rangle_{A_c}|z\rangle_{B_c}
		|\varphi_z\rangle_{A_0B_0},
		\label{eq:arbitrary-code-vector}
	\end{equation}
	with $\sum_z\langle\varphi_z|\varphi_z\rangle=1$.

Tracing out $B$ in Eq.~\eqref{eq:arbitrary-code-vector}
and using the orthogonality of the states $|z\rangle_{B_c}$,
we obtain
\begin{equation}
	\widetilde\psi_A
	=
	\tau_{A_s}\otimes
	\sum_{z\in\mathbb F_3^t}
	|z\rangle\langle z|_{A_c}\otimes\sigma_z,
	\label{eq:marginal-A}
\end{equation}
where
\[
\tau_{A_s}
=\operatorname{Tr}_{B_s}
|\mathrm{fix}\rangle\langle\mathrm{fix}|,
\qquad
\sigma_z
=\operatorname{Tr}_{B_0}
|\varphi_z\rangle\langle\varphi_z|.
\]
Let $p_z=\operatorname{Tr}\sigma_z$, so that $\sum_zp_z=1$.
Since $\tau_{A_s}\in\STAB$ and each $|z\rangle_{A_c}$
is a stabilizer state, convexity and invariance under
adjoining stabilizer ancillas give
\[
V(\widetilde\psi_A)
\leq
\sum_{z:p_z>0}p_zV(\sigma_z/p_z)
\leq a,
\]
where the last inequality follows because each normalized
block acts on $a$ qutrits.

By local Clifford invariance, $V(\psi_A)\leq a$.
Similarly, $V(\psi_B)\leq b$. Hence,
using Eq.~\eqref{eq:reference-count},
\[
V(\psi_A)+V(\psi_B)
\leq a+b
\leq a+b+t
=k
=\nu(\psi_{AB}),
\]
which proves Eq.~\eqref{eq:pure-strong-superadditivity}.
	
	Let $\rho_{AB}$ be an arbitrary mixed state and consider any
	pure-state ensemble
	\[
	\rho_{AB}=\sum_i p_i\psi_i,
	\qquad
	\psi_i=|\psi_i\rangle\langle\psi_i|.
	\]
	Its marginals satisfy
	\[
	\rho_A=\sum_i p_i\psi_{i,A},
	\qquad
	\rho_B=\sum_i p_i\psi_{i,B}.
	\]
	Since $V$ is a convex roof, it is convex. Hence
	\[
	\begin{aligned}
		V(\rho_A)+V(\rho_B)
		&\leq
		\sum_i p_i\bigl[V(\psi_{i,A})+V(\psi_{i,B})\bigr]\\
		&\leq
		\sum_i p_i\nu(\psi_i),
	\end{aligned}
	\]
	where the second inequality follows from the pure-state result
	proved above. This holds for every pure-state ensemble of
	$\rho_{AB}$. Taking the infimum over all such ensembles yields
	\[
	V(\rho_A)+V(\rho_B)
	\leq
	\inf_{\rho_{AB}=\sum_i p_i\psi_i}
	\sum_i p_i\nu(\psi_i)
	=V(\rho_{AB}).
	\]
	This establishes Eq.~\eqref{eq:strong-superadditivity}.
\end{proof}

\par
\endgroup

\begingroup

\section{\textbf{Phase-space estimates for random mixed states}}
\label{app:random-phase-space}

We prove the two phase-space estimates used in
Theorem~\ref{thm:random-cpwp-distillation}. The ensembles and moment
condition are those of Sec.~\ref{subsec:random-models}. Throughout,
\[
 \begin{gathered}
 d=3^N,\qquad K=3^{N-\ell_N},\\
 {c=K/d=3^{-\ell_N}},
 \end{gathered}
\]
and $\mathcal V=\mathbb F_3^{2N}$ denotes the phase space.
We use the spectral estimates in Lemma~\ref{lem:random-spectrum}.

For a scalar random variable $Y$ with law $\mu$, write
\[
 \|Y\|_{L_p(\mu)}=(\mathbb E_\mu|Y|^p)^{1/p},\qquad 1\leq p<\infty.
\]
We omit the law when it is clear; $L_p(H)$ and $L_p(C)$ refer to
the Haar and circuit ensembles. For matrices, $\|X\|_q$ is the
unnormalized Schatten $q$ norm and $\|X\|_\infty$ the operator norm;
in particular, $\|X\|_2$ is the Hilbert--Schmidt norm. Thus
\[
 \bigl\|\|X\|_q\bigr\|_{L_p(\mu)}
 =\bigl(\mathbb E_\mu\|X\|_q^p\bigr)^{1/p}.
\]
The matrix-norm index $q$ and the probability-moment order $p$ are
independent.

\subsection{\textbf{Centered residual bound}}
\label{app:centered-residual}

Let
\begin{equation}
 a=e^{-1/2},\qquad g(x)=\sin x-ax,
 \label{eq:residual-centered-function}
\end{equation}
and define, for any state $\rho$,
\begin{equation}
 \begin{aligned}
 x_v&=\sqrt{dK}\,\Tr(A_v\rho)-{\sqrt{c}},\\
 B_\rho&=\frac {\sqrt{c}}d\sum_{v\in\mathcal V}\sin(x_v)A_v,\\
 R_\rho&=\frac {\sqrt{c}}d\sum_{v\in\mathcal V}g(x_v)A_v.
 \end{aligned}
 \label{eq:residual-construction}
\end{equation}
Phase-point orthogonality and $\sum_v A_v=dI$ give
\begin{equation}
 \begin{aligned}
 \frac {\sqrt{c}}d\sum_v x_vA_v&=K\rho-cI,\\
 R_\rho&=B_\rho-a(K\rho-cI),\\
 \Tr(A_vB_\rho)&={\sqrt{c}}\sin(x_v).
 \end{aligned}
 \label{eq:random-centered-identities}
\end{equation}
In particular, $\max_v|\Tr(A_vB_\rho)|\leq {\sqrt{c}}$ deterministically.
The choice of $a$ ensures that, for a real standard Gaussian $Z$,
\begin{equation}
 \mathbb E g(Z)=\mathbb E g'(Z)=\mathbb E g''(Z)=0.
 \label{eq:residual-three-cancellations}
\end{equation}
Indeed, $g$ and $g''$ are odd, whereas
$\mathbb E g'(Z)=\mathbb E\cos Z-a=e^{-1/2}-a$.

\begin{lemma}

\label{lem:random-centered-residual}
For either ensemble under the assumptions of
Lemma~\ref{lem:random-spectrum} {and with $\ell_N=o(N)$}, the residual in
Eq.~\eqref{eq:residual-construction} satisfies
\begin{equation}
 \Pr\!\left\{\|R_{\rho_N}\|_\infty\leq M_*{\sqrt{c}}\right\}
 \longrightarrow1
 \label{eq:random-centered-residual-bound}
\end{equation}
for an absolute constant $M_*>0$.
\end{lemma}

\begin{proof}

Fix the even moment order $m=2N$. Constants denoted by $C$ are
absolute and may increase between estimates.

\medskip
\noindent\emph{1. Bounding the normalization error \(R_{\rho_H}-R_X\).}
For the Haar-induced ensemble, write $\rho_H=W/\Tr W$, where
$W=GG^\dagger$ and $G$ is a $d\times K$ complex Ginibre matrix.
Set $\tau=\Tr W/(dK)$ and define
\begin{equation}
 \begin{aligned}
 X_v&=\frac{\Tr[A_v(W-KI)]}{\sqrt{dK}}
      =\tau(x_v+{\sqrt{c}})-{\sqrt{c}},\\
 R_X&=\frac {\sqrt{c}}d\sum_{v\in\mathcal V}g(X_v)A_v.
 \end{aligned}
 \label{eq:residual-unnormalized}
\end{equation}

Since \(X_v-x_v=(\tau-1)(x_v+{\sqrt{c}})\) and \(g\) is \((1+a)\)-Lipschitz, phase-point orthogonality gives
\[
\begin{aligned}
	\|R_{\rho_H}-R_X\|_2^2
	&=\frac{{c}}{d}\sum_v|g(x_v)-g(X_v)|^2\\
	&\le(1+a)^2|\tau-1|^2
	\frac{{c}}{d}\sum_v|x_v+{\sqrt{c}}|^2\\
	&=(1+a)^2|\tau-1|^2\|K\rho_H\|_2^2.
\end{aligned}
\]
thus we obtain,
\begin{equation}
	\|R_{\rho_H}-R_X\|_2
	\leq(1+a)|\tau-1|\,\|K\rho_H\|_2.
	\label{eq:residual-normalization-error}
\end{equation}

The gamma distribution of $dK\tau$ yields
$\|\tau-1\|_{L_m}\leq C(\sqrt{m/(dK)}+m/(dK))$.
Moreover, $\tau$ and $\rho_H$ are independent. Using
$\|K\rho_H\|_2^2\leq K\|K\rho_H\|_\infty$ and
Eq.~\eqref{eq:random-refined-haar-moment} at order $m/2$ gives
$\|\|K\rho_H\|_2\|_{L_m}\leq C\sqrt K$.
Consequently,
\begin{equation}
	\bigl\|\|R_{\rho_H}-R_X\|_2\bigr\|_{L_m}
	\leq C\sqrt{m/d}=o({\sqrt{c}}),
	\label{eq:residual-normalization-moment}
\end{equation}
because $\sqrt{m/d}/{\sqrt{c}}=\sqrt{m/K}\to0$.
Since the coordinates $X_v$ are dependent. Let $(Z_v)_{v\in\mathcal V}$ be independent
standard real Gaussian variables and define
\begin{equation}
	R_Z=\frac {\sqrt{c}}d\sum_{v\in\mathcal V}g(Z_v)A_v.
	\label{eq:residual-gaussian-model}
\end{equation}
We will compare the trace moments of $R_X$ and $R_Z$.

\medskip
\noindent\emph{2. \(\bigl||M_Q|_\infty\bigr|_{L_p}\leq C\).}
For a fixed subset $Q\subseteq\mathcal V$, define
\begin{equation}
 \begin{aligned}
 M_Q&=\frac1d\sum_{v\notin Q}g(Z_v)A_v,\\
 \end{aligned}
 \label{eq:residual-MQ}
\end{equation}
We prove that
\begin{equation}
 \bigl\|\|M_Q\|_\infty\bigr\|_{L_p}\leq C,
 \qquad 1\leq p\leq m=2N,
 \label{eq:residual-MQ-moment}
\end{equation}
with an absolute constant independent of $Q$, $p$, and $N$.
Write $M_Z:=M_\varnothing$.

We first estimate the Gaussian matrix
\[
 {L_Z}=\frac1d\sum_{v\in\mathcal V}Z_vA_v.
\]
For any unit vector $\psi\in\mathbb C^d$, phase-point orthogonality
gives
\[
 \sum_v|\langle\psi,A_v\psi\rangle|^2=d.
\]
Hence $\langle\psi,{L_Z}\psi\rangle$ is a centered real Gaussian
variable with variance $1/d$. Choose a $1/4$-net $\mathcal N$ of
the unit sphere with $|\mathcal N|\leq9^{2d}$. Since ${L_Z}$ is
Hermitian,
\[
 \|{L_Z}\|_\infty
 \leq2\max_{\psi\in\mathcal N}|\langle\psi,{L_Z}\psi\rangle|.
\]
The Gaussian tail bound and a union bound yield
\[
 \Pr\{\|{L_Z}\|_\infty>t\}
 \leq2\exp\!\left(2d\ln9-\frac{dt^2}{8}\right).
\]
For $t\geq\sqrt{32\ln9}$, this is at most $2e^{-dt^2/16}$.
Integrating the tail bound gives
\[
 \bigl\|\|{L_Z}\|_\infty\bigr\|_{L_p}
 \leq C\left(1+\sqrt{\frac pd}\right)\leq C,
\]
for $1\leq p\leq m$, where the last inequality uses
$m=2N\leq d=3^N$.

We next compare $M_Z$ with ${L_Z}$. Set $r_v=|Z_v|$ and write
$Z_v=\varepsilon_vr_v$, where
\[
 \Pr(\varepsilon_v=1)=\Pr(\varepsilon_v=-1)=\frac12.
\]
The variables $\varepsilon_v$ are mutually independent and
independent of all $r_v$. Since $g$ is odd,
\[
 g(Z_v)=\varepsilon_vg(r_v).
\]
Fix all $r_v$ and all $\varepsilon_w$ with $w\ne v$.
Writing $B$ for the sum of the other terms, the average over
$\varepsilon_v$ is $F(g(r_v))$, where
\[
 F(t):=\frac12\left(
 \left\|B+\frac td A_v\right\|_\infty^p
 +
 \left\|B-\frac td A_v\right\|_\infty^p
 \right).
\]
For $p\geq1$, $F$ is even and convex, so $F(t)\leq F(s)$ whenever
$|t|\leq s$. Since
\[
 |g(r_v)|\leq(1+a)r_v,
\]
replacing $g(r_v)$ by $(1+a)r_v$ does not decrease this average.
Applying this comparison successively to every $v$, and then
taking expectation over all $r_v$, gives
\[
 \mathbb E\|M_Z\|_\infty^p
 \leq(1+a)^p\mathbb E\|{L_Z}\|_\infty^p
 \leq C^p.
\]

Finally, for any fixed $Q$, independence and
$\mathbb E g(Z_v)=0$ imply
\[
 M_Q=\mathbb E\!\left[
 M_Z\,\middle|\,(Z_v)_{v\notin Q}
 \right].
\]
Indeed, the terms indexed by $Q$ have conditional mean zero,
and the remaining terms are fixed by the conditioning.
Conditional Jensen's inequality therefore gives
\[
 \begin{aligned}
 \mathbb E\|M_Q\|_\infty^p
 &\leq\mathbb E\!\left[
 \mathbb E\!\left[
 \|M_Z\|_\infty^p
 \,\middle|\,(Z_v)_{v\notin Q}
 \right]\right]\\
 &=\mathbb E\|M_Z\|_\infty^p\leq C^p.
 \end{aligned}
\]
Taking $p$th roots proves the claimed bound uniformly in $Q$.

In particular, since $R_Z={\sqrt{c}}M_Z$ and $m$ is even,
\begin{equation}
 \begin{aligned}
 \mathbb E\Tr R_Z^m
 &\leq d\,\mathbb E\|R_Z\|_\infty^m\\
 &\leq d(C{\sqrt{c}})^m.
 \end{aligned}
 \label{eq:residual-gaussian-trace-moment}
\end{equation}

\medskip
\noindent\emph{3. Expanding the moment difference \(E_m\).}
Define
\begin{equation}
 E_m:=\mathbb E\Tr R_X^m-\mathbb E\Tr R_Z^m.
 \label{eq:residual-moment-difference}
\end{equation}
The first expansion is simply the distributive law for the matrix
power. Keeping the order of the matrices unchanged gives
\begin{equation}
 R_X^m=\left(\frac {\sqrt{c}}d\right)^m
 \sum_{\boldsymbol v\in\mathcal V^m}
 \left[\prod_{j=1}^m g(X_{v_j})\right]A_{v_1}\cdots A_{v_m}.
 \label{eq:residual-matrix-power-expansion}
\end{equation}
Taking the trace and expectation, and subtracting the analogous
formula for $R_Z$, yields
\begin{equation}
 \begin{aligned}
 E_m&=\left(\frac {\sqrt{c}}d\right)^m
 \sum_{\boldsymbol v\in\mathcal V^m}
 \Tr(A_{v_1}\cdots A_{v_m})\Delta_{\boldsymbol v}^g,\\
 \Delta_{\boldsymbol v}^g
 &:=\mathbb E\prod_{j=1}^m g(X_{v_j})
   -\mathbb E\prod_{j=1}^m g(Z_{v_j}).
 \end{aligned}
 \label{eq:residual-trace-comparison}
\end{equation}
The $X_v$ need not be independent for this identity.

The second expansion concerns $\Delta_{\boldsymbol v}^g$ for a phase word $\boldsymbol v=(v_1,\ldots,v_m)$.
Put $S_{\boldsymbol v}(z)=\sum_{j=1}^m z_jA_{v_j}$ and
$\Gamma_{ij}=\delta_{v_i v_j}$. Gaussian integration over the columns
of $G$ gives
\begin{equation}
 \begin{aligned}
 \mathcal M_X(z)&:=\mathbb E\exp\!\left(\sum_jz_jX_{v_j}\right)\\
 &=e^{-{\sqrt{c}}\sum_jz_j}
   \det\!\left(I-\frac{S_{\boldsymbol v}(z)}{\sqrt{dK}}\right)^{-K}\\
 &=\mathcal M_Z(z)e^{\mathcal Q_{\boldsymbol v}(z)},\qquad
 \mathcal M_Z(z)=e^{z^{\mathsf T}\Gamma z/2}.
 \end{aligned}
 \label{eq:residual-mgf}
\end{equation}
The second line follows from Gaussian integration over the $K$
independent columns of $G$, with the factor
$e^{-{\sqrt{c}}\sum_j z_j}$ accounting for the centering in $X_{v_j}$.
The third line follows from the convergent expansion
\[
-\mathop{{\ln}}\nolimits\det\!\left(I-\frac{S_{\boldsymbol v}(z)}{\sqrt{dK}}\right)
=\sum_{q\geq1}
\frac{\Tr[S_{\boldsymbol v}(z)^q]}{q(dK)^{q/2}},
\]
valid for $z$ sufficiently close to zero.
Using $\Tr A_v=1$, the linear contribution to $\mathop{{\ln}}\nolimits \mathcal M_X(z)$ is
\[
\frac{K}{\sqrt{dK}}\Tr S_{\boldsymbol v}(z)
={\sqrt{c}}\sum_j z_j,
\]
which cancels the centering term $-{\sqrt{c}}\sum_j z_j$.
Using $\Tr(A_vA_w)=d\,\delta_{vw}$, the quadratic contribution is
\[
\frac{1}{2d}\Tr[S_{\boldsymbol v}(z)^2]
=\frac12\sum_{i,j}z_i z_j\delta_{v_i v_j}
=\frac12 z^{\mathsf T}\Gamma z.
\]
The remaining terms, of degree at least three, are collected in
$\mathcal Q_{\boldsymbol v}(z)$:
\begin{equation}
 \begin{aligned}
 \mathcal Q_{\boldsymbol v}(z)
 &=\sum_{q\geq3}\frac{K\Tr[S_{\boldsymbol v}(z)^q]}{q(dK)^{q/2}}\\
 &=\sum_{q\geq3}\sum_{i_1,\ldots,i_q=1}^m
 b_{q;i_1,\ldots,i_q}(\boldsymbol v)z_{i_1}\cdots z_{i_q},\\
 b_{q;i_1,\ldots,i_q}(\boldsymbol v)
 &=\frac{K}{q(dK)^{q/2}}
   \Tr(A_{v_{i_1}}\cdots A_{v_{i_q}}).
 \end{aligned}
 \label{eq:residual-Q}
\end{equation}
Here $j$ and the $i_k$ index positions in the word; $v_j$ is the
phase label at position $j$. 

For each $s\geq1$, expanding the product gives
\[
\begin{aligned}
	\frac{\mathcal Q_{\boldsymbol v}(z)^s}{s!}
	={}&
	\sum_{q_1,\ldots,q_s\geq3}
	\sum_{\substack{
			i_{\beta,k}\in\{1,\ldots,m\}\\
			1\leq\beta\leq s,\;1\leq k\leq q_\beta}}
	\frac1{s!}\\
	&\quad\times
	\prod_{\beta=1}^s
	\left[
	b_{q_\beta;i_{\beta,1},\ldots,i_{\beta,q_\beta}}
	(\boldsymbol v)
	\prod_{k=1}^{q_\beta}z_{i_{\beta,k}}
	\right].
\end{aligned}
\]
To specify a pattern $\mathcal P$, choose $s\geq1$ and block
lengths $q_1,\ldots,q_s\geq3$, and then choose an ordered tuple
$(i_{\beta,1},\ldots,i_{\beta,q_\beta})
\in\{1,\ldots,m\}^{q_\beta}$ for each block $\beta$.
The pattern records these choices as the ordered list
\[
\mathcal P=
\bigl(
(q_1;i_{1,1},\ldots,i_{1,q_1}),
\ldots,
(q_s;i_{s,1},\ldots,i_{s,q_s})
\bigr).
\]
Define its coefficient by
\[
c_{\mathcal P}(\boldsymbol v)
:=
\frac1{s!}\prod_{\beta=1}^s
b_{q_\beta;i_{\beta,1},\ldots,i_{\beta,q_\beta}}
(\boldsymbol v).
\]
Collecting equal powers of the parameters and summing over $s$
therefore gives
\[
\sum_{s\geq1}\frac{\mathcal Q_{\boldsymbol v}(z)^s}{s!}
=
\sum_{\mathcal P}c_{\mathcal P}(\boldsymbol v)
\prod_{j=1}^m z_j^{n_j(\mathcal P)}.
\]
Here the sum runs over all patterns, and $n_j(\mathcal P)$
counts the total number of occurrences of position $j$ among
all selected tuples, including repetitions within a block.
In particular, $n_j(\mathcal P)=0$ if position $j$ is never selected.

Substituting this expansion into
Eq.~\eqref{eq:residual-mgf} and using
$\mathcal M_Z(z)=\mathbb E\prod_j e^{z_jZ_{v_j}}$ gives
\[
\begin{aligned}
	\mathcal M_X(z)-\mathcal M_Z(z)
	&=\mathcal M_Z(z)\sum_{s\geq1}\frac{\mathcal Q_{\boldsymbol v}(z)^s}{s!}\\
	&=\sum_{\mathcal P}c_{\mathcal P}(\boldsymbol v)
	\mathbb E\prod_{j=1}^m
	\left[z_j^{n_j(\mathcal P)}e^{z_jZ_{v_j}}\right],
\end{aligned}
\]
Using
\[
g(x)=\frac{e^{ix}-e^{-ix}}{2i}
-a\left.\partial_{z_j}e^{z_jx}\right|_{z_j=0},
\]
termwise parameter evaluation and differentiation send
$z_j^n e^{z_jx}=\partial_x^n e^{z_jx}$ to $g^{(n)}(x)$.
Applying this in each $z_j$ gives
\begin{equation}
 \Delta_{\boldsymbol v}^g
 =\sum_{\mathcal P}c_{\mathcal P}(\boldsymbol v)
       \mathbb E\prod_{j=1}^m
          g^{(n_j(\mathcal P))}(Z_{v_j}).
 \label{eq:residual-delta-expansion}
\end{equation}
Let
\begin{equation}
 L(\mathcal P)=\sum_{\beta=1}^s q_\beta
 =\sum_{j=1}^m n_j(\mathcal P)
\end{equation}

These expansions are absolutely convergent. Indeed, uniformly in
$n\geq0$, $\|g^{(n)}(Z)\|_{L_m}\leq C\sqrt m$, so every derivative
product has absolute expectation at most $(C\sqrt m)^m$.
The crude bound $|\Tr(A_{v_{i_1}}\cdots A_{v_{i_q}})|\leq d$ gives,
for fixed $\boldsymbol v$,
\[
 \begin{aligned}
 \sum_{\mathcal P}|c_{\mathcal P}(\boldsymbol v)|
 &\leq e^{B_m}-1,\\
 B_m&=Kd\sum_{q\geq3}\frac1q
            \left(\frac m{\sqrt{dK}}\right)^q<\infty.
 \end{aligned}
\]
Also $2m<\sqrt{dK}$ for sufficiently large $N$, so the logarithmic
determinant series converges normally on a polydisc containing all
parameter evaluations used above. This justifies termwise parameter
differentiation and rearrangement of the series.

Substituting Eq.~\eqref{eq:residual-delta-expansion} into
Eq.~\eqref{eq:residual-trace-comparison} and interchanging the finite
phase-word sum with the absolutely convergent pattern sum yields
\begin{equation}
 \begin{aligned}
 E_m={}&\left(\frac {\sqrt{c}}d\right)^m
 \sum_{\mathcal P}\sum_{\boldsymbol v\in\mathcal V^m}
 c_{\mathcal P}(\boldsymbol v)
 \Tr(A_{v_1}\cdots A_{v_m})\\
 &\hspace{15mm}\times
 \mathbb E\prod_{j=1}^m g^{(n_j(\mathcal P))}(Z_{v_j}).
 \end{aligned}
 \label{eq:residual-full-comparison-sum}
\end{equation}
We next estimate the coefficient and the Gaussian factors in this
sum, keeping the sum over untouched labels inside the matrix trace.

\medskip
\noindent\emph{4. Bounding \(\lvert c_{\mathcal P}(\boldsymbol v)\rvert\).}
By the identity
\cite[Eq.~(2.12)]{GrossNezamiWalter2021},
\[
|\Tr(A_{w_1}A_{w_2}A_{w_3})|=1.
\]
For $q\geq4$, unitarity gives
$|\Tr(A_{w_1}\cdots A_{w_q})|\leq d$.
Thus Eq.~\eqref{eq:residual-Q} yields
\[
\begin{aligned}
	|b_{3;i_1,i_2,i_3}|
	&\leq \frac{K}{3(dK)^{3/2}}
	\leq K^{-3/2},\\
	|b_{q;i_1,\ldots,i_q}|
	&\leq \frac{Kd}{q(dK)^{q/2}}
	\leq K^{-q/2},\qquad q\geq4,
\end{aligned}
\]
where the last inequalities use $d\geq1$ and $K\leq d$.
Since $K=d{c}$, we obtain the uniform bound
\begin{equation}
	|b_{q;i_1,\ldots,i_q}|
	\leq K^{-q/2}
	=d^{-q/2}{c^{-q/2}},\qquad q\geq3.
	\label{eq:residual-block-coefficient-bound}
\end{equation}
Multiplying over the blocks and using $L=\sum_{\beta=1}^s q_\beta$
gives
\begin{equation}
	|c_{\mathcal P}(\boldsymbol v)|
	\leq\frac1{s!}d^{-L/2}{c^{-L/2}}
	\leq d^{-L/2}{c^{-L/2}}.
	\label{eq:residual-coefficient-bound}
\end{equation}

\medskip
\noindent\emph{5. Bounding the Gaussian factors after partial summation.}
Fix a pattern $\mathcal P$ and abbreviate
$n_j=n_j(\mathcal P)$ and $L=L(\mathcal P)$.
For a phase word, define
\begin{equation}
	Q=\{v_j:n_j>0\},
	\qquad
	J=\{j:v_j\in Q\}.
	\label{eq:residual-touched-sets}
\end{equation}
Thus $Q$ contains the labels at differentiated positions, whereas
$J$ contains all occurrences of those labels, including positions
with $n_j=0$.

We regroup the sum in
Eq.~\eqref{eq:residual-full-comparison-sum}.
For a fixed pattern $\mathcal P$, each choice of \(v_1,\ldots,v_m\) determines
$Q=\{v_j:n_j>0\}$ and $J=\{j:v_j\in Q\}$.
Once $Q$, $J$, and the values $(v_j)_{j\in J}$ are fixed,
the remaining labels range  over $\mathcal V\setminus Q$.
Thus
\begingroup
\begin{equation}
\begin{aligned}
        &E_m=
        \sum_{\mathcal P}\sum_{Q\subseteq\mathcal V}
        \sum_{\{j:n_j>0\}\subseteq J}
        \sum_{\substack{(v_j)_{j\in J}\in Q^J\\
        \{v_j:n_j>0\}=Q}}\\
        &\quad\times\Biggl[
        \left(\frac {\sqrt{c}}d\right)^m
        \sum_{\substack{v_j\in\mathcal V\setminus Q\\j\notin J}}
        c_{\mathcal P}(\boldsymbol v)\\
        &\qquad\times\Tr(A_{v_1}\cdots A_{v_m})
        \mathbb E\prod_{j=1}^m g^{(n_j)}(Z_{v_j})
        \Biggr].
    \end{aligned}
    \label{eq:residual-regrouped-sum}
\end{equation}
\par\endgroup
We now estimate the bracketed contribution for fixed
$\mathcal P$, $Q$, $J$, and $(v_j)_{j\in J}$.

\begingroup
For these fixed labels, $c_{\mathcal P}(\boldsymbol v)$ is
independent of the remaining labels, since every position on which
it depends has $n_j>0$ and hence belongs to $J$.
The Gaussian variables indexed by $Q$ are independent of those
indexed by $\mathcal V\setminus Q$, and $n_j=0$ for $j\notin J$.
Therefore
\begin{equation}
    \begin{aligned}
        &\mathbb E\prod_{j=1}^m g^{(n_j)}(Z_{v_j})\\
        &\quad=
        \left(\mathbb E\prod_{j\in J}g^{(n_j)}(Z_{v_j})\right)
        \left(\mathbb E\prod_{j\notin J}g(Z_{v_j})\right).
    \end{aligned}
    \label{eq:residual-fixed-label-factorization}
\end{equation}
\par\endgroup

We first bound $
\mathbb E\!\left[\prod_{j\in J}g^{(n_j)}(Z_{v_j})\right]
$ involving positions in $J$.
For $n=0$, $|g(x)|\leq1+a|x|$; for $n=1$,
$|g'(x)|\leq1+a$; and for $n\geq2$,
$|g^{(n)}(x)|\leq1$.
Hence
$\|g^{(n)}(Z)\|_{L_p}\leq C\sqrt p$
uniformly in $n\geq0$ and $p\geq1$.
H\"older's inequality yields
\begin{equation}
	\begin{aligned}
		\left|
		\mathbb E\prod_{j\in J}g^{(n_j)}(Z_{v_j})
		\right|
		&\leq
		\prod_{j\in J}
		\|g^{(n_j)}(Z_{v_j})\|_{L_{|J|}}
		\\
		&\leq
		(C\sqrt{|J|})^{|J|}
		\leq
		(C\sqrt m)^{|J|}.
	\end{aligned}
	\label{eq:residual-touched-bound}
\end{equation}

If a label in $Q$ occurs only once, at position $j$, and
$n_j\in\{1,2\}$, then the corresponding term in
Eq.~\eqref{eq:residual-fixed-label-factorization} vanishes
by Eq.~\eqref{eq:residual-three-cancellations}.
Consequently, if $h$ labels in $Q$ occur exactly once,
every nonzero contribution satisfies
\begin{equation}
	3h\leq L,
	\qquad
	|Q|\leq L.
	\label{eq:residual-label-counts}
\end{equation}
The second inequality holds because every label in $Q$
receives at least one derivative.

We next consider the remaining Gaussian factor$\mathbb E\prod_{j\notin J}g(Z_{v_j})$. We bound its contribution to
Eq.~\eqref{eq:residual-fixed-label-factorization}
by keeping the trace weights and summing over the
unfixed labels $v_j\in\mathcal V\setminus Q$.
For the fixed labels at positions in $J$, set
\[
B_j=
\begin{cases}
	A_{v_j},&j\in J,\\
	M_Q,&j\notin J.
\end{cases}
\]
Substituting Eq.~\eqref{eq:residual-MQ} and expanding
the matrix product gives
\begingroup
\[
    \begin{aligned}
        &B_1\cdots B_m\\
        &\quad=
        \frac1{d^{m-|J|}}
        \sum_{\substack{v_j\in\mathcal V\setminus Q\\j\notin J}}
        \left(\prod_{j\notin J}g(Z_{v_j})\right)
        A_{v_1}\cdots A_{v_m}.
    \end{aligned}
\]
\par\endgroup
Taking the trace and expectation therefore yields
\[
\begin{aligned}
	\mathbb E\Tr(B_1\cdots B_m)
	&=
	\frac1{d^{m-|J|}}
	\sum_{\substack{v_j\in\mathcal V\setminus Q\\j\notin J}}
	\Tr(A_{v_1}\cdots A_{v_m})
	\\
	&\qquad\times
	\mathbb E\prod_{j\notin J}g(Z_{v_j}).
\end{aligned}
\]
Using
$|\Tr B|\leq d\|B\|_\infty$,
submultiplicativity of the operator norm, and
$\|A_v\|_\infty=1$, we obtain
\begin{equation}
	\begin{aligned}
		&\Biggl|
		\frac1{d^{m-|J|}}
		\sum_{\substack{v_j\in\mathcal V\setminus Q\\j\notin J}}
		\Tr(A_{v_1}\cdots A_{v_m})
		\\
		&\hspace{22mm}\times
		\mathbb E\prod_{j\notin J}g(Z_{v_j})
		\Biggr|
		\\
		&\qquad\leq
		d\,\mathbb E\|M_Q\|_\infty^{m-|J|}
		\leq dC^{m-|J|},
	\end{aligned}
	\label{eq:residual-untouched-partial-sum-bound}
\end{equation}
where the last inequality follows from
Eq.~\eqref{eq:residual-MQ-moment}.
When $|J|=m$, no labels remain to be summed, and the bound
is simply $d$.

Combining
Eqs.~\eqref{eq:residual-fixed-label-factorization},
\eqref{eq:residual-touched-bound},
\eqref{eq:residual-untouched-partial-sum-bound}, and
\eqref{eq:residual-coefficient-bound}, we obtain
\begingroup
\begin{equation}
\begin{aligned}
        &\Biggl|
        \left(\frac{\sqrt{c}}d\right)^m
        \sum_{\substack{v_j\in\mathcal V\setminus Q\\j\notin J}}
        c_{\mathcal P}(\boldsymbol v)\\
        &\quad\times\Tr(A_{v_1}\cdots A_{v_m})
        \mathbb E\prod_{j=1}^m g^{(n_j)}(Z_{v_j})
        \Biggr|\\
        &\quad\leq
        c^{m/2}d^{-|J|}(d^{-L/2}c^{-L/2})\\
        &\qquad\times(C\sqrt m)^{|J|}\,dC^{m-|J|}\\
        &\quad=
        d(C\sqrt{c})^m m^{|J|/2}
        d^{-|J|-L/2}c^{-L/2}.
    \end{aligned}
\label{eq:residual-product-partial-sum-bound}
\end{equation}
\par\endgroup

\medskip
\noindent\emph{6. Combining the estimates and counting the terms.}

We now estimate the outer sums in
Eq.~\eqref{eq:residual-regrouped-sum},
starting with the sum over the label assignments,
\[
\sum_{\substack{
		(v_j)_{j\in J}\in Q^J\\
		\{v_j:n_j>0\}=Q}}.
\]
For fixed $\mathcal P$, $Q$, and $J$, each of the
$|J|$ positions takes a label from $Q$.
By Eq.~\eqref{eq:residual-product-partial-sum-bound},
this sum therefore contributes at most
\[
|Q|^{|J|}
d(C{\sqrt{c}})^m m^{|J|/2}
d^{-|J|-L/2}{c^{-L/2}}.
\]

Next, consider the sum over the positions,
\[
\sum_{\{j:n_j>0\}\subseteq J}.
\]
For a fixed value of $|J|$, there are at most
$\binom m{|J|}\leq m^{|J|}$ choices of $J$.
Since $|Q|\leq|J|\leq m$, the combined number of
position choices and label assignments is at most
\[
\binom m{|J|}|Q|^{|J|}
\leq m^{2|J|}.
\]
Thus, for fixed $Q$ and $|J|$, the two sums above
contribute at most
\[
d(C{\sqrt{c}})^m m^{5|J|/2}
d^{-|J|-L/2}{c^{-L/2}}.
\]

We then consider the sum over label sets,
\[
\sum_{Q\subseteq\mathcal V}.
\]
For a fixed value of $|Q|$, there are
$\binom{d^2}{|Q|}\leq d^{2|Q|}$ choices of $Q$.
Consequently, for a fixed pattern $\mathcal P$
and fixed cardinalities $|Q|$ and $|J|$,
the three label sums contribute at most
\begin{equation}
	d(C{\sqrt{c}})^m m^{5|J|/2}
	d^{2|Q|-|J|-L/2}{c^{-L/2}}.
	\label{eq:residual-fixed-cardinalities}
\end{equation}

If $h$ labels in $Q$ occur exactly once, the
other $|Q|-h$ labels each occupy at least two positions.
Therefore,
\[
|J|\geq h+2(|Q|-h)=2|Q|-h.
\]
Using Eq.~\eqref{eq:residual-label-counts}, we obtain
\begin{equation}
	|Q|\leq L,\qquad
	2|Q|-|J|\leq h\leq L/3.
	\label{eq:residual-admissible-cardinalities}
\end{equation}

For sufficiently large $N$, $m^{5/2}\leq d$.
If $|J|\geq2|Q|$, then
\[
m^{5|J|/2}d^{2|Q|-|J|}
=
m^{5|Q|}
\left(\frac{m^{5/2}}d\right)^{|J|-2|Q|}
\leq m^{5|Q|}.
\]
If $|J|<2|Q|$, then
\[
m^{5|J|/2}d^{2|Q|-|J|}
\leq m^{5|Q|}d^{2|Q|-|J|}.
\]
Combining these two cases and using
Eq.~\eqref{eq:residual-admissible-cardinalities}
gives
\begin{equation}
	\begin{aligned}
		m^{5|J|/2}d^{2|Q|-|J|}
		&\leq
		m^{5|Q|}d^{\max\{2|Q|-|J|,0\}}\\
		&\leq m^{5L}d^{L/3}.
	\end{aligned}
	\label{eq:residual-counting-gain}
\end{equation}

There are at most $m^2$ possible pairs of cardinalities
$(|Q|,|J|)$, since $1\leq|Q|\leq|J|\leq m$.
Substituting Eq.~\eqref{eq:residual-counting-gain}
into Eq.~\eqref{eq:residual-fixed-cardinalities}
and summing over these pairs shows that, for a fixed
pattern $\mathcal P$ of total derivative order $L$,
all three label sums contribute at most
\[
d(C{\sqrt{c}})^m m^2
d^{-L/2}{c^{-L/2}}m^{5L}d^{L/3}.
\]

Finally, consider the sum over patterns,
\[
\sum_{\mathcal P}.
\]
We group the patterns by their total derivative order
$L\geq3$.
For fixed $L$, the ordered block lengths
$q_1,\ldots,q_s$ satisfy
$q_\beta\geq3$ and $\sum_{\beta=1}^s q_\beta=L$.
There are at most $2^L$ such lists, since there are
$2^{L-1}$ ordered compositions of $L$ into positive
integers.
For each list, the $L$ derivative indices have at most
$m^L$ choices.
Thus there are at most $(2m)^L$ patterns of total
derivative order $L$.

Applying the triangle inequality to the outer sums in
Eq.~\eqref{eq:residual-regrouped-sum} therefore yields
\begingroup
\begin{equation}
\begin{aligned}
        &|E_m|\\
        &\quad\leq d(C\sqrt{c})^m m^2
        \sum_{L\geq3}(2m)^L d^{-L/2}c^{-L/2}
        m^{5L}d^{L/3}\\
        &\quad=d(C\sqrt{c})^m m^2
        \sum_{L\geq3}
        \left(\frac{2m^6}{d^{1/6}\sqrt{c}}\right)^L.
    \end{aligned}
\label{eq:residual-error-series}
\end{equation}
\par\endgroup
Set
\begin{equation}
	\theta_N
	:=
	\frac{2m^6}{d^{1/6}{\sqrt{c}}}
	=
	2(2N)^6\,3^{-N/6+\ell_N/2}.
	\label{eq:residual-series-ratio}
\end{equation}
Since $\ell_N=o(N)$, we have
\begingroup
\[
    \begin{aligned}
        \ln\theta_N
        &=\ln2+6\ln(2N)-\frac{N\ln3}{6}
        +\frac{\ell_N\ln3}{2}\\
        &=-\frac{N\ln3}{6}+o(N),
    \end{aligned}
\]
\par\endgroup
and hence $\theta_N\to0$.
In particular, $\theta_N<1$ for all sufficiently large $N$.
Summing the geometric series in
Eq.~\eqref{eq:residual-error-series} gives
\begin{equation}
	|E_m|
	\leq
	d(C{\sqrt{c}})^m
	\frac{m^2\theta_N^3}{1-\theta_N}.
	\label{eq:residual-error-geometric}
\end{equation}
Moreover,
\[
\begin{aligned}
	m^2\theta_N^3
	&=
	8(2N)^{20}\,3^{-N/2+3\ell_N/2}\\
	&=
	\exp\!\left[-\frac{N\ln3}{2}+o(N)\right]
	\longrightarrow0.
\end{aligned}
\]
Since $1-\theta_N\to1$, Eq.~\eqref{eq:residual-error-geometric}
therefore implies
\begin{equation}
	|E_m|\leq o(1)\,d(C{\sqrt{c}})^m.
	\label{eq:residual-comparison-error}
\end{equation}

By the definition of $E_m$,
\[
\mathbb E\Tr R_X^m
=
\mathbb E\Tr R_Z^m+E_m.
\]
Combining Eq.~\eqref{eq:residual-comparison-error}
with the Gaussian moment bound in
Eq.~\eqref{eq:residual-gaussian-trace-moment},
and taking $C$ large enough to cover both bounds, yields
\[
\begin{aligned}
	\mathbb E\Tr R_X^m
	&\leq
	\mathbb E\Tr R_Z^m+|E_m|\\
	&\leq
	[1+o(1)]\,d(C{\sqrt{c}})^m.
\end{aligned}
\]
Thus there is an absolute constant ${K_R}$ such that,
for all sufficiently large $N$,
\begin{equation}
	\mathbb E\Tr R_X^{2N}
	\leq d({K_R}{\sqrt{c}})^{2N}.
	\label{eq:residual-X-moment}
\end{equation}

Since $R_X$ is Hermitian and $2N$ is even,
\[
\|R_X\|_\infty^{2N}\leq\Tr R_X^{2N}.
\]
Markov's inequality and
Eq.~\eqref{eq:residual-X-moment} therefore give
\begin{equation}
	\begin{aligned}
		\Pr\{\|R_X\|_\infty>M{\sqrt{c}}\}
		&\leq
		\frac{\mathbb E\Tr R_X^{2N}}{(M{\sqrt{c}})^{2N}}\\
		&\leq
		3^N\left(\frac{{K_R}}{M}\right)^{2N}\\
		&=
		\left[3\left(\frac{{K_R}}{M}\right)^2\right]^N.
	\end{aligned}
	\label{eq:residual-X-tail}
\end{equation}
Choosing any fixed $M>\sqrt3\,{K_R}$ makes the
right-hand side tend to zero exponentially in $N$.

Combining Eqs.~\eqref{eq:residual-X-tail} and
\eqref{eq:residual-normalization-moment}, and using
Markov's inequality together with
$\|\cdot\|_\infty\leq\|\cdot\|_2$, we obtain, for any fixed
$M>\sqrt{3}\,{K_R}$,
\begingroup
\begin{equation}
\begin{aligned}
        &\Pr_{\mathrm H}\{\|R_\rho\|_\infty>(M+1){\sqrt{c}}\}\\
        &\quad\leq\Pr\{\|R_X\|_\infty>M{\sqrt{c}}\}\\
        &\qquad+\Pr\{\|R_\rho-R_X\|_\infty>{\sqrt{c}}\}\\
        &\quad\leq
        \left[3\left(\frac{{K_R}}{M}\right)^2\right]^N
        +\frac{\mathbb E\|R_\rho-R_X\|_2^{2N}}{{c^N}}\\
        &\quad\leq
        \left[3\left(\frac{{K_R}}{M}\right)^2\right]^N
        +\left(\frac{C\sqrt{2N/d}}{{\sqrt{c}}}\right)^{2N}\\
        &\quad=
        \left[3\left(\frac{{K_R}}{M}\right)^2\right]^N
        +\left(C\sqrt{\frac{2N}{K}}\right)^{2N}\\
        &\quad\longrightarrow0.
    \end{aligned}
    \label{eq:residual-haar-tail}
\end{equation}
\par\endgroup
Here we used $d{c}=K=3^{N-\ell_N}$ and $\ell_N=o(N)$.
Thus $\|R_\rho\|_\infty=O({\sqrt{c}})$ with high probability
under the Haar-induced ensemble.

\medskip
\noindent\emph{7. Transfer to the circuit ensemble.}
\begingroup
Haar concentration \cite[Corollary~17]{Meckes2013}, applied to
\[
    U\longmapsto
    \sqrt{dK}\langle0|U^\dagger(A_v\otimes I_E)U|0\rangle-{\sqrt{c}},
\]
gives
\begin{equation}
    \|x_v\|_{L_p(H)}\leq C\sqrt p,
    \qquad p\geq1.
    \label{eq:residual-phase-moments}
\end{equation}
Indeed, this function has mean zero and Lipschitz constant at most
$2\sqrt{dK}$ for the Hilbert--Schmidt metric on $\mathrm U(dK)$.
\par\endgroup

Let $\Psi$ denote the pure state density operator
$\Psi_H$ or $\Psi_C$, with $\rho=\Tr_E\Psi$.
Recalling that
$x_v=\sqrt{dK}\Tr(A_v\rho)-{\sqrt{c}}$, define
\[
C_v:=\sqrt{dK}(A_v\otimes I_E)-{\sqrt{c}}I_{dK}.
\]
We have
\[
x_v=\Tr(C_v\Psi),
\qquad
\|C_v\|_\infty
\leq\sqrt{dK}+{\sqrt{c}}\leq2d,
\]
where we used $K\leq d$, ${c=K/d}$, and
$\|A_v\|_\infty=1$.
Consequently,
\[
x_v^{2N}
=[\Tr(C_v\Psi)]^{2N}
=\Tr(C_v^{\otimes 2N}\Psi^{\otimes 2N}).
\]
For all sufficiently large $N$, we have $2N\leq t_N$.
Applying Eq.~\eqref{eq:random-moment-transfer} and then
Eq.~\eqref{eq:residual-phase-moments} therefore gives
\[
\begin{aligned}
	\mathbb E_C|x_v|^{2N}
	&\leq
	\mathbb E_H|x_v|^{2N}
	+\delta_N\|C_v\|_\infty^{2N}\\
	&\leq
	(C\sqrt{2N})^{2N}+\delta_N(2d)^{2N}.
\end{aligned}
\]
Taking the absolute constant $B$ sufficiently large that
$B>3\sqrt{2}\,C$, Markov's inequality and a union bound
over the $d^2$ labels yield
\begingroup
\begin{equation}
\begin{aligned}
        &\Pr_C\left\{\max_{v\in\mathcal V}|x_v|>B\sqrt N\right\}\\
        &\quad\leq\frac1{(B\sqrt N)^{2N}}
        \sum_{v\in\mathcal V}\mathbb E_C|x_v|^{2N}\\
        &\quad\leq
        d^2\left(\frac{\sqrt2\,C}{B}\right)^{2N}
        +\delta_N\frac{d^2(2d)^{2N}}{(B\sqrt N)^{2N}}\\
        &\quad=
        \left(\frac{3\sqrt2\,C}{B}\right)^{2N}
        +\exp[-N^5+O(N^2)]\\
        &\quad\longrightarrow0.
    \end{aligned}
    \label{eq:residual-circuit-phase-event}
\end{equation}
\par\endgroup
where we used $d=3^N$ and $\delta_N=e^{-N^5}$.
Thus $\max_{v\in\mathcal V}|x_v|\leq B\sqrt N$
with probability $1-o(1)$ under the circuit ensemble.

Choose an even integer $r_0$ satisfying
$Am\leq r_0<Am+2$, where $m=2N$ and $A$ is a sufficiently
large fixed constant. Let $p_{r_0}$ be the Taylor polynomial
of $\sin$ of degree $r_0-1$, and define
\[
R_\rho^{(r_0)}
:=\frac {\sqrt{c}}d\sum_{v\in\mathcal V}
[p_{r_0}(x_v)-ax_v]A_v.
\]
Taylor's theorem and Eq.~\eqref{eq:residual-phase-moments}
give
\[
\begin{aligned}
	\|\sin(x_v)-p_{r_0}(x_v)\|_{L_m(H)}
	&\leq
	\frac{\|x_v\|_{L_{mr_0}(H)}^{r_0}}{r_0!}\\
	&\leq
	\left(Ce\sqrt{\frac m{r_0}}\right)^{r_0}
	\leq2^{-r_0},
\end{aligned}
\]
where the last inequality holds for sufficiently large $A$.

The linear terms cancel in
$R_\rho-R_\rho^{(r_0)}$.
Using phase-point orthogonality, we obtain
\begingroup
\[
    \begin{aligned}
        &\|R_\rho-R_\rho^{(r_0)}\|_2^2\\
        &\quad=\frac{{c}}{d^2}
        \sum_{v,w\in\mathcal V}\Tr(A_vA_w)\\
        &\qquad\times
        [\sin(x_v)-p_{r_0}(x_v)]
        [\sin(x_w)-p_{r_0}(x_w)]\\
        &\quad=\frac{{c}}{d}
        \sum_{v\in\mathcal V}|\sin(x_v)-p_{r_0}(x_v)|^2.
    \end{aligned}
\]
\par\endgroup
Minkowski's inequality in $L_{m/2}(H)$ therefore yields
\begin{equation}
	\begin{aligned}
		&\left(
		\mathbb E_H
		\|R_{\rho_H}-R_{\rho_H}^{(r_0)}\|_2^m
		\right)^{1/m}\\
		&\quad=
		\frac {\sqrt{c}}{\sqrt d}
		\left\|
		\sum_{v\in\mathcal V}
		|\sin(x_v)-p_{r_0}(x_v)|^2
		\right\|_{L_{m/2}(H)}^{1/2}\\
		&\quad\leq
		\frac {\sqrt{c}}{\sqrt d}
		\left(
		\sum_{v\in\mathcal V}
		\|\sin(x_v)-p_{r_0}(x_v)\|_{L_m(H)}^2
		\right)^{1/2}\\
		&\quad\leq
		{\sqrt{c}}\sqrt d\,2^{-r_0}
		=o({\sqrt{c}}),
	\end{aligned}
	\label{eq:residual-haar-polynomial-error}
\end{equation}
where we used $|\mathcal V|=d^2$.
The last step follows from $d=3^N$ and $r_0\geq2AN$,
after increasing $A$ if necessary.

Combining Eqs.~\eqref{eq:residual-X-moment},
\eqref{eq:residual-normalization-moment}, and
\eqref{eq:residual-haar-polynomial-error}, and applying
Minkowski's inequality under the preceding Wishart coupling,
we obtain
\[
\left(
\mathbb E_H
\|R_{\rho_H}^{(r_0)}\|_\infty^m
\right)^{1/m}
\leq d^{1/m}{K_R}{\sqrt{c}}+o({\sqrt{c}})
\leq C_P{\sqrt{c}}.
\]
Here $C_P$ is an absolute constant, since
$d^{1/m}=\sqrt3$.
As $m$ is even, this implies
\begin{equation}
	\begin{aligned}
		\mathbb E_H
		\Tr[(R_{\rho_H}^{(r_0)})^m]
		&\leq
		d\,\mathbb E_H
		\|R_{\rho_H}^{(r_0)}\|_\infty^m\\
		&\leq d(C_P{\sqrt{c}})^m.
	\end{aligned}
	\label{eq:residual-haar-polynomial-moment}
\end{equation}

We next transfer this polynomial moment to the circuit
ensemble. Using the previously defined $C_v$ and $\Psi$,
for arbitrary phase-point labels $v_1,\ldots,v_q$ we have
\[
\prod_{a=1}^q x_{v_a}
=
\Tr\!\left[
(C_{v_1}\otimes\cdots\otimes C_{v_q})
\Psi^{\otimes q}
\right].
\]
Equation~\eqref{eq:random-moment-transfer}, together with
$\|C_v\|_\infty\leq2d$, therefore gives
\begingroup
\begin{equation}
\begin{aligned}
        &\left|
        \mathbb E_C\prod_{a=1}^q x_{v_a}
        -\mathbb E_H\prod_{a=1}^q x_{v_a}
        \right|\\
        &\quad\leq\delta_N(2d)^q,
        \qquad 1\leq q\leq t_N.
    \end{aligned}
    \label{eq:residual-circuit-mixed-moments}
\end{equation}
\par\endgroup

Write
\[
p_{r_0}(x)-ax
=\sum_{j=0}^{r_0-1}\beta_jx^j,
\]
and define
\begingroup
\[
    \begin{aligned}
        \mathcal A_{r_0}
        &:=\sum_{j=0}^{r_0-1}|\beta_j|(2d)^j\\
        &\leq2(r_0+1)(1+2d)^{r_0}
        =\exp[O(N^2)].
    \end{aligned}
\]
\par\endgroup
Expanding the matrix product gives
\[
\begin{aligned}
	\Tr[(R_\rho^{(r_0)})^m]
	&=
	\left(\frac {\sqrt{c}}d\right)^m
	\sum_{v_1,\ldots,v_m\in\mathcal V}
	\Tr(A_{v_1}\cdots A_{v_m})\\
	&\quad\times
	\sum_{0\leq j_1,\ldots,j_m<r_0}
	\left(\prod_{b=1}^m\beta_{j_b}\right)
	\prod_{b=1}^m x_{v_b}^{j_b}.
\end{aligned}
\]
Each monomial has coordinate degree at most
\[
\sum_{b=1}^m j_b
\leq m(r_0-1)=O(N^2)\leq t_N
\]
for all sufficiently large $N$.
Applying Eq.~\eqref{eq:residual-circuit-mixed-moments}
term by term, and using
$|\Tr(A_{v_1}\cdots A_{v_m})|\leq d$, yields
\[
\begin{aligned}
	&\left|
	\mathbb E_C\Tr[(R_{\rho_C}^{(r_0)})^m]
	-
	\mathbb E_H\Tr[(R_{\rho_H}^{(r_0)})^m]
	\right|\\
	&\quad\leq
	\delta_N
	\left(\frac {\sqrt{c}}d\right)^m
	d^{2m}d
	\left(
	\sum_{j=0}^{r_0-1}|\beta_j|(2d)^j
	\right)^m\\
	&\quad=
	\delta_N{c^{m/2}} d^{m+1}\mathcal A_{r_0}^m.
\end{aligned}
\]
Combining this estimate with
Eq.~\eqref{eq:residual-haar-polynomial-moment} gives
\[
\mathbb E_C\Tr[(R_{\rho_C}^{(r_0)})^m]
\leq
d(C_P{\sqrt{c}})^m
+\delta_N{c^{m/2}} d^{m+1}\mathcal A_{r_0}^m.
\]
Since $R_{\rho_C}^{(r_0)}$ is Hermitian and $m$ is even,
Markov's inequality implies, for any fixed
$M>\sqrt3\,C_P$,
\begingroup
\begin{equation}
\begin{aligned}
        &\Pr_C\{\|R_{\rho_C}^{(r_0)}\|_\infty>M{\sqrt{c}}\}\\
        &\quad\leq
        \frac{\mathbb E_C\Tr[(R_{\rho_C}^{(r_0)})^m]}{(M{\sqrt{c}})^m}\\
        &\quad\leq d(C_P/M)^m
        +\frac{\delta_Nd^{m+1}\mathcal A_{r_0}^m}{M^m}\\
        &\quad=o(1).
    \end{aligned}
    \label{eq:residual-circuit-moment}
\end{equation}
\par\endgroup
Here the first term tends to zero because
$d=3^N$ and $m=2N$, while the second tends to zero because
\[
\log(d^{m+1}\mathcal A_{r_0}^m)=O(N^3),
\qquad
\delta_N=e^{-N^5}.
\]

It remains to remove the polynomial approximation.
On the event $\max_v|x_v|\leq B\sqrt N$, which has
probability $1-o(1)$ by
Eq.~\eqref{eq:residual-circuit-phase-event},
the same phase-point orthogonality calculation gives
\begin{equation}
	\begin{aligned}
		\|R_{\rho_C}-R_{\rho_C}^{(r_0)}\|_\infty
		&\leq
		\|R_{\rho_C}-R_{\rho_C}^{(r_0)}\|_2\\
		&\leq
		{\sqrt{c}}\sqrt d\,
		\max_v|\sin(x_v)-p_{r_0}(x_v)|\\
		&\leq
		{\sqrt{c}}\sqrt d\,\frac{(B\sqrt N)^{r_0}}{r_0!}\\
		&\leq
		{\sqrt{c}}\sqrt d
		\left(\frac{eB\sqrt N}{r_0}\right)^{r_0}
		=o({\sqrt{c}}),
	\end{aligned}
	\label{eq:residual-circuit-polynomial-error}
\end{equation}
where the last step uses $d=3^N$ and $r_0=\Theta(N)$.
For all sufficiently large $N$, this error is at most ${\sqrt{c}}$
on the coordinate event.
Combining Eqs.~\eqref{eq:residual-circuit-moment},
\eqref{eq:residual-circuit-phase-event}, and
\eqref{eq:residual-circuit-polynomial-error}, we conclude that
\begingroup
\begin{equation}
\begin{aligned}
        &\Pr_C\{\|R_{\rho_C}\|_\infty>(M+1){\sqrt{c}}\}\\
        &\quad\leq
        \Pr_C\{\|R_{\rho_C}^{(r_0)}\|_\infty>M{\sqrt{c}}\}\\
        &\qquad+\Pr_C\{\max_v|x_v|>B\sqrt N\}\\
        &\quad\longrightarrow0.
    \end{aligned}
    \label{eq:residual-circuit-tail}
\end{equation}
\par\endgroup
Together with Eq.~\eqref{eq:residual-haar-tail},
this proves the claimed bound for both ensembles
with a common absolute constant $M_*$.

\end{proof}

\subsection{\textbf{A witness for $\ell_N=O(1)$}}
\label{app:bounded-witness}

Assume $0\leq\ell_N\leq L$ for a fixed $L<\infty$. Thus,
\begingroup
\begin{equation}
 3^{-L}\leq c=K/d\leq1,\qquad K\geq3^{-L}d.
 \label{eq:bounded-regime}
\end{equation}
\par\endgroup
\begingroup
For either ensemble, recall the centered coordinate and define
the Hermitian witness
\begin{equation}
 \begin{aligned}
 x_v&=\sqrt{dK}\,\Tr(A_v\rho)-\sqrt c,\\
 h(x)&=\frac{(1-\cos x)(1-\sin x)}4,\\
 F_\rho&=\frac1d\sum_{v\in\mathcal V}h(x_v)A_v.
 \end{aligned}
 \label{eq:bounded-witness}
\end{equation}
\par\endgroup
We first establish the three properties needed to turn $F_\rho$
into a distillation test. All constants below are independent of $N$.

\begin{lemma}[\textbf{A bounded witness with negative expectation}]

\label{lem:bounded-witness}
For every state $\rho$,
\[
0\leq\Tr(A_vF_\rho)\leq1\qquad\text{for every }v.
\]
There exists $M_L\geq1$, depending only on $L$, such that for either
the Haar-induced ensemble or the circuit ensemble satisfying
Eq.~\eqref{eq:random-moment-condition}, with probability tending to one,
\begin{equation}
 \begin{gathered}
 \Tr(\rho_NF_{\rho_N})\leq-\delta,\\
 \|F_{\rho_N}\|_\infty\leq M_L.
 \end{gathered}
 \label{eq:bounded-witness-properties}
\end{equation}
\begingroup
Here
\[
 \delta=\frac{2e^{-1/2}-1-e^{-2}}8>0.
\]
Positivity follows from $e^{-1/2}>3/5$ and $e^{-2}<1/7$.
\par\endgroup
\end{lemma}

\begin{proof}

Phase-point orthogonality gives
\[
{\Tr(A_vF_\rho)=h(x_v)\in[0,1].}
\]
This holds for every state $\rho$ and every phase point $v$.

\medskip
\begingroup
\noindent\emph{1. \(\Tr(\rho_NF_{\rho_N})\leq-\delta\) for Haar states.} Write
$S_\rho:=\Tr(\rho F_\rho)$. By definition,
\begin{equation}
 S_\rho=\frac1{\sqrt c\,d^2}
 \sum_v(x_v+\sqrt c)h(x_v).
 \label{eq:bounded-witness-identities}
\end{equation}
{\textbf{We use the unnormalized coordinate $\boldsymbol{X_v}$ from Eq.~\eqref{eq:residual-unnormalized}.}}
For the Haar ensemble, write
\[
 \begin{aligned}
 \rho_H&=\frac{GG^\dagger}{\Tr(GG^\dagger)},
 &\tau&=\frac{\Tr(GG^\dagger)}{dK},\\
 {\boldsymbol{X_v}}&=\frac{\Tr(A_vGG^\dagger)}{\sqrt{dK}}-\sqrt c.
 \end{aligned}
\]
Here $G$ is a standard complex Gaussian $d\times K$ matrix, and
\[
 {\boldsymbol{X_v}}+\sqrt c=\tau(x_v+\sqrt c).
\]
In an eigenbasis of $A_v$,
\[
 {\boldsymbol{X_v}}=\frac1{\sqrt{dK}}
 \sum_{i=1}^d\sum_{j=1}^K
 a_i\bigl(|G_{ij}|^2-1\bigr),
\]
where $a_i\in\{-1,1\}$ are the eigenvalues of $A_v$.
Gaussian unitary invariance preserves the independence and standard
complex Gaussian law of the entries. The centered summands have
unit variance and uniformly bounded third absolute moments.
The Lyapunov central limit theorem therefore gives
${\boldsymbol{X_v}}\xrightarrow{\mathrm d}Z\sim N(0,1)$.
Since $\tau\to1$ in probability,
\[
 x_v=\frac{{\boldsymbol{X_v}}+\sqrt c}{\tau}-\sqrt c
 \xrightarrow{\mathrm d}Z.
\]
These conclusions hold uniformly for $c\in[3^{-L},1]$.

The function $h$ is bounded and continuous. Moreover,
Eq.~\eqref{eq:residual-phase-moments} implies
\[
 \sup_N\mathbb E_H|x_vh(x_v)|^2
 \leq\sup_N\mathbb E_H|x_v|^2<\infty.
\]
Thus $x_vh(x_v)$ is uniformly integrable. Consequently,
\[
 \begin{aligned}
 \mathbb E_Hh(x_v)&=\mathbb Eh(Z)+o(1),\\
 \mathbb E_H[x_vh(x_v)]&=\mathbb E[Zh(Z)]+o(1).
 \end{aligned}
\]
All $x_v$ have the same marginal law. Combining
these limits with Eq.~\eqref{eq:bounded-witness-identities} and
$c\geq3^{-L}$ gives
\begin{equation}
 \mathbb E_H S_{\rho_H}=s(c)+o(1),
 \label{eq:bounded-haar-mean}
\end{equation}
where the Gaussian identities
\[
 \begin{aligned}
 \mathbb Eh(Z)&=\frac{1-e^{-1/2}}4\quad
 \mathbb E[Zh(Z)]&=\frac{e^{-2}-e^{-1/2}}4
 \end{aligned}
\]
yield
\begin{equation}
 \begin{aligned}
 s(c)&:=\mathbb Eh(Z)+\frac{\mathbb E[Zh(Z)]}{\sqrt c}\\
 &=\frac{1-e^{-1/2}}4
   +\frac{e^{-2}-e^{-1/2}}{4\sqrt c}\\
 &\leq\frac{1-2e^{-1/2}+e^{-2}}4=-2\delta.
 \end{aligned}
 \label{eq:bounded-gaussian-expectation}
\end{equation}
The inequality uses $0<\sqrt c\leq1$ and
$e^{-2}-e^{-1/2}<0$.

To control the variance, set
\[
 q_c(x)=(x+\sqrt c)h(x),\qquad
 D_\rho=\frac1d\sum_vq_c'(x_v)A_v.
\]
Differentiating Eq.~\eqref{eq:bounded-witness-identities} in a
Hermitian direction $\Delta\rho$ gives
$\mathrm dS_\rho[\Delta\rho]=\Tr(D_\rho\Delta\rho)$.
Since $|q_c'(x)|\leq C(1+|x+\sqrt c|)$, Parseval gives
\[
 \begin{aligned}
 \|D_\rho\|_2^2
 &=\frac1d\sum_v|q_c'(x_v)|^2\\
 &\leq\frac Cd\sum_v\bigl[1+(x_v+\sqrt c)^2\bigr]\\
 &=Cd\bigl(1+K\Tr\rho^2\bigr).
 \end{aligned}
\]
For the Haar unitary representation
$\rho_H=\Tr_E(U|0\rangle\langle0|U^\dagger)$, differentiation
in $U$ bounds the Hilbert--Schmidt gradient of $S_{\rho_H}$ by
$2\|D_{\rho_H}\|_2$.
The Poincar\'e inequality implied by the log-Sobolev inequality
in \cite[Theorem~15]{Meckes2013} therefore yields
\[
 \begin{aligned}
 \operatorname{Var}_H(S_{\rho_H})
 &\leq\frac C{dK}\mathbb E_H\|D_{\rho_H}\|_2^2\\
 &\leq\frac CK\bigl(1+K\mathbb E_H\Tr\rho_H^2\bigr)
 \leq\frac CK,
 \end{aligned}
\]
where the last inequality uses the Haar second-moment identity
\[
 K\mathbb E_H\Tr\rho_H^2
 =\frac{K(d+K)}{dK+1}\leq2.
\]
Chebyshev's inequality now gives, for $t>0$,
\begin{equation}
 \Pr_H\{|S_{\rho_H}-\mathbb E_HS_{\rho_H}|>t\}
 \leq\frac C{Kt^2}.
 \label{eq:bounded-scalar-concentration}
\end{equation}
By Eqs.~\eqref{eq:bounded-haar-mean}
and~\eqref{eq:bounded-gaussian-expectation}, the mean is at most
$-3\delta/2$ for all sufficiently large $N$. Taking $t=\delta/2$
therefore gives
\[
 \Pr_H\{S_{\rho_H}>-\delta\}
 \leq\frac{4C}{K\delta^2}\longrightarrow0.
\]
\par\endgroup

\medskip
\noindent\emph{2. \( \|F_{\rho_N}\|_\infty\leq M_L\) for Haar states.}
We first replace the random denominator $\Tr(GG^\dagger)$
in $\rho_H$ by its mean $dK$, and define
\begingroup
\[
 \begin{aligned}
 \widehat F&=\frac1d\sum_vh({\boldsymbol{X_v}})A_v,\\
 {\boldsymbol{X_v}}&=\frac{\Tr(A_vGG^\dagger)}{\sqrt{dK}}-\sqrt c.
 \end{aligned}
\]
By unitary invariance of $G$, the expectation
$\mu_N:=\mathbb E_Hh({\boldsymbol{X_v}})$ does not depend on $v$.
Since $\sum_vA_v=dI$ and $0\leq h\leq1$,
\par\endgroup
\[
\mathbb E_H\widehat F=\mu_NI,\qquad 0\leq\mu_N\leq1.
\]
Also $\|\widehat F\|_\infty\leq\|\widehat F\|_2\leq\sqrt d$.
The Lipschitz property of $h$ and Parseval give directly
\begin{equation}
 \begin{aligned}
 &\|\widehat F(G)-\widehat F(G')\|_2\\
 &\quad\leq\frac{\|h'\|_\infty}{\sqrt{dK}}
                \|GG^\dagger-G'G'^\dagger\|_2\\
 &\quad\leq\frac{\|h'\|_\infty}{\sqrt{dK}}
       (\|G\|_\infty+\|G'\|_\infty)\|G-G'\|_2.
 \end{aligned}
 \label{eq:bounded-lipschitz}
\end{equation}
Gaussian singular-value bounds imply
\[
 {\mathcal E}=\{\|G\|_\infty\leq C_0\sqrt d\},\qquad
 \Pr({\mathcal E}^c)\leq2e^{-c_0d}.
\]
Hence, for each unit vector $\psi$, the function
$G\mapsto\langle\psi,\widehat F(G)\psi\rangle$ is
$C/\sqrt K$-Lipschitz on ${\mathcal E}$.
\begingroup
Apply the local concentration inequality
\cite[Lemma~5.9]{Guionnet2009} directly on ${\mathcal E}$.
Since the scalar function is bounded by $\sqrt d$,
$0\in{\mathcal E}$, and $\mathbb E\|G\|_2^2=dK$, the mean and
threshold corrections in that lemma have total size at most
\[
 \begin{aligned}
 &2\sqrt d\,\Pr({\mathcal E}^c)
   +\frac C{\sqrt K}\mathbb E
       [\|G\|_2\mathbf1_{{\mathcal E}^c}]\\
 &\qquad\leq C\sqrt d\,\Pr({\mathcal E}^c)^{1/2}=o(1),
 \end{aligned}
\]
uniformly in $\psi$, by Cauchy--Schwarz and the Gaussian tail bound.
Absorbing this vanishing shift into the constants yields
\par\endgroup
\[
\Pr\!\left(
\{|\langle\psi,(\widehat F-\mu_NI)\psi\rangle|>s\}
\cap{\mathcal E}
\right)
\leq 2e^{-c_1Ks^2}
\]
for $s\geq1$ and all sufficiently large $N$.
 We now use a finite net to obtain a bound
uniform over all unit vectors.
Choose a $1/4$-net $\mathcal N$ of the unit sphere in
$\mathbb C^d$ with $|\mathcal N|\leq9^{2d}$.
The standard net estimate~\cite[Sec.~5.2.2]{Vershynin2012} gives
\[
\|\widehat F-\mu_NI\|_\infty
\leq2\max_{\psi\in\mathcal N}
|\langle\psi,(\widehat F-\mu_NI)\psi\rangle|.
\]
Separate the common bad event ${\mathcal E}^c$ before summing over
the net. This gives
\begin{equation}
 \begin{aligned}
 &\Pr\{\|\widehat F-\mu_NI\|_\infty>2s\}\\
 &\quad\leq\Pr({\mathcal E}^c)\\
 &\qquad+\sum_{\psi\in\mathcal N}
 \Pr\!\left(
 \{|\langle\psi,(\widehat F-\mu_NI)\psi\rangle|>s\}
 \cap{\mathcal E}\right)\\
 &\quad\leq2e^{-c_0d}+2\exp(2d\ln9-c_1Ks^2).
 \end{aligned}
 \label{eq:bounded-net-tail}
\end{equation}
Since ${K\geq3^{-L}d}$, a sufficiently large fixed $s=s_L$ makes
this probability $O(e^{-c_Ld})$. Together with
$\|\widehat F\|_\infty\leq\sqrt d$, this yields
\begin{equation}
 \bigl\|\|\widehat F\|_\infty\bigr\|_{L_p(H)}\leq {K_L},
 \qquad 2\leq p\leq2N,
 \label{eq:bounded-unnormalized-moments}
\end{equation}
because $d^{p/2}e^{-c_Ld}=o(1)$ uniformly in this range.

\begingroup
We now return to $F_{\rho_H}$ and take the even moment order
$m=2N$. Using
\[
 {\boldsymbol{X_v}}-x_v=(\tau-1)(x_v+\sqrt c),
\]
Parseval gives
\par\endgroup
\[
 \|F_{\rho_H}-\widehat F\|_2
 \leq\|h'\|_\infty|\tau-1|\sqrt{dK}\,\|\rho_H\|_2.
\]
The radius $\tau$ is independent of $\rho_H$ and satisfies
$\|\tau-1\|_{L_m}\leq C(\sqrt{m/(dK)}+m/(dK))$.
Since $\|K\rho_H\|_2^2\leq K\|K\rho_H\|_\infty$,
Eq.~\eqref{eq:random-haar-operator-moment} with $q=m/2$ gives
\[
 \begin{aligned}
 \bigl\|\|\rho_H\|_2\bigr\|_{L_m(H)}
 &\leq\frac{1+\sqrt c+\sqrt{m/(2d)}}{\sqrt K}\leq\frac3{\sqrt K},
 \end{aligned}
\]
where $c\leq1$ and {$m=2N\leq d$}.
Consequently,
\begin{equation}
 \bigl\|\|F_{\rho_H}-\widehat F\|_2\bigr\|_{L_m(H)}
 \leq C\sqrt{m/K}=o(1),
 \label{eq:bounded-radial-moment}
\end{equation}
where $C$ is an absolute constant. Together with
Eq.~\eqref{eq:bounded-unnormalized-moments}, this proves the Haar
high-probability norm bound and, after increasing ${K_L}$ if needed,
\begin{equation}
 \mathbb E_H\Tr F_{\rho_H}^{m}\leq d{K_L}^{m}.
 \label{eq:bounded-haar-trace-moment}
\end{equation}

\medskip
\noindent\emph{3. Transfer of both bounds to circuit states.}
Take $m=2N$ and $r=\lceil Am\rceil$ for a sufficiently large
fixed constant $A$. Let $p_r$ be the Taylor polynomial of $h$
{at zero}, of degree at most $r-1$, and define
\[
F_\rho^P=\frac1d\sum_v p_r(x_v)A_v,
\qquad S_\rho^P=\operatorname{Tr}(\rho F_\rho^P).
\]
\begingroup
The expansion
\[
 h(x)=\frac14-\frac14\cos x-\frac14\sin x+\frac18\sin2x
\]
shows that, for $r\geq1$,
\[
 \|h^{(r)}\|_\infty\leq\frac12+\frac{2^r}{8}\leq2^r.
\]
Combining this bound with Eq.~\eqref{eq:residual-phase-moments},
Taylor's theorem, phase-point orthogonality, and Minkowski's
inequality gives
\par\endgroup
\[
\begin{aligned}
 &\|{F_{\rho_H}}-{F_{\rho_H}^P}\|_\infty^2
 \leq \operatorname{Tr}[({F_{\rho_H}}-{F_{\rho_H}^P})^2]\\
 &\quad=\frac1{d^2}\sum_{v,v'}
       \operatorname{Tr}(A_vA_{v'})\\
 &\qquad\quad{}\times
       [h(x_v)-p_r(x_v)]
       [h(x_{v'})-p_r(x_{v'})]\\
 &\quad=\frac1d\sum_v|h(x_v)-p_r(x_v)|^2.
\end{aligned}
\]
Here we used $\|X\|_\infty^2\leq\operatorname{Tr}(X^2)$
for Hermitian $X$ and phase-point orthogonality
$\operatorname{Tr}(A_vA_{v'})=d\,\delta_{v,v'}$. Consequently,
\[
\begin{aligned}
 &\left(\mathbb E_H
   \|{F_{\rho_H}}-{F_{\rho_H}^P}\|_\infty^m\right)^{1/m}\\
 &\quad\leq\frac1{\sqrt d}
   \left\|\sum_v|h(x_v)-p_r(x_v)|^2
   \right\|_{L_{m/2}(H)}^{1/2}\\
 &\quad\leq\frac1{\sqrt d}
   \left(\sum_v
   \|h(x_v)-p_r(x_v)\|_{L_m(H)}^2\right)^{1/2}\\
 &\quad\leq\sqrt d\,\max_v
   \|h(x_v)-p_r(x_v)\|_{L_m(H)}\\
 &\quad\leq\sqrt d\left(2eC\sqrt{m/r}\right)^r\\
 &\quad\leq\sqrt d\,2^{-r}=o(1).
\end{aligned}
\]
The second inequality in the latter display follows from
Minkowski's inequality in $L_{m/2}$, valid since $m\geq2$,
and the next inequality uses the $d^2$ phase points.

\begingroup
Combining this approximation estimate with
Eq.~\eqref{eq:bounded-haar-trace-moment} gives
\[
 \begin{aligned}
 \bigl\|\|{F_{\rho_H}^P}\|_\infty\bigr\|_{L_m(H)}
 &\leq\bigl(\mathbb E_H\Tr {F_{\rho_H}}^m\bigr)^{1/m}+o(1)\\
 &\leq\sqrt3\,{K_L}+o(1)\leq C_P,
 \end{aligned}
\]
for a constant $C_P$ independent of $N$. In particular,
$\mathbb E_H\Tr[({F_{\rho_H}^P})^m]\leq dC_P^m$.
\par\endgroup

Write $p_r(x)=\sum_{j<r}\beta_jx^j$ and set
\[
\mathcal A_r
:=\sum_{j<r}|\beta_j|(2d)^j
\leq r(1+4d)^r=\exp[O(N^2)].
\]
Since
$x_v=\operatorname{Tr}[(\sqrt{dK}A_v-{\sqrt{c}}I)\rho]$
and $\|\sqrt{dK}A_v-{\sqrt{c}}I\|_\infty\leq2d$,
the moment condition gives, for arbitrary phase-point labels,
\[
\begin{aligned}
	&\left|
	\mathbb E_C\prod_{a=1}^{q}x_{v_a}
	-\mathbb E_H\prod_{a=1}^{q}x_{v_a}
	\right|\\
	&\qquad\leq\delta_N(2d)^q,
	\qquad q\leq t_N.
\end{aligned}
\]

Expanding $\operatorname{Tr}[(F_\rho^P)^m]$ and applying
this estimate term by term, the $d^{2m}$ phase-label tuples,
the prefactor $d^{-m}$, and
$|\operatorname{Tr}(A_{v_1}\cdots A_{v_m})|\leq d$
give
\[
\begin{aligned}
	&\mathbb E_C\operatorname{Tr}[({F_{\rho_C}^P})^m]\\
	&\quad\leq
	\mathbb E_H\operatorname{Tr}[({F_{\rho_H}^P})^m]
	+\delta_Nd^{m+1}\mathcal A_r^m\\
	&\quad\leq dC_P^m+\exp[-N^5+O(N^3)].
\end{aligned}
\]
Here the maximal coordinate degree is
$m(r-1)=O(N^2)\leq t_N$.

For the scalar signal, observe that
\[
S_\rho^P
=\frac1{{\sqrt{c}}d^2}\sum_v(x_v+{\sqrt{c}})p_r(x_v).
\]
\begingroup
By Eq.~\eqref{eq:bounded-haar-mean}
and the variance bound proved in Step~1,
$\|{S_{\rho_H}}-s(c)\|_{L_2(H)}=o(1)$. Since $\rho_H$ is a state and
$m\geq2$, the preceding approximation bound gives
\[
 \begin{aligned}
 &\|{S_{\rho_H}^P}-s(c)\|_{L_2(H)}\\
 &\quad\leq\|{S_{\rho_H}}-s(c)\|_{L_2(H)}\\
 &\qquad+\bigl\|\|{F_{\rho_H}^P}-{F_{\rho_H}}\|_\infty\bigr\|_{L_m(H)}\\
 &\quad=o(1).
 \end{aligned}
\]
The polynomial $(S_\rho^P-s(c))^2$ has coordinate degree at most
$2r$.
\par\endgroup Applying the same mixed-moment estimate
to its expansion yields
\begingroup
\[
    \begin{aligned}
        &\mathbb E_C({S_{\rho_C}^P}-s(c))^2\\
        &\quad\leq\mathbb E_H({S_{\rho_H}^P}-s(c))^2\\
        &\qquad+\delta_N
        \left(\frac{2d+\sqrt{c}}{\sqrt{c}}\mathcal A_r
        +|s(c)|\right)^2\\
        &\quad\leq o(1)+\exp[-N^5+O(N^2)]=o(1),
    \end{aligned}
\]
\par\endgroup
where we used ${\sqrt{c}}\geq3^{-L/2}$, $|{s(c)}|=O_L(1)$,
and $2r=O(N)\leq t_N$.

\begingroup
For any fixed $M>\max\{1,\sqrt3\,C_P\}$, Markov's inequality gives
\[
 \Pr_C\{\|{F_{\rho_C}^P}\|_\infty>M\}
 \leq d(C_P/M)^m+o(1)=o(1).
\]
Since $s(c)\leq-2\delta$, Chebyshev's inequality gives
\[
 \begin{aligned}
 \Pr_C\{{S_{\rho_C}^P}>-3\delta/2\}
 &\leq\frac4{\delta^2}\mathbb E_C({S_{\rho_C}^P}-s(c))^2\\
 &=o(1).
 \end{aligned}
\]
It remains to remove the polynomial approximation.
\par\endgroup

Finally, Eq.~\eqref{eq:residual-circuit-phase-event} gives
\[
\max_v|x_v|\leq B\sqrt N
\]
with probability tending to one. On this event
\[
\begin{aligned}
	&\|{F_{\rho_C}}-{F_{\rho_C}^P}\|_\infty\\
	&\quad\leq\sqrt d\,\max_v|h(x_v)-p_r(x_v)|\\
	&\quad\leq\sqrt d\,\frac{(2B\sqrt N)^r}{r!}\\
	&\quad\leq\sqrt d
	\left(\frac{2eB\sqrt N}{r}\right)^r
	=o(1),
\end{aligned}
\]
where the last step uses $d=3^N$ and $r=\Theta(N)$.

Recall that
\[
\begin{aligned}
	{S_{\rho_C}}&=\operatorname{Tr}(\rho_C{F_{\rho_C}}),\\
	{S_{\rho_C}^P}&=\operatorname{Tr}(\rho_C{F_{\rho_C}^P}).
\end{aligned}
\]
and hence
\[
\begin{aligned}
	|{S_{\rho_C}}-{S_{\rho_C}^P}|&=
	\left|\operatorname{Tr}
	[\rho_C({F_{\rho_C}}-{F_{\rho_C}^P})]\right|\\
	&\leq\|{F_{\rho_C}}-{F_{\rho_C}^P}\|_\infty=o(1).
\end{aligned}
\]
\begingroup
For all sufficiently large $N$, this error is at most
$\min\{1,\delta/2\}$ on the coordinate event.
Intersecting with the previously established events
$\|{F_{\rho_C}^P}\|_\infty\leq M$ and ${S_{\rho_C}^P}\leq-3\delta/2$, we obtain
\[
 \begin{aligned}
 \|{F_{\rho_C}}\|_\infty&\leq M+1=:M_L,\\
 \Tr(\rho_C{F_{\rho_C}})&={S_{\rho_C}}\leq-\frac{3\delta}{2}
                         +\frac{\delta}{2}=-\delta
 \end{aligned}
\]
with probability tending to one.
\par\endgroup
\end{proof}

\endgroup

\begingroup

\section{\textbf{Conversion between pure magic states}}
\label{app:pure-currency}

We establish the pure-state conversion rates and the comparison
bounds used to complete Theorem~\ref{thm:typical}.

\begin{lemma}[\textbf{Positive finite conversion rates}]

\label{lem:pure-currency-conversions}
Let $\phi$ be {a pure magic state} on a finite number
of qutrits. Then
\begin{equation}
 \begin{gathered}
 0<D_{\CPWP}^{\Strange}(\phi)
 \leq C_{\CPWP}^{\Strange}(\phi)<\infty,\\
 0<D_{\CPWP}^{\phi}(\Strange)
 \leq C_{\CPWP}^{\phi}(\Strange)<\infty.
 \end{gathered}
 \label{eq:pure-currency-positive-rates}
\end{equation}
\end{lemma}

\begin{proof}

Let $d_\phi$ be the dimension of $\phi$. By the discrete Hudson
theorem~\cite{Gross2006}, some phase point satisfies
$\Tr(A_{v_0}\phi)=-\zeta<0$. Define
\[
E=\frac{(d_\phi+1)I-A_{v_0}}{d_\phi+2}.
\]
Since $A_{v_0}^2=I$, this is an effect. Phase-point orthogonality
and $\Tr A_v=1$ give
\begin{equation}
 \begin{aligned}
 p:=\Tr(\phi E)&=\frac{d_\phi+1+\zeta}{d_\phi+2},\\
 \Tr(A_vE)&=\frac{d_\phi+1-d_\phi\delta_{v,v_0}}{d_\phi+2}.
 \end{aligned}
 \label{eq:pure-currency-test}
\end{equation}
Thus every phase-point response lies in $[0,z]$, where
$z=(d_\phi+1)/(d_\phi+2)<p$.
Proposition~\ref{prop:cpwp-distillation} implies
\[
D_{\CPWP}^{\Strange}(\phi)
\geq\frac{d_{\mathrm{bin}}(p\|z)}{\ln(5/3)}>0.
\]
Since $\phi$ is pure, the formation bound
Eq.~\eqref{eq:cpwp-cost} gives
\[
C_{\CPWP}^{\Strange}(\phi)
\leq\frac{\log_2d_\phi}{2\log_2(5/3)}<\infty.
\]
Together with $D\leq C$, these bounds prove the first line of
Eq.~\eqref{eq:pure-currency-positive-rates}.
Reparametrizing the same asymptotic conversion tasks gives
\begin{equation}
 \begin{aligned}
 C_{\CPWP}^{\phi}(\Strange)
 &=\frac1{D_{\CPWP}^{\Strange}(\phi)},\\
 D_{\CPWP}^{\phi}(\Strange)
 &=\frac1{C_{\CPWP}^{\Strange}(\phi)}.
 \end{aligned}
 \label{eq:pure-currency-reciprocals}
\end{equation}
The second line follows.
\end{proof}
We now examine the relation between $\eta_{\CPWP}^{\phi}(\rho)$ and
$\eta_{\CPWP}^{\Strange}(\rho)$.

Concatenating conversion protocols gives
$R_{\CPWP}(\omega\to\tau)\geq
R_{\CPWP}(\omega\to\sigma)R_{\CPWP}(\sigma\to\tau)$~\cite[Eq.~(53)]{KuroiwaYamasaki2020},
which, applied to paths through $\Strange$ and through $\rho$, yields
\begin{equation}
 \begin{aligned}
  C_{\CPWP}^{\phi}(\Strange)D_{\CPWP}^{\Strange}(\rho)
  &\leq C_{\CPWP}^{\phi}(\rho)\\
  &\leq C_{\CPWP}^{\phi}(\Strange)C_{\CPWP}^{\Strange}(\rho),\\
  D_{\CPWP}^{\phi}(\Strange)D_{\CPWP}^{\Strange}(\rho)
  &\leq D_{\CPWP}^{\phi}(\rho)\\
  &\leq D_{\CPWP}^{\phi}(\Strange)C_{\CPWP}^{\Strange}(\rho).
 \end{aligned}
 \label{eq:currency-rate-sandwiches}
\end{equation}
Dividing the lower (upper) bound on the distillable yield by the
upper (lower) bound gives
\begin{equation}
	\begin{aligned}
		\eta_{\CPWP}^{\phi}(\Strange)
		\eta_{\CPWP}^{\Strange}(\rho)
		&\leq \eta_{\CPWP}^{\phi}(\rho)\\
		&\leq
		\frac{\eta_{\CPWP}^{\phi}(\Strange)}
		{\eta_{\CPWP}^{\Strange}(\rho)}.
	\end{aligned}
	\label{eq:currency-efficiency-sandwich}
\end{equation}
\begin{lemma}[\textbf{Strict gap for pure magic currencies}]

\label{lem:pure-currency-strict-gap}
Let $\phi$ be {a pure magic state} on finitely many
qutrits. Define its mana by
$\mathcal M(\phi)=\log_2\|\phi\|_{W,1}$, where
$\|X\|_{W,1}=\sum_v|W_X(v)|$. Then
\begin{equation}
 \mathcal M(\phi)<\frac{\nu(\phi)}{2}\log_2 3.
 \label{eq:pure-currency-strict-mana}
\end{equation}
{Consequently, with the same constant $\alpha$,}
\begin{equation}
 \begin{aligned}
 D_{\CPWP}^{\Strange}(\phi)
 &\leq\frac{\mathcal M(\phi)}{\log_2(5/3)}
 <\alpha\nu(\phi),\\
 C_{\CPWP}^{\phi}(\Strange)
 &>\frac1{\alpha\nu(\phi)}.
 \end{aligned}
 \label{eq:pure-currency-strict-gap}
\end{equation}
\end{lemma}

\begin{proof}

By magic compression, as in Eq.~\eqref{eq:currency-decoding},
a Clifford unitary maps $\phi$ to a tensor product of pure
stabilizer ancillas and a pure state $\widetilde\phi$ on $\nu(\phi)$
qutrits. Clifford conjugation permutes the Wigner coefficients,
and a stabilizer ancilla has Wigner trace norm one. Thus
$\mathcal M(\phi)=\mathcal M(\widetilde\phi)$.
Set ${d_{\widetilde\phi}}=3^{\nu(\phi)}>1$. Phase-point orthogonality and purity give
\[
 \sum_v W_{\widetilde\phi}(v)^2=\frac1{{d_{\widetilde\phi}}},
 \qquad
 \|\widetilde\phi\|_{W,1}\leq\sqrt{{d_{\widetilde\phi}}}.
\]
The second inequality is Cauchy--Schwarz over the ${d_{\widetilde\phi}}^2$
phase points. We show that equality is impossible.

Suppose equality held. Then every Wigner coefficient would
have the same absolute value, so
\[
 W_{\widetilde\phi}(v)=s_v{d_{\widetilde\phi}}^{-3/2},
 \qquad s_v\in\{-1,1\}.
\]
Using $\widetilde\phi=\sum_vW_{\widetilde\phi}(v)A_v$
and $\sum_vA_v={d_{\widetilde\phi}} I$, define
\begin{equation}
 B:=\sum_{v:s_v=-1}A_v
 =\frac{{d_{\widetilde\phi}} I-{d_{\widetilde\phi}}^{3/2}\widetilde\phi}{2}.
 \label{eq:flat-wigner-algebraic-obstruction}
\end{equation}
All entries of $A_v$ lie in
$\mathbb Z[{\omega_3}]$, where ${\omega_3}=e^{2\pi i/3}$.
The same therefore holds for $B$.
Thus $B$ has algebraic-integer entries, and all its
eigenvalues are algebraic integers. However,
Eq.~\eqref{eq:flat-wigner-algebraic-obstruction} shows that
$B$ has the rational eigenvalue ${d_{\widetilde\phi}}/2$ on the
nonzero subspace orthogonal to $\widetilde\phi$.
A rational algebraic integer must be an integer, whereas
${d_{\widetilde\phi}}/2$ is not an integer because ${d_{\widetilde\phi}}$ is odd.
This contradiction proves
$\|\widetilde\phi\|_{W,1}<\sqrt{{d_{\widetilde\phi}}}$ and hence
Eq.~\eqref{eq:pure-currency-strict-mana}.

Applying the argument of \cite[Proposition~21]{Wang2019}
to the Strange target, using the additivity of min-thauma
and $\theta_{\min}(\Strange)=\log_2(5/3)$
\cite[Propositions~1--2]{Wang2020}, gives
\[
D_{\CPWP}^{\Strange}(\phi)
\leq \frac{\theta_{\max}(\phi)}{\log_2(5/3)}
\leq \frac{\mathcal M(\phi)}{\log_2(5/3)}.
\]
Here $\theta_{\max}$ denotes the max-thauma, and the
second inequality follows from its definition.
Combining this bound with
Eq.~\eqref{eq:pure-currency-strict-mana} and
Eq.~\eqref{eq:pure-currency-reciprocals}
proves Eq.~\eqref{eq:pure-currency-strict-gap}.
\end{proof}

\par\endgroup

\bibliography{references}

\end{document}